\documentclass[twocolumn,eqsecnum,nofootinbib,showpacs,preprintnumbers,widetext]{revtex4}

\usepackage{amsmath}
\usepackage{amssymb}

\usepackage{latexsym}
\usepackage{amsmath,amsfonts}
\usepackage{times}
\newcommand{\rf}[1]{(\ref{#1})}

\newcommand{\mc}{\multicolumn}

\def\be{\begin{equation}}
\def\ee{\end{equation}}
\def\beq{\begin{eqnarray}}
\def\eeq{\end{eqnarray}}

\def\parline{\,\partial\kern -0.55em /\,\,}

\def\half{{\frac{1}{2}}}

\def\DD{{\cal D}}

\def\LL{{\cal L}}

\def\PP{{\cal P}}

\def\TT{{\cal T}}

\def\phik{|\phi\rangle}
\def\phibr{\langle\phi|}

\def\xik{|\xi\rangle}

\def\xibr{\langle\xi|}

\def\mubf{{\boldsymbol{\mu}}}

\def\smponetwo{{\scriptscriptstyle [1,2]}}

\def\Iwt{\widetilde{I}}

\def\ewt{\widetilde{e}}
\def\cwt{\widetilde{c}}

\def\alpar{\alpha\partial}
\def\albpar{\bar\alpha\partial}

\def\Cb{\bar{C}}

\def\eb{\bar{e}}

\def\Cbf{{\bf C}}
\def\Obf{{\bf O}}

\def\eff{{\rm eff}}
\def\tr{{\rm tr}}
\def\sh{{\rm sh}}
\def\st{{\rm st}}
\def\cur{{\rm cur}}

\def\sigg{u}

\begin{document}

\preprint{arXiv: 0907.4678 [hep-th]; FIAN-TD-2009-12}

\title{  Gauge invariant two-point vertices of shadow fields,
AdS/CFT, and conformal fields }

\author{ R.R. Metsaev}

\email{metsaev@lpi.ru}

\affiliation{ Department of Theoretical Physics, P.N. Lebedev Physical
Institute, Leninsky prospect 53,  Moscow 119991, Russia}

%\vspace*{10mm}
\begin{abstract}
In the framework of gauge invariant Stueckelberg approach, totally symmetric
arbitrary spin shadow fields in flat space-time of dimension greater than or
equal to four are studied. Gauge invariant two-point vertices for such shadow
fields are obtained. We demonstrate that, in Stueckelberg gauge frame, these
gauge invariant vertices become the standard two-point vertices of CFT.
Light-cone gauge two-point vertices of the shadow fields are also obtained.
AdS/CFT correspondence for the shadow fields and the non-normalizable
solutions of free massless totally symmetric arbitrary spin AdS fields is
studied. AdS fields are considered in a modified de Donder gauge and this
simplifies considerably the study of AdS/CFT correspondence. We demonstrate
that the bulk action, when it is evaluated on solution of the Dirichlet
problem, leads to the two-point gauge invariant vertex of shadow field. Also
we shown that the bulk action evaluated on solution of the Dirichlet problem
leads to new description of conformal fields. The new description involves
Stueckelberg gauge symmetries and gives simple higher-derivative Lagrangian
for the conformal arbitrary spin field. In the Stueckelberg gauge frame, our
Lagrangian becomes the standard Lagrangian of conformal field. Light-cone
gauge Lagrangian of the arbitrary spin conformal field is also obtained.
\end{abstract}


\maketitle

\section{ Introduction}

The present paper is a sequel to our paper \cite{Metsaev:2008fs} where gauge
invariant approach to $CFT$ was developed. Brief review of our results in
Ref.\cite{Metsaev:2008fs} may be found in Sec. \ref{sec03} in this paper. In
space-time of dimension $d\geq 4$, fields of $CFT$ can be separated into two
groups: conformal currents and shadow fields. This is to say that field
having Lorentz algebra spin $s$ and conformal dimension $\Delta = s+d-2$,
is referred to as conformal current with canonical dimension,%
\footnote{We note that conformal currents with $s=1$, $\Delta= d-1$ and
$s=2$, $\Delta = d$, correspond to conserved vector current and conserved
traceless rank-2 tensor field (energy-momentum tensor) respectively.
Conserved conformal currents can be built from massless scalar, spinor and
spin-1 fields (see e.g. \cite{Konstein:2000bi}). Discussion of higher-spin
conformal conserved charges bilinear in $4d$ massless fields of arbitrary
spins may be found in \cite{Gelfond:2006be}.}
while field having Lorentz algebra spin $s$ and dual conformal dimension
$\Delta = 2-s$ is referred to as shadow field.%
\footnote{ It is the shadow fields that are used to discuss conformal
invariant equations of motion and Lagrangian formulations (see e.g. Refs.
\cite{Fradkin:1985am}-\cite{Boulanger:2001he}). Discussion of equations for
mixed-symmetry conformal fields with discrete $\Delta$ may be found in
\cite{Shaynkman:2004vu}.}
We remind that in the framework of $AdS/CFT$ correspondence
\cite{Maldacena:1997re}, the conformal currents and shadow fields manifest
themselves in two related ways at least. First, the conformal currents appear
as boundary values of {\it normalazible} solutions of equations of motion for
bulk fields of $AdS$ supergravity theories, while the shadow fields appear as
boundary values of {\it non-normalazible} solutions of equations of motion
for bulk fields of $AdS$ supergravity theories (see
e.g. \cite{Balasubramanian:1998sn}-\cite{Metsaev:2005ws}%
\footnote{In earlier literature, discussion of shadow field dualities may be
found in \cite{Petkou:1994ad,Petkou:1996jc}.}).
Second, the conformal currents, which are dual to string theory states, can
be built in terms of fields of supersymmetric Yang-Mills (SYM) theory. In
view of these relations to suprgravity/superstring in $AdS$ background and
SYM theory we think that various alternative formulations of the conformal
currents and shadow fields will be useful to understand string/gauge theory
dualities better.

In our approach, starting with the field content of the standard formulation
of currents (and shadow fields), we introduce additional field degrees of
freedom (D.o.F), i.e., we extend space of fields entering the standard
conformal field theory. We note that these additional field D.o.F are similar
to the ones used in the gauge invariant Stueckelberg formulation of massive
fields. Therefore such additional field D.o.F are referred to as Stueckelberg
fields. As is well known, the Stueckelberg approach turned out be successful
for study of theories involving massive fields. This is to say that all
covariant formulations of string theories are realized by using Stueckelberg
gauge symmetries. Therefore we expect that use of the Stueckelberg fields in
$CFT$ might be useful for the study of various aspects of $AdS/CFT$
dualities.

In this paper we develop further our approach initiated in
Ref.\cite{Metsaev:2008fs}. As in Ref.\cite{Metsaev:2008fs}, we discuss
bosonic arbitrary spin conformal currents and shadow fields in space-time of
dimension $d \geq 4$. Our results in this paper can be summarized as follows.

i) Using shadow field gauge symmetries found in Ref.\cite{Metsaev:2008fs}, we
obtain the two-point gauge invariant vertex for the arbitrary spin-$s$ shadow
field. Imposing some gauge condition, which we refer to as Stueckelberg
gauge, we demonstrate that our vertex is reduced to the two-point vertex
appearing in the standard approach to $CFT$. Imposing light-cone gauge on the
shadow field, we obtain light-cone gauge fixed two-point vertex. As usually,
a kernel of our shadow field vertex gives a correlation function of the
conformal current.

 ii) We study $AdS/CFT$ correspondence for massless arbitrary spin-$s$
 $AdS$ field and boundary spin-$s$ shadow field. Namely, using the modified de
 Donder gauge condition for $AdS$ field, we demonstrate that the two-point
 gauge invariant vertex of the shadow field does indeed emerge from massless
 $AdS$ field action when it is evaluated on solution of the Dirichlet
 problem. $AdS$ field action evaluated on solution of
 the Dirichlet problem will be referred to as effective action in this paper.
 We show that use of the modified de Donder gauge provides considerable
 simplification when computing the effective action.

 iii) We show that the effective action of $AdS$ massless field leads to new
 interesting description of conformal field. As compared to the standard
 approaches to conformal fields \cite{Fradkin:1985am,Segal:2002gd}, our
 approach involves additional field D.o.F. and the respective additional gauge
 symmetries which are realized as the Stueckelberg gauge symmetries. We obtain
 very simple higher-derivative Lagrangian for conformal arbitrary spin field.
 Using the Stueckelberg gauge frame, we demonstrate that our Lagrangian is reduced
 to the standard Lagrangian of the conformal field. We also obtain light-cone
 gauge fixed Lagrangian of the conformal field.

The rest of the paper  is organized as follows.

In Sec. \ref{sec02},  we summarize the notation used in this paper and review
the standard approach to the conformal currents and shadow fields. Also we
briefly review the gauge invariant approach developed in
Ref.\cite{Metsaev:2008fs}.

In Sec. \ref{man02-sec-02x}, we  start with  the examples of low-spin,
$s=1,2$, shadow fields. For these shadow fields, we obtain the gauge
invariant two-point vertices. Using the Stueckelberg gauge frame, we show how
our gauge invariant vertices are related to the vertices appearing in the
standard approach to $CFT$. Light-cone gauge fixed vertices are also
obtained.

Section \ref{man02-sec04} is devoted to the study of the two-point vertex for
the arbitrary spin-$s$ shadow field. In this section, we generalize results
obtained in Sec. \ref{man02-sec-02x} to the case of the arbitrary spin shadow
field.

In Sec. \ref{secAdS/CFT}, we study $AdS/CFT$ correspondence for low-spin
$AdS$ massless fields and boundary shadow fields. One of remarkable features
of the modified de Donder gauge is that the computation of the effective
action for massless arbitrary spin-$s$, $s\geq 1$, $AdS$ field subject to the
modified de Donder gauge is similar to the computation of the effective
action for a massive scalar $AdS$ field. Therefore we begin with brief review
of the computation of the effective action for the massive scalar field.
After that we proceed with the discussion of the effective actions for the
massless spin $s=1,2$, $AdS$ fields. We demonstrate that these effective
actions coincide with the respective gauge invariant two-point vertices for
the spin $s=1,2$ shadow fields.

Section \ref{man02-sec-06} is devoted to the study of $AdS/CFT$
correspondence for massless arbitrary spin-$s$ $AdS$ field and boundary
arbitrary spin-$s$ shadow field. In this section we generalize results
obtained in Sec. \ref{secAdS/CFT} to the case of arbitrary spin fields.

In Sec. \ref{man02-sec-07}, we deal with conformal fields. We  start with the
examples of low-spin, $s=1,2$, conformal fields. For these fields, we discuss
our new Lagrangian and show how this Lagrangian, taken in the Stueckelberg
gauge frame, is reduced to the standard Lagrangian. Light-cone gauge fixed
Lagrangian is also obtained. After that we discuss generalization of these
results to the case of arbitrary spin-$s$ conformal field.

Section \ref{conl-sec-01} summarizes our conclusions and suggests directions
for future research.

We collect various technical details in five appendices. In Appendix
\ref{man02-app-01}, we study restrictions imposed on the shadow field
two-point vertex by the Poincar\'e algebra symmetries, dilatation symmetry,
and the shadow field gauge symmetries. We demonstrate that these restrictions
allow us to determine the vertex uniquely. Invariance of the two-point gauge
invariant vertex under the conformal boost transformations is demonstrated in
Appendix \ref{app-04082009-01}. In Appendix \ref{man02-app-04}, we present
details of the derivation of the effective action. In Appendix
\ref{man02-app-02}, we derive $CFT$ adapted Lagrangian for massless spin-1
and spin-2 fields in $AdS_{d+1}$. In Appendix \ref{man02-app-03}, we discuss
some details of the derivation of normalization factor in the Dirichlet
problem.

\section{Preliminaries}
\label{sec02}

\subsection{Notation}

Our conventions are as follows. $x^a$ denotes coordinates in $d$-dimensional
flat space-time, while $\partial_a$ denotes derivatives with respect to
$x^a$, $\partial_a \equiv \partial /
\partial x^a$. Vector indices of the Lorentz algebra $so(d-1,1)$ take
the values $a,b,c,e=0,1,\ldots ,d-1$. We use mostly positive flat metric
tensor $\eta^{ab}$. To simplify our expressions we drop $\eta_{ab}$ in scalar
products, i.e., we use $X^aY^a \equiv \eta_{ab}X^a Y^b$.

We use a set of the creation operators $\alpha^a$, $\alpha^z$, and the
respective set of annihilation operators $\bar{\alpha}^a$, $\bar{\alpha}^z$,
\beq
\label{04092008-03} && [\bar{\alpha}^a,\alpha^b]=\eta^{ab}\,, \qquad
[\bar\alpha^z,\alpha^z]=1\,,
\\[5pt]
\label{04092008-04} && \bar\alpha^a |0\rangle = 0\,,\qquad  \bar\alpha^z
|0\rangle = 0\,,
\\[5pt]
\label{04092008-05} && \alpha^{a\dagger} = \bar\alpha^a\,, \qquad
\alpha^{z\dagger} = \bar\alpha^z \,.
\eeq
These operators will often be referred to as oscillators in what follows.%
\footnote{ We use oscillator formulation to handle the many indices appearing
for tensor fields (for recent discussion of oscillator formulation see
\cite{Boulanger:2008up}.) In a proper way, oscillators arise in the framework
of world-line approach to higher-spin fields (see e.g.
\cite{Bastianelli:2007pv,Cherney:2009mf}).}
The oscillators $\alpha^a$, $\bar\alpha^a$ and $\alpha^z$, $\bar\alpha^z$,
transform in the respective vector and scalar representations of the
$so(d-1,1)$ Lorentz algebra. Throughout this paper we use operators
constructed out of the derivatives and the oscillators,
\beq
& \Box=\partial^a\partial^a\,,\quad \alpha\partial
=\alpha^a\partial^a\,,\quad \bar\alpha\partial =\bar\alpha^a\partial^a\,, &
\\[3pt]
& \alpha^2 = \alpha^a\alpha^a\,,\qquad \bar\alpha^2 =
\bar\alpha^a\bar\alpha^a\,, &
\\[3pt]
\label{man02-25072009-01} & N_\alpha \equiv \alpha^a \bar\alpha^a  \,,
\qquad
N_z \equiv \alpha^z \bar\alpha^z \,, &
\\[5pt]
\label{18052008-08} & \Pi^\smponetwo \equiv 1 -\alpha^2\frac{1}{2(2N_\alpha
+d)}\bar\alpha^2\,. &
\eeq

\subsection{Global conformal symmetries }

In $d$-dimensional flat space-time, the conformal algebra $so(d,2)$ consists
of translation generators $P^a$, dilatation generator $D$, conformal boost
generators $K^a$, and generators of the $so(d-1,1)$ Lorentz algebra $J^{ab}$.
We assume the following normalization for commutators of the conformal
algebra:
\beq
\label{ppkk}
&& {}[D,P^a]=-P^a\,, \hspace{0.5cm}  {}[P^a,J^{bc}]=\eta^{ab}P^c
-\eta^{ac}P^b, \qquad
\\
&& [D,K^a]=K^a\,, \hspace{0.7cm} [K^a,J^{bc}]=\eta^{ab}K^c - \eta^{ac}K^b,
\qquad
\\[5pt]
\label{pkjj} && \hspace{1.5cm} {}[P^a,K^b]=\eta^{ab}D - J^{ab}\,,
\\
&& \hspace{1.5cm} [J^{ab},J^{ce}]=\eta^{bc}J^{ae}+3\hbox{ terms} \,.
\eeq

Let $\phik$ denotes conformal current (or shadow field) in flat space-time of
dimension $d\geq 4$. Under conformal algebra symmetries the $\phik$
transforms as
\be \label{04092008-01} \delta_{\hat{G}} \phik  = \hat{G} \phik \,, \ee
where realization of the conformal algebra generators $\hat{G}$ in
terms of differential operators takes the form
\beq
\label{conalggenlis01} && P^a = \partial^a \,,
\\[3pt]
\label{conalggenlis02} && J^{ab} = x^a\partial^b -  x^b\partial^a + M^{ab}\,,
\\[3pt]
\label{conalggenlis03} && D = x\partial  + \Delta\,,
\\[3pt]
\label{conalggenlis04} && K^a = K_{\Delta,M}^a + R^a\,,
\eeq
and we use the notation
\beq
\label{kdelmdef01} && K_{\Delta,M}^a \equiv -\frac{1}{2}x^2\partial^a + x^a D
+ M^{ab}x^b\,,
\eeq
\be x\partial \equiv x^a \partial^a \,, \qquad x^2 = x^a x^a\,.\ee
In \rf{conalggenlis02}-\rf{conalggenlis04}, $\Delta$ is operator of conformal
dimension, $M^{ab}$ is spin operator of the Lorentz algebra,
\be  [M^{ab},M^{ce}]=\eta^{bc}M^{ae}+3\hbox{ terms} \,. \ee
For arbitrary spin conformal currents and shadow field studied in this paper,
oscillator representation of the $M^{ab}$ takes the form
\be \label{mabdef0001} M^{ab} \equiv \alpha^a \bar\alpha^b -
\alpha^b\bar\alpha^a\,.\ee
$R^a$ is operator depending, in general, on derivatives with respect to
space-time coordinates%
\footnote{For the conformal currents and shadow fields studied in this paper,
the operator $R^a$ does not depend on derivatives. Dependence on derivatives
of $R^a$ appears e.g., in ordinary-derivative approach to conformal fields
\cite{Metsaev:2007fq}.}
and not depending on space-time coordinates $x^a$, $[P^a,R^b]=0$. In the
standard formulation of the conformal currents and shadow fields, the
operator $R^a$ is equal to zero, while, in the gauge invariant approach, the
operator $R^a$ turns out be nontrivial \cite{Metsaev:2008fs}.

\subsection{ Standard approach to conformal currents and shadow fields}

We begin with brief review of the standard approach to conformal currents and
shadow fields. To keep our presentation as simple as possible we restrict our
attention to the case of arbitrary spin {\it totally symmetric} conformal
currents and shadow fields which have the appropriate canonical conformal
dimensions given below. In this section we recall main facts of conformal
field theory about these currents and shadow fields.

{\bf Conformal current with the canonical conformal dimension}. Consider
totally symmetric rank-$s$ tensor field $T^{a_1\ldots a_s}$ of the Lorentz
algebra $so(d-1,1)$. The field is referred to as spin-$s$ {\it conformal
current with canonical dimension} if $T^{a_1\ldots a_s}$ satisfies the
constraints
\be  \label{16052008-13} T^{aaa_3\ldots a_s}=0\,, \qquad
\partial^a T^{aa_2\ldots a_s}=0
\ee
and has the conformal dimension\!
\footnote{ The fact that expression in r.h.s. of \rf{candim} is the lowest
energy value of totally symmetric spin-$s$ massless fields propagating in
$AdS_{d+1}$ space was demonstrated in Ref.\cite{Metsaev:1994ys}.
Generalization of relation \rf{candim} to mixed-symmetry fields in $AdS$ may
be found in Ref.\cite{Metsaev:1995re}.}
\be \label{candim} \Delta = s + d - 2\,, \ee
which is referred to as {\it the canonical conformal dimension of spin-$s$
conformal current}. Taking into account that the operator $R^a$ of the
conformal current $T^{a_1\ldots a_s}$ is equal to zero, using the well-known
spin operator $M^{ab}$ of the totally symmetric traceless current
$T^{a_1\ldots a_s}$ and $\Delta$ in \rf{candim}, one can make sure that
constraints \rf{16052008-13} are invariant under conformal algebra
transformations \rf{04092008-01}.

{\bf Shadow field with the canonical conformal dimension}. Consider totally
symmetric rank-$s$ tensor field $\Phi^{a_1\ldots a_s}$ of the Lorentz algebra
$so(d-1,1)$. The field $\Phi^{a_1\ldots a_s}$ is referred to as shadow field
if it meets the following requirements:

{\bf i}) The field $\Phi^{a_1\ldots a_s}$ is traceless,
\be \label{16052008-10} \Phi^{aaa_3\ldots a_s}=0 \,. \ee

{\bf ii}) The field $\Phi^{a_1\ldots a_s}$ transforms under the conformal
algebra symmetries so that the following two point current-shadow field
interaction vertex
\be \label{16052008-12} \LL = \frac{1}{s!}\, \Phi^{a_1\ldots a_s}
T^{a_1\ldots a_s} \ee
is invariant (up to total derivative) under conformal algebra
transformations.

We now note that:
\\
{\bf i}) Conformal dimension of the spin-$s$ shadow field given by
\be\label{candimsh} \Delta = 2 -s\,, \ee
is referred to as {\it the canonical conformal dimension of spin-$s$ shadow
field}. The operator $R^a$ of the shadow field $\Phi^{a_1\ldots a_s}$ is
equal to zero.
\\
{\bf ii})  Divergence-free constraint \rf{16052008-13} and requirement for
the vertex $\LL$ to be invariant imply that the shadow field is defined by
module of gauge transformation
\be \label{16052008-11} \delta \Phi^{a_1 \ldots a_s} = \Pi^\tr
\partial^{(a_1} \xi^{a_2\ldots a_s)}\,, \ee
where $\xi^{a_1\ldots a_{s-1}}$ is traceless parameter of gauge
transformation and the projector $\Pi^\tr$ is inserted to respect
tracelessness constraint \rf{16052008-10}.

\subsection{ Gauge invariant approach to conformal currents and shadow fields
} \label{sec03}

We now briefly review the gauge invariant approach to conformal currents and
shadow fields developed in Ref.\cite{Metsaev:2008fs}. The gauge invariant
approach to the conformal currents and shadow fields can be summarized as
follows.

{\bf i}) To discuss the arbitrary spin-$s$ conformal current and spin-$s$
shadow field we use the respective totally symmetric $so(d-1,1)$ Lorentz
algebra tensor fields $\phi_\cur^{a_1\ldots a_{s'}}$ and $\phi_\sh^{a_1\ldots
a_{s'}}$, where $s'=0,1,\ldots,s$. For $s'\geq 4$, these fields are
restricted to be double-traceless,

\be \label{man02-15072009-02} \phi_\cur^{aabba_5\ldots a_{s'}}=0\,, \qquad
\phi_\sh^{aabba_5\ldots a_{s'}}=0\,.\ee
Conformal dimension of the field $\phi_\cur^{a_1\ldots a_{s'}}$ is equal to
$s'+d-2$, while conformal dimension of the field $\phi_\sh^{a_1\ldots
a_{s'}}$ is equal to $2-s'$.

{\bf ii}) On space of the fields $\phi_\cur^{a_1\ldots a_{s'}}$  (and
separately on space of the fields $\phi_\sh^{a_1\ldots a_{s'}}$), we
introduce new differential constraints, gauge transformations, and conformal
algebra transformations.

{\bf iii}) The new differential constraints are invariant under the gauge
transformations and the conformal algebra transformations.

{\bf iv})  The gauge symmetries and the new differential constraints make it
possible to match our approach and the standard one, i.e., by appropriate
gauge fixing of the Stueckelberg fields and by solving some differential
constraints (Stueckelberg gauge frame) we obtain standard formulation of the
conformal currents and shadow fields.

For the spin-$s$ conformal current, use of the Stueckelberg gauge frame leads
to $\phi_\cur^{a_1\ldots a_{s'}}=0$ for $s'=0,1,\ldots,s-1$ and
divergence-free and tracelessness constraint for $\phi_\cur^{a_1\ldots a_s}$,
\be
\partial^a \phi_\cur^{aa_2\ldots a_s} = 0 \,, \qquad \phi_\cur^{aaa_3\ldots a_s}=
0\,. \ee
We see that, in the Stueckelberg gauge frame, our field $
\phi_\cur^{a_1\ldots a_s}$ can be identified with the current $T^{a_1\ldots
a_s}$ \rf{16052008-13}, i.e. our approach reduces to the standard one.

For the spin-$s$ shadow field, use of the Stueckelberg gauge frame leads to
tracelessness constraint for the field $\phi_\sh^{a_1\ldots a_{s}}$,
\be \phi_\sh^{aaa_3\ldots a_{s}}= 0\,, \ee
and this field is not subject to any differential constraint. Also, the
Stueckelberg gauge frame makes it possible to express the fields
$\phi_\sh^{a_1\ldots a_{s'}}$, $s'=0,1,\ldots,s-1$, in terms of the field
$\phi_\sh^{a_1\ldots a_{s}}$, i.e., we are left with one traceless field
$\phi_\sh^{a_1 \ldots a_s}$, which can be identified with the shadow field
$\Phi^{a_1\ldots a_s}$ of the standard approach to $CFT$.

Summary of the gauge invariant approach to the low-spin, $s=1,2$ conformal
currents and shadow fields is given in Table I.

\newpage
\begin{widetext}
{\sf Table I. In the Table, we present the field contents, conformal
dimensions, differential constraints, gauge transformations and
transformations under operator $R^a$ entering the gauge invariant approach of
the low-spin, $s=1,2$ conformal currents and shadow fields. The operators
$R^a$ enter conformal boost transformations given in \rf{conalggenlis04}.}

{\small
\begin{center}
\begin{tabular}{|c|c|c|l|l|}
\hline &&& & \\[-3mm]
Field  & \ Conf.  & Differential  & \hspace{2cm} Gauge & \hspace{1cm} Action
of
\\
cont.   & dim. & constraint & \hspace{1.5cm} transformation & \hspace{1cm}
operator $R^a$
\\
&&&&
\\ \hline
\mc{5}{|c|}{} \\[-3mm]
\mc{5}{|c|}{{\bf spin-1 current}}
\\ [1mm]\hline
&&&&
\\[-3mm]
$\phi_\cur^a$ & $d-1$ && $\hspace{1cm} \delta \phi_\cur^a =  \partial^a
\xi_\cur$ & $R^a\phi_\cur^b = (2-d)\eta^{ab}\phi_\cur$
\\ [2mm] \cline{1-2} \cline{4-5}
\\[-7mm]
&&$\partial^a \phi_\cur^a + \Box \phi_\cur = 0$ &&
\\
$ \phi_\cur$ & $d-2$ & &  $\hspace{1cm} \delta \phi_\cur =  - \xi_\cur$ &
$R^a\phi_\cur = 0$
\\[-3mm]
&&&&
\\ \hline
\mc{5}{|c|}{} \\[-3mm]
\mc{5}{|c|}{ {\bf spin-1 shadow field }}
\\ [1mm]\hline
&&&&
\\[-3mm]
$\phi_\sh^a$ & $1$ && $\hspace{1cm} \delta \phi_\sh^a =  \partial^a \xi_\sh$
& $ R^a \phi_\sh^b = 0$
\\ [2mm] \cline{1-2} \cline{4-5}
\\[-7mm]
&&$\partial^a \phi_\sh^a + \phi_\sh = 0$ &&
\\
$ \phi_\sh$ & $2$ & &  $\hspace{1cm} \delta \phi_\sh =  - \Box \xi_\sh$ & $
R^a \phi_\sh = (d-2)\phi_\sh^a$
\\[2mm]\hline
\mc{5}{|c|}{} \\[-3mm]
\mc{5}{|c|}{ {\bf spin-2 current }}
\\ [1mm]\hline
&&&&
\\[-2mm]
$\phi_\cur^{ab}$ & $d$ & $ \ \partial^b \phi_\cur^{ab} - \half \partial^a
\phi_\cur^{bb} + \Box \phi_\cur^a = 0$
& $\ \ \delta \phi_\cur^{ab} =\partial^a \xi_{ cur}^b + \partial^b \xi_\cur^a
$ & $R^a \phi_\cur^{bc} = 2\eta^{bc}\phi_\cur^a $
\\[3mm]
&& $ \partial^a \phi_\cur^a + \half \phi_\cur^{aa} + \sigg \Box \phi_\cur = 0
$ &\hspace{1cm} $ + \frac{2}{d-2}\eta^{ab}\Box \xi_\cur $ & $ \hspace{1.2cm}-
d (\eta^{ab} \phi_\cur^c + \eta^{ac}\phi_\cur^b)$
\\[3mm] \cline{1-2} \cline{4-5}
&& $ \sigg \equiv \sqrt{2} \Bigl(\frac{d-1}{d-2}\Bigr)^{1/2} $ &&
\\[-3mm]
$\phi_\cur^a$ & $d-1$ &  & $\ \ \delta \phi_\cur^a = \partial^a \xi_\cur -
\xi_\cur^a  $  & $R^a \phi_\cur^b =  -\sqrt{2(d-1)(d-2)}
\eta^{ab}\phi_\cur^{}$
\\[3mm]\cline{1-2} \cline{4-5}
&&&&
\\[-3mm]
$ \phi_\cur$ & $ d-2 $ & & $\ \ \delta \phi_\cur = - \sigg \xi_\cur $  & $R^a
\phi_\cur  =0$
\\[2mm]\hline
\mc{5}{|c|}{} \\[-3mm]
\mc{5}{|c|}{ {\bf spin-2 shadow field }}
\\ [1mm]\hline
&&&&
\\[-2mm]
$\phi_\sh^{ab}$ & $0$ & $ \ \ \partial^b \phi_\sh^{ab} - \half \partial^a
\phi_\sh^{bb} + \phi_\sh^a = 0$
& $\ \ \delta \phi_\sh^{ab} =\partial^a \xi_\sh^b +
\partial^b \xi_\sh^a + \frac{2}{d-2}\eta^{ab}\xi_\sh$  & $R^a \phi_\sh^{bc}
=0$
\\[3mm] \cline{1-2}\cline{4-5}
&&&&
\\[-3mm]
$\phi_\sh^a$ & $1$ & $ \partial^a \phi_\sh^a + \half \Box \phi_\sh^{aa} +
\sigg \phi_\sh = 0 $  & $\ \ \delta \phi_\sh^a = \partial^a \xi_\sh - \Box
\xi_\sh^a  $  & $R^a \phi_\sh^b =  d\phi_\sh^{ab} -\eta^{ab}\phi_\sh^{cc}$
\\[3mm]\cline{1-2}\cline{4-5}
&&&&
\\[-3mm]
$ \phi_\sh$ & $ 2 $ & $ u \equiv \Bigl(2\frac{d-1}{d-2}\Bigr)^{1/2} $ & $\ \
\delta \phi_\sh = - \sigg \Box \xi_\sh $  & $R^a \phi_\sh  =
\sqrt{2(d-1)(d-2)} \phi_\sh^a$
\\[2mm]\hline
\end{tabular}
\end{center}
}
\end{widetext}

\newpage
{}~\vspace{5cm}

\begin{widetext}
{\sf Table II. In the Table, we present the field contents, conformal
dimensions, differential constraints, gauge transformations and the operators
$R^a$ entering the gauge invariant approach of arbitrary spin-$s$ conformal
currents and shadow fields. The operators $R^a$ enter conformal boost
transformations given in \rf{conalggenlis04}.}

{\small
\begin{center}
\begin{tabular}{|c|c|l|c|l|}
\hline &&&& \\[-3mm]
Field  & Conformal  & \hspace{1.5cm} Differential & \hspace{0.7cm} Gauge &
\hspace{1.5cm} Operator
\\
content   & dimension & \hspace{1.5cm}constraint  & \hspace{0.7cm}
transformation & \hspace{2cm} $R^a$
\\
&&&&
\\ \hline
\mc{5}{|c|}{} \\[-3mm]
\mc{5}{|c|}{{\bf spin-$s$ current}}
\\ [1mm]\hline
&&&&
\\[-2mm]
&  & $ \ \  \bar{C}_\cur|\phi_\cur\rangle  =  0\,, $
& $\hspace{-1.5cm} \delta |\phi_\cur\rangle  =  \Bigl( \alpar - e_{1,\cur} $
& $\ R^a = \bar{r} \Bigl(
\alpha^a-\alpha^2\frac{1}{2N_\alpha+d-2}\bar\alpha^a $
\\[3mm]
&&&&
\\[-3mm]
$|\phi_\cur\rangle$ & $\hspace{-0.5cm} s+d $ & $ \ \ \bar{C}_\cur = \albpar -
\half \alpar \bar\alpha^2 $ & $\hspace{1cm}   - \frac{\alpha^2}{2s + d- 6
-2N_z} \eb_{1,\cur} \Box \Bigr) |\xi_\cur \rangle\,, $ & $ \hspace{0.4cm} +
\alpha^2 \frac{2}{(2N_\alpha+d -2)(2N_\alpha+d)} \bar{C}_\perp^a\Bigr),$
\\[3mm]
&&&&
\\[-3mm]
& $ -2-N_z $ & $ \ \ \ \ - \eb_{1,\cur} \Pi^\smponetwo  \Box +  \half
e_{1,\cur} \bar\alpha^2 $ & $ e_{1,\cur} = \alpha^z \ewt_1\,,\quad
\eb_{1,\cur} =  - \ewt_1 \bar\alpha^z$ & $ \ \bar{r} \equiv -
\sqrt{2s+d-4-N_z}$
\\[2mm]
&&&& $ \hspace{0.5cm}\times \sqrt{2s+d-4-2N_z}\bar\alpha^z $
\\[-1mm]
&&&&
\\ \hline
\mc{5}{|c|}{} \\[-3mm]
\mc{5}{|c|}{{\bf spin-$s$ shadow field}}
\\ [1mm]\hline
&&&&
\\[-2mm]
&  &  $ \ \  \bar{C}_\sh |\phi_\sh \rangle  =  0\,, $
& $\hspace{-1.5cm} \delta |\phi_\sh\rangle  =  \Bigl( \alpar - e_{1,\sh} \Box
$ & $\ R^a  =  r \Bigl( \bar\alpha^a - \alpha^a \frac{1}{2N_\alpha+d }
\bar\alpha^2\Bigr),$
\\[3mm]
&&&&
\\[-3mm]
$|\phi_\sh\rangle$ & $ 2-s+N_z $ & $ \ \ \bar{C}_\sh  = \albpar - \half
\alpar \bar\alpha^2 $ & $\hspace{1cm}   - \frac{\alpha^2}{2s + d- 6
-2N_z}\eb_{1,\sh}\Bigr) |\xi_\sh \rangle\,, $ & $\ r \equiv \alpha^z
\sqrt{2s+d-4-N_z}$
\\[3mm]
&&&&
\\[-3mm]
&  &  $ \ \ \ \ - \eb_{1,\sh} \Pi^\smponetwo +  \half e_{1,\sh} \bar\alpha^2
\Box $ & $ e_{1,\sh} = \alpha^z \ewt_1\,,\quad
\eb_{1,\sh} =  -  \ewt_1 \bar\alpha^z$ & $ \hspace{0.4cm} \times
\sqrt{2s+d-4-2N_z}$
\\ [2mm] \hline
\mc{5}{|c|}{} \\[-2mm]
\mc{5}{|c|}{
$\bar{C}_\perp^a \equiv \bar\alpha^a - \frac{1}{2} \alpha^a \bar\alpha^2$,
\qquad
$\Pi^\smponetwo \equiv 1 -\alpha^2 \frac{1}{2(2N_\alpha
+d)}\bar\alpha^2$,\qquad
$\ewt_1 \equiv \Bigl(\frac{2s+d-4-N_z}{2s+d-4-2N_z}\Bigr)^{1/2} $ }
\\ [2mm]\hline
\end{tabular}
\end{center}
}

\end{widetext}

To simplify presentation of the gauge invariant approach to the arbitrary
spin-$s$ conformal current and shadow field we use oscillators
\rf{04092008-03} and introduce the respective ket-vectors $|\phi_\cur\rangle$
and $|\phi_\sh\rangle$ defined by
\beq
&& \label{phikdef01}   |\phi_\cur\rangle \equiv \sum_{s'=0}^s
\alpha_z^{s-s'}|\phi_{\cur,\, s'}\rangle \,,
\nonumber\\[5pt]
&& |\phi_{\cur,\, s'}\rangle \equiv
\frac{\alpha^{a_1} \ldots \alpha^{a_{s'}}}{s'!\sqrt{(s-s')!}}
\, \phi_\cur^{a_1\ldots a_{s'}} |0\rangle\,,
\eeq
\beq \label{phikdef01sh}
&&  |\phi_\sh\rangle \equiv \sum_{s'=0}^s \alpha_z^{s-s'}|\phi_{\sh,\,
s'}\rangle \,,
\nonumber\\[5pt]
&& |\phi_{\sh,\, s'}\rangle \equiv
\frac{\alpha^{a_1} \ldots \alpha^{a_{s'}}}{s'!\sqrt{(s-s')!}}
\, \phi_\sh^{a_1\ldots a_{s'}} |0\rangle\,.
\eeq

To describe gauge symmetries of the arbitrary spin-$s$ conformal current and
spin-$s$ shadow field we use gauge transformation parameters which are the
respective totally symmetric $so(d-1,1)$ Lorentz algebra tensor fields
$\xi_\cur^{a_1\ldots a_{s'}}$ and $\xi_\sh^{a_1\ldots a_{s'}}$, where
$s'=0,1,\ldots,s-1$. For $s'\geq 2$, these gauge transformation parameters
are restricted to be traceless,
\be \xi_\cur^{aaa_3\ldots a_{s'}}=0\,, \qquad \xi_\sh^{aaa_3\ldots
a_{s'}}=0\,.\ee
Again, to simplify presentation of result, we collect the gauge
transformation parameters in the respective ket-vectors $|\xi_\cur\rangle$
and $|\xi_\sh\rangle$ defined by
\beq \label{man02-22072009-01}
&&  |\xi_\cur\rangle \equiv \sum_{s'=0}^{s-1} \alpha_z^{s-1-s'}|\xi_{\cur,\,
s'}\rangle \,,
\nonumber\\[5pt]
&& |\xi_{\cur,\,s'}\rangle \equiv
\frac{\alpha^{a_1} \ldots \alpha^{a_{s'}}}{s'!\sqrt{(s -1 - s')!}}
\, \xi_\cur^{a_1\ldots a_{s'}} |0\rangle\,,\qquad
\eeq
\beq  \label{man02-22072009-02}
&&  |\xi_\sh\rangle \equiv \sum_{s'=0}^{s-1} \alpha_z^{s-1-s'}|\xi_{\sh,\,
s'}\rangle \,,
\nonumber\\[5pt]
&& |\xi_{\sh,\,s'}\rangle \equiv
\frac{\alpha^{a_1} \ldots \alpha^{a_{s'}}}{s'!\sqrt{(s -1 - s')!}}
\, \xi_\sh^{a_1\ldots a_{s'}} |0\rangle\,.\qquad
\eeq
With these conventions for the ket-vectors, summary of our study of gauge
invariant approach to the arbitrary spin-$s$ conformal current and shadow
field is given in Table II.

%%%%%%%%%%%%%%%%%%%%%%%%%%%%%%%%%%%%%%%%%%%%%%%%%%%%%%%%%%%%%%%%%%%%%%%%%%
%%%%%%%%%%%%%%%%%%%%%%%%%%%%%%%%%%%%%%%%%%%%%%%%%%%%%%%%%%%%%%%%%%%%%%%%%%
\section{Gauge invariant two-point vertex for low-spin shadow
fields}\label{man02-sec-02x}
%%%%%%%%%%%%%%%%%%%%%%%%%%%%%%%%%%%%%%%%%%%%%%%%%%%%%%%%%%%%%%%%%%%%%%%%%%
%%%%%%%%%%%%%%%%%%%%%%%%%%%%%%%%%%%%%%%%%%%%%%%%%%%%%%%%%%%%%%%%%%%%%%%%%%

We now discuss shadow field two-point vertex. In the gauge invariant
approach, the vertex is determined by requiring the vertex to be invariant
under gauge transformations of the shadow field. Also, the vertex should be
invariant under conformal algebra transformations. We consider the two-point
vertices for spin-1, spin-2, and arbitrary spin-$s$ shadow fields in turn.

\subsection{ Gauge invariant two-point vertex for spin-1 shadow
field }\label{man02-sec-02}

We begin with the discussion of the two-point vertex for the spin-1 shadow
field. This simplest example demonstrates all characteristic features of our
approach. In the gauge invariant approach, the spin-1 shadow field is
described by vector field $\phi_\sh^a$ and scalar field $\phi_\sh$ (see Table
I). The gauge invariant two-point vertex we find takes the form
\beq \label{manus2009-02-01}
&& \hspace{-1.2cm}  \Gamma  = \int d^dx_1 d^dx_2 \Gamma_{12} \,,
\\[7pt]
\label{manus2009-02-02} && \hspace{-1.2cm} \Gamma_{12} =
\frac{\phi_\sh^a(x_1) \phi_\sh^a(x_2)}{2|x_{12}|^{2(d-1)}} +
\frac{1}{4(d-2)^2} \frac{\phi_\sh(x_1) \phi_\sh(x_2)}{|x_{12}|^{2(d-2)}},
\eeq
\be \label{manus2009-02-03}
|x_{12}|^2 \equiv x_{12}^a x_{12}^a\,, \qquad x_{12}^a = x_1^a - x_2^a\,.
\ee
One can check that this vertex is invariant under the gauge transformations
of the spin-1 shadow field given in Table I. The vertex is obviously
invariant with respect to the Poincar\'e algebra and dilatation symmetries.
Also, using the operator $R^a$ given in Table I, we check that the vertex is
invariant under the conformal boost transformations.

The kernel of the vertex $\Gamma$ is related to a two-point correlation
function of the spin-$1$ conformal current. In our approach, the spin-$1$
conformal current is described by gauge fields $\phi_\cur^a$, $\phi_\cur$
(see Table I). Therefore, in order to discuss correlation functions of the
fields $\phi_\cur^a$, $\phi_\cur$ in a proper way, we should impose gauge
condition on the fields $\phi_\cur^a$, $\phi_\cur$.%
\footnote{ We note that, in the gauge invariant approach, correlation
functions of the conformal current can be studied without gauge fixing. To do
that one needs to construct gauge invariant field strengths for the gauge
potentials $\phi_\cur^a$, $\phi_\cur$. Study of field strengths for the
conformal currents is beyond the scope of this paper.}
The two-point correlation functions of gauge fixed fields $\phi_\cur^a$,
$\phi_\cur$ are obtained from the kernel of the two-point vertex $\Gamma$
taken to be in appropriate gauge frame.  To explain what has just been said
we discuss two gauge conditions which can be used for studying the
correlations functions - Stueckelberg gauge and light-cone gauge. We would
like to discuss these gauges because of the following reasons.

i) As we have said, the Stueckelberg gauge reduces our approach to the
standard formulation of $CFT$. Therefore the use of the Stueckelberg gauge
allows us to demonstrate how the standard two-point correlation function of
the spin-1 conformal current is obtained from the kernel of our gauge
invariant two-point vertex $\Gamma$.

ii) Motivation for considering the light-cone gauge vertex cames from
conjectured duality of the free large $N$ conformal ${\cal N}=4$ SYM theory
and the theory of massless higher-spin
$AdS$ fields \cite{hagsun}.%
\footnote{ Discussion of this theme in the context of various limits in $AdS$
superstring may be found in \cite{Tseytlin:2002gz}-\cite{Bonelli:2003kh}.}
On the one hand, one expects that massless higher-spin $AdS$ fields appear in
the tensionless limit of $AdS$ string. On the other hand, by analogy with
flat space, we expect that a quantization of the Green-Schwarz $AdS$
superstring \cite{Metsaev:1998it} will be straightforward only in the
light-cone gauge \cite{Metsaev:2000yf,Metsaev:2000yu}. As we shall
demonstrate in Sec. \ref{secAdS/CFT}, correlation function of the arbitrary
spin-$s$ conformal current is obtained from the effective action of massless
arbitrary spin-$s$ $AdS$ field. Therefore it seems that from the stringy
perspective of $AdS/CFT$ correspondence, light-cone approach to $CFT$ is the
fruitful direction to go.

We now discuss the Stueckelberg gauge and light-cone gauge in turn.

{\bf Stueckelberg gauge frame two-point vertex}. We begin with the discussion
of Stueckelberg gauge fixed two-point vertex of the spin-1 shadow field, i.e.
we relate our vertex with the one in the standard approach to $CFT$. In
general, Stueckelberg gauge frame is achieved through the use of the
Stueckelberg gauge and differential constraints. The gauge invariant approach
to the spin-1 shadow field does not involve the Stueckelberg gauge
symmetries. This implies that we should just solve the differential
constraint which is given in Table I, $\partial^a \phi_\sh^a + \phi_\sh = 0$.
Solution to this constraint is obvious, $ \phi_\sh=-
\partial^a \phi_\sh^a $. Plugging this $\phi_\sh$ into \rf{manus2009-02-02}
and ignoring the total derivative, we find that two-point density
$\Gamma_{12}$ \rf{manus2009-02-02} takes  the following form:
\beq
\label{manus2009-02-04}
&& \Gamma_{12}^{{\rm Stuck.g.fram}}  =  k_1 \Gamma_{12}^{{\rm stand}}\,,
\\[10pt]
\label{manus2009-02-04x} && \qquad  \Gamma_{12}^{{\rm stand}}  =
\frac{\phi_\sh^a(x_1) O_{12}^{ab}\phi_\sh^b(x_2)}{|x_{12}|^{2(d-1)}}\,,
\\[5pt]
\label{manus2009-02-05} && \qquad O_{12}^{ab}  \equiv \eta^{ab} -
\frac{2x_{12}^a x_{12}^b}{|x_{12}|^2}\,,
\\[5pt]
\label{manus2009-02-06} &&  \qquad k_1 \equiv \frac{d-1}{2(d-2)}\,,
\eeq
where $\Gamma_{12}^{{\rm stand}}$ \rf{manus2009-02-04x} stands for the
two-point vertex of the spin-1 shadow field in the standard approach to
$CFT$. From \rf{manus2009-02-04}, we see that our gauge invariant vertex
taken to be in the Stueckelberg gauge frame coincides, up to normalization
factor $k_1$, with the two-point vertex in the standard approach to $CFT$. As
is well known, $\Gamma_{12}^{{\rm stand}}$ \rf{manus2009-02-04x} is invariant
under gauge transformation appearing in the standard approach to $CFT$,
\be \delta \phi_\sh^a =\partial^a \xi_\sh\,.\ee
The kernel of vertex $\Gamma^{{\rm stand}}$ \rf{manus2009-02-04x} defines a
correlation function of the spin-1 conformal current taken to be in the
Stueckelberg gauge frame. This is to say that the gauge invariant approach to
the spin-1 conformal current involves the Stueckelberg gauge symmetry
associated with the gauge transformation parameter $\xi_\cur$ (see Table I).
This symmetry allows us to gauge away the scalar field $\phi_\cur$ by
imposing the Stueckelberg gauge $\phi_\cur=0$. Thus, in the Stueckelberg
gauge frame, we are left with divergence-free vector field $\phi_\cur^a$.
Two-point correlation function of this vector field is given by the kernel of
vertex $\Gamma^{{\rm stand}}$ \rf{manus2009-02-04x},
\be \langle \phi_\cur^a(x_1),\phi_\cur^b(x_2)\rangle =
\frac{O_{12}^{ab}}{|x_{12}|^{2(d-1)}}\,. \ee

{\bf  Light-cone gauge two-point vertex}. Light-cone gauge frame is achieved
through the use of the light-cone gauge and differential constraints. Taking
into account the gauge transformation of the field $\phi_\sh^a$ (see Table
I), we impose the light-cone gauge condition,%
\footnote{ In light-cone frame, space-time coordinates are decomposed as
$x^a= x^+, x^-,x^i$, where light-cone coordinates in $\pm$ directions are
defined as $x^\pm=(x^{d-1} \pm x^0)/\sqrt{2}$ and $x^+$ is taken to be a
light-cone time. $so(d-2)$ algebra vector indices take values $i,j =1,\ldots,
d-2$. We adopt the conventions: $\partial^i=\partial_i\equiv\partial/\partial
x^i$, $\partial^\pm=\partial_\mp \equiv
\partial/\partial x^\mp$.}
\be \label{man02-16072009-01} \phi_\sh^+ = 0\,.\ee
Plugging this gauge condition in the differential constraint for spin-1
shadow field (see Table I) we obtain solution for $\phi_\sh^-$,
\be \label{man02-16072009-02} \phi_\sh^- = -
\frac{\partial^j}{\partial^+}\phi_\sh^j - \frac{1}{\partial^+}\phi_\sh\,. \ee
We see that we are left with vector field $\phi_\sh^i$ and scalar field
$\phi_\sh$. These fields constitute the field content of the light-cone gauge
frame. Note that, in contrast to the Stueckelberg gauge frame, the scalar
field $\phi_\sh$ becomes an independent field D.o.F in the light-cone gauge
frame.

Using \rf{man02-16072009-01} in \rf{manus2009-02-02} leads to light-cone
gauge fixed vertex
\be \label{man02-16072009-03}
\Gamma_{12}^{({\rm l.c.})} =  \frac{\phi_\sh^i(x_1)
\phi_\sh^i(x_2)}{2|x_{12}|^{2(d-1)}} + \frac{1}{4(d-2)^2} \frac{\phi_\sh(x_1)
\phi_\sh(x_2)}{|x_{12}|^{2(d-2)}}\,.
\ee
We note that, as in the case of gauge invariant vertex \rf{manus2009-02-02},
light-cone vertex \rf{man02-16072009-03} is diagonal with respect to the
fields $\phi_\sh^i$ and $\phi_\sh$. Note however that, in contrast to the
gauge invariant vertex, the light-cone vertex is constructed out of the
fields which are not subject to any constraints.

The kernel of the light-cone vertex gives two-point correlation function of
spin-1 conformal current taken to be in the light-cone gauge. This is to say
that, using the gauge symmetry of the spin-1 conformal current (see Table I),
we impose light-cone gauge on the field $\phi_\cur^a$,
\be \phi_\cur^+ =0 \,.\ee
Using this gauge condition in the differential constraint for the conformal
spin-1 current (see Table I), we find
\be \phi_\cur^- = - \frac{\partial^j}{\partial^+}\phi_\cur^j -
\frac{\Box}{\partial^+}\phi_\cur\,. \ee
We see that we are left with vector field $\phi_\cur^i$ and scalar field
$\phi_\cur$. These fields constitute the field content of the light-cone
gauge frame. Defining two-point correlation functions of the fields
$\phi_\cur^i$, $\phi_\cur$ in a usual way,
\beq
&& \langle \phi_\cur^i(x_1), \phi_\cur^j(x_2)\rangle =
\frac{\delta^2\Gamma^{({\rm l.c.})}}{\delta \phi_\sh^i(x_1)\delta
\phi_\sh^j(x_2)}\,,\qquad
\\[5pt]
&& \langle \phi_\cur(x_1), \phi_\cur(x_2)\rangle =
\frac{\delta^2\Gamma^{({\rm l.c.})}}{\delta \phi_\sh(x_1)\delta
\phi_\sh(x_2)}\,,
\eeq
and using \rf{man02-16072009-03}, we obtain the two-point light-cone gauge
correlation functions of the spin-1 conformal field,
\beq
&& \langle \phi_\cur^i(x_1), \phi_\cur^j(x_2)\rangle = \delta^{ij}
|x_{12}|^{-2(d-1)}\,,
\\[5pt]
&& \langle \phi_\cur(x_1), \phi_\cur(x_2)\rangle = \frac{1}{2(d-2)^2}
|x_{12}|^{-2(d-2)}\,.\qquad\quad
\eeq

%%%%%%%%%%%%%%%%%%%%%%%%%%%%%%%%%%%%%%%%%%%%%%%%%%%%%%%%%%%%%%%%%%%%%
%%%%%%%%%%%%%%%%%%%%%%%%%%%%%%%%%%%%%%%%%%%%%%%%%%%%%%%%%%%%%%%%%%%%%%%
\subsection{ Gauge invariant two-point vertex for spin-2 shadow
field } \label{man02-sec-01}
%%%%%%%%%%%%%%%%%%%%%%%%%%%%%%%%%%%%%%%%%%%%%%%%%%%%%%%%%%%%%%%%%%%%%%%
%%%%%%%%%%%%%%%%%%%%%%%%%%%%%%%%%%%%%%%%%%%%%%%%%%%%%%%%%%%%%%%%%%%%%%%

We proceed with the discussion of two-point vertex for spin-2 shadow field.
As compared to the spin-1 shadow field, this important case demonstrates some
new features of our approach. In the gauge invariant approach, the spin-2
shadow field is described by rank-2 tensor field $\phi_\sh^{ab}$, vector
field $\phi_\sh^a$, and scalar field $\phi_\sh$. Note that the tensor field
$\phi_\sh^{ab}$ is not traceless. The gauge invariant two-point vertex we
find takes the form given \rf{manus2009-02-01}, where the two-point density
$\Gamma_{12}$ is given by
\beq \label{manus2009-02-07}
\Gamma_{12} &  = & \frac{1}{4|x_{12}|^{2d}}
\Bigl(\phi_\sh^{ab}(x_1)\phi_\sh^{ab}(x_2) - \half
\phi_\sh^{aa}(x_1)\phi_\sh^{bb}(x_2)\Bigr)
\nonumber\\[5pt]
& +  & \frac{1}{4d(d-1)}
\frac{\phi_\sh^a(x_1)\phi_\sh^a(x_2)}{|x_{12}|^{2(d-1)}}
\nonumber\\[5pt]
& +  & \frac{1}{8d(d-1)(d-2)^2}
\frac{\phi_\sh(x_1)\phi_\sh(x_2)}{|x_{12}|^{2(d-2)}}\,.
\eeq
One can check that this vertex is invariant under the gauge transformations
of the spin-2 shadow field given in Table I. Also, using the operator $R^a$
given in Table I, we check that the vertex is invariant under conformal boost
transformations. Remarkable feature of the vertex is its diagonal form with
respect to the fields $\phi_\sh^{ab}$, $\phi_\sh^a$, and $\phi_\sh$.

We now discuss Stueckelberg gauge and light-cone gauge fixed vertices in
turn.

{\bf Stueckelberg gauge frame two-point vertex}. As we have said, the
standard approach to $CFT$ is obtained from our approach by using the
Stueckelberg gauge frame. Therefore to illustrate our approach, we begin with
the discussion of Stueckelberg gauge fixed two-point vertex of the spin-2
shadow field. The Stueckelberg gauge frame is achieved through the use of the
Stueckelberg gauge and differential constraints. From the gauge
transformations of the spin-2 shadow field given in Table I, we see that the
trace of the rank-2 tensor field $\phi_\sh^{ab}$ transforms as the
Stueckelberg field, i.e., $\phi_\sh^{aa}$ can be gauged away via the
Stueckelberg gauge fixing,
\be \label{manus2009-02-08}
\phi_\sh^{aa} = 0\,. \ee
Gauge condition \rf{manus2009-02-08} leads to traceless field $\phi_\sh^{ab}$
which can be identified with the shadow field of the standard approach to
$CFT$. Taking into account \rf{manus2009-02-08}, we solve the differential
constraints for the spin-2 shadow field (see Table I) to express the fields
$\phi_\sh^a$ and $\phi_\sh$ in terms of the traceless tensor field
$\phi_\sh^{ab}$,
\beq
\label{manus2009-02-09} \phi_\sh^a & = & -\partial^b\phi_\sh^{ab} \,,
\\[5pt]
\label{manus2009-02-10} \phi_\sh & = & \frac{1}{u} \partial^a
\partial^b\phi_\sh^{ab} \,,
\\[5pt]
\label{manus2009-02-11} && u \equiv \Bigl(2\frac{d-1}{d-2}\Bigr)^{1/2}\,.
\eeq
Relations \rf{manus2009-02-08}-\rf{manus2009-02-10} provide complete
description of the Stueckelberg gauge frame for the spin-2 shadow field.
Plugging \rf{manus2009-02-08}-\rf{manus2009-02-10} in \rf{manus2009-02-07}
and ignoring the total derivative, we find that our two-point density
$\Gamma_{12}$ \rf{manus2009-02-07} takes  the following form:
\beq \label{manus2009-02-12}
&& \Gamma_{12}^{{\rm Stuck.g.fram}}  =   k_2 \Gamma_{12}^{{\rm stand}}\,,
\\[5pt]
\label{manus2009-02-12x} && \qquad \Gamma_{12}^{{\rm stand}}  =
\phi_\sh^{a_1a_2}(x_1) \frac{ O_{12}^{a_1b_1} O_{12}^{a_2b_2}}{|x_{12}|^{2d}}
\phi_\sh^{b_1b_2}(x_2) \,,\qquad \quad
\\[5pt]
\label{manus2009-02-13} && \qquad k_2 \equiv  \frac{d+1}{4(d-1)}\,,
\eeq
where $O_{12}^{ab}$ is defined in \rf{manus2009-02-05}, while
$\Gamma_{12}^{{\rm stand}}$ \rf{manus2009-02-12x} stands for the two-point
vertex of the spin-2 shadow field in the standard approach to $CFT$. From
\rf{manus2009-02-12}, we see that our gauge invariant vertex taken to be in
the Stueckelberg gauge frame coincides, up to normalization factor $k_2$,
with the two-point vertex in the standard approach to $CFT$. We note that
$\Gamma_{12}^{{\rm stand}}$ \rf{manus2009-02-12x} is invariant under gauge
transformation appearing in the standard approach to $CFT$,
\be \delta \phi_\sh^{ab}  = \partial^a\xi_\sh^b  + \partial^b\xi_\sh^a
-\frac{2}{d}\eta^{ab} \partial^c\xi_\sh^c\,. \ee

The kernel of vertex $\Gamma^{{\rm stand}}$ \rf{manus2009-02-12x} defines
two-point correlation function of the spin-2 conformal current taken to be in
the Stueckelberg gauge frame. This is to say that the gauge invariant
approach to the spin-2 conformal current involves the Stueckelberg gauge
symmetries associated with the gauge transformation parameters $\xi_\cur^a$,
$\xi_\cur$ (see Table I). These symmetries allow us to gauge away the vector
field $\phi_\cur^a$ and the scalar field $\phi_\cur$ by imposing the
Stueckelberg gauge $\phi_\cur^a=0$, $\phi_\cur=0$. After that, the
differential constraints lead to traceless divergence-free tensor field
$\phi_\cur^{ab}$. Two-point correlation function of this tensor field is
given by the kernel of vertex $\Gamma^{{\rm stand}}$ \rf{manus2009-02-12x}.

{\bf Light-cone gauge two-point vertex}. We now consider light-cone gauge
fixed vertex. In our approach, the light-cone gauge frame is achieved through
the use of the light-cone gauge and differential constraints. Taking into
account the gauge transformations of the fields $\phi_\sh^{ab}$, $\phi_\sh^a$
(see Table I), we impose the light-cone gauge condition,
\be \label{man02-16072009-04} \phi_\sh^{+a} = 0\,, \qquad \phi_\sh^+ = 0 \,.
\ee
Plugging this gauge condition in the differential constraints for the spin-2
shadow field (see Table I) we find ,
\beq \label{man02-16072009-05}
&& \phi_\sh^{ii} = 0 \,,
\\[5pt]
\label{man02-16072009-06} && \phi_\sh^{-i} = -
\frac{\partial^j}{\partial^+}\phi_\sh^{ij} -
\frac{1}{\partial^+}\phi_\sh^i\,,
\\[5pt]
\label{man02-16072009-07} && \phi_\sh^- = -
\frac{\partial^j}{\partial^+}\phi_\sh^j - \frac{u}{\partial^+}\phi_\sh\,,
\\[5pt]
\label{man02-16072009-08} && \phi_\sh^{--} =
\frac{\partial^i\partial^j}{\partial^+\partial^+}\phi_\sh^{ij}+
\frac{2\partial^i}{\partial^+\partial^+}\phi_\sh^i +
\frac{u}{\partial^+\partial^+}\phi_\sh\,.\qquad\quad
\eeq
We see that we are left with the $so(d-2)$ algebra traceless rank-2 tensor
field $\phi_\sh^{ij}$, vector field $\phi_\sh^i$, and scalar field
$\phi_\sh$. These fields constitute the field content of the light-cone gauge
frame. Note that, in contrast to the Stueckelberg gauge frame, the vector
field $\phi_\sh^i$ and the scalar field $\phi_\sh$ become independent field
D.o.F in the light-cone gauge frame.

Using \rf{man02-16072009-04},\rf{man02-16072009-05} in \rf{manus2009-02-07}
leads to light-cone gauge fixed vertex
\beq \label{man02-16072009-09}
\Gamma_{12}^{({\rm l.c.})} &  = & \frac{1}{4|x_{12}|^{2d}}
\phi_\sh^{ij}(x_1)\phi_\sh^{ij}(x_2)
\nonumber\\[5pt]
& +  & \frac{1}{4d(d-1)}
\frac{\phi_\sh^i(x_1)\phi_\sh^i(x_2)}{|x_{12}|^{2(d-1)}}
\nonumber\\[5pt]
& +  & \frac{1}{8d(d-1)(d-2)^2}
\frac{\phi_\sh(x_1)\phi_\sh(x_2)}{|x_{12}|^{2(d-2)}}\,.\qquad
\eeq
We see that, as in the case of gauge invariant vertex \rf{manus2009-02-07},
light-cone vertex \rf{man02-16072009-09} is diagonal with respect to the
fields $\phi_\sh^{ij}$, $\phi_\sh^i$ and $\phi_\sh$. Note however that, in
contrast to the gauge invariant vertex, the light-cone vertex is constructed
out of the fields which are not subject to any differential constraints.

The kernel of light-cone vertex \rf{man02-16072009-09} gives two-point
correlation function of the spin-2 conformal current taken to be in the
light-cone gauge. This is to say that using the gauge transformations of the
fields $\phi_\cur^{ab}$, $\phi_\cur^a$ (see Table I), we impose the
light-cone gauge condition,
\be \phi_\cur^{+a} =0 \,,\qquad \phi_\cur^+ =0 \,.\ee
Using this gauge condition in the differential constraints for the conformal
spin-2 current (see Table I), we find
\beq
&&\phi_\cur^{ii}=0\,,
\\[5pt]
&& \phi_\cur^{-i} = - \frac{\partial^j}{\partial^+}\phi_\cur^{ij} -
\frac{\Box}{\partial^+}\phi_\cur^i\,,
\\[5pt]
&& \phi_\cur^- = - \frac{\partial^j}{\partial^+}\phi_\cur^j -
\frac{u\Box}{\partial^+}\phi_\cur\,,
\\[5pt]
&& \phi_\cur^{--} =
\frac{\partial^i\partial^j}{\partial^+\partial^+}\phi_\cur^{ij}+
\frac{2\Box\partial^i}{\partial^+\partial^+}\phi_\cur^i +
\frac{u\Box^2}{\partial^+\partial^+}\phi_\cur\,.\qquad
\eeq
We see that we are left with traceless rank-2 tensor field $\phi_\cur^{ij}$,
vector field $\phi_\cur^i$ and scalar field $\phi_\cur$. These fields
constitute the field content of the light-cone gauge frame.

Defining two-point correlation functions of the fields $\phi_\cur^{ij}$,
$\phi_\cur^i$, $\phi_\cur$ as the second functional derivative of $\Gamma$
with respect to the fields $\phi_\sh^{ij}$, $\phi_\sh^i$, $\phi_\sh$ we
obtain
\beq
&&   \langle \phi_\cur^{ij}(x_1), \phi_\cur^{kl}(x_2)\rangle = \half
|x_{12}|^{-2d} \Pi^{ij;kl}\,,
\\[7pt]
&& \langle \phi_\cur^i(x_1), \phi_\cur^j(x_2)\rangle =
\frac{|x_{12}|^{-2(d-1)} }{2d(d-1)} \delta^{ij}\,,
\\[5pt]
&&  \langle \phi_\cur(x_1), \phi_\cur(x_2)\rangle =
\frac{|x_{12}|^{-2(d-2)}}{4d(d-1)(d-2)^2}\,, \qquad
\eeq
where we use the notation
\be
\qquad \Pi^{ij;kl} = \half\Bigl(\delta^{ik}\delta^{jl} +
\delta^{il}\delta^{jk} - \frac{2}{d-2} \delta^{ij}\delta^{kl}\Bigr) \,. \ee

%%%%%%%%%%%%%%%%%%%%%%%%%%%%%%%%%%%%%%%%%%%%%%%%%%%%%%%%%%%%%%%%%%%%%
%%%%%%%%%%%%%%%%%%%%%%%%%%%%%%%%%%%%%%%%%%%%%%%%%%%%%%%%%%%%%%%%%%%%%%%
\section{ Gauge invariant two-point vertex for arbitrary spin shadow
field }\label{man02-sec04}
%%%%%%%%%%%%%%%%%%%%%%%%%%%%%%%%%%%%%%%%%%%%%%%%%%%%%%%%%%%%%%%%%%%%%%%
%%%%%%%%%%%%%%%%%%%%%%%%%%%%%%%%%%%%%%%%%%%%%%%%%%%%%%%%%%%%%%%%%%%%%%%

We now study two-point vertex for arbitrary spin-$s$ shadow field. In the
gauge invariant approach, the spin-$s$ shadow field is described by totally
symmetric $so(d-1,1)$ Lorentz algebra tensor fields $\phi_\sh^{a_1\ldots
a_{s'}}$, where $s'=0,1,\ldots,s$. For $s'\geq 4$, these fields are
restricted to be double-traceless \rf{man02-15072009-02}. To discuss the
two-point vertex it is convenient to collect the fields $\phi_\sh^{a_1\ldots
a_{s'}}$ into ket -vector $|\phi_\sh\rangle$ as in \rf{phikdef01sh}. In terms
of the ket-vector $|\phi_\sh\rangle$ we find concise expression for the
two-point density $\Gamma_{12}$,
\beq
\label{11062009-10}  \Gamma_{12} &  = &  \half \langle\phi_\sh(x_1)|
\frac{\mubf f_\nu}{ |x_{12}|^{2\nu + d }} |\phi_\sh (x_2)\rangle \,,
\\[5pt]
\label{11062009-10x1} && \mubf \equiv 1-\frac{1}{4}\alpha^2\bar\alpha^2 \,,
\\[5pt]
&& \label{29102008-01} f_\nu = \frac{\Gamma(\nu + \frac{d}{2})\Gamma(\nu +
1)}{4^{\nu_s- \nu} \Gamma(\nu_s + \frac{d}{2})\Gamma(\nu_s + 1)} \,,
\\[5pt]
\label{29102008-01xxx1} &&  \nu = s +\frac{d-4}{2} -N_z\,,
\\[5pt]
\label{29102008-01xxx2} && \nu_s \equiv s +\frac{d-4}{2}\,,
\eeq
where $N_z$ is defined in \rf{man02-25072009-01}. We note that vertex
$\Gamma_{12}$ \rf{11062009-10} is invariant under gauge transformation of the
shadow field $|\phi_\sh\rangle$ provided the shadow field satisfies the
differential constraint. The gauge transformation and the differential
constraint are given in Table II. The vertex is obviously invariant under the
Poincar\'e algebra and dilatation symmetries. Details of the derivation of
vertex $\Gamma_{12}$ \rf{11062009-10} may be found in Appendix
\ref{man02-app-01}. Invariance of vertex \rf{11062009-10} under the conformal
boost transformations is demonstrated in Appendix \ref{app-04082009-01}.

To illustrate structure of the vertex $\Gamma_{12}$ we note that, in terms of
the tensor fields $\phi_\sh^{a_1\ldots a_{s'}}$, vertex $\Gamma_{12}$,
\rf{11062009-10} can be represented as
\beq \label{man02-17072009-03}
&&  \hspace{-1cm} \Gamma_{12} = \sum_{s'=0}^s \Gamma_{12}^{(s')}\,,
\\[5pt]
\label{man02-17072009-04} &&  \hspace{-1cm} \Gamma_{12}^{(s')} =
\frac{w_{s'}}{2|x_{12}|^{2(s'+d-2)}}\Bigr( \phi_\sh^{a_1\ldots a_{s'}}
(x_1)\phi_\sh^{a_1\ldots a_{s'}}(x_2)
\nonumber\\[5pt]
&&  \hspace{-0.2cm} - \frac{s'(s'-1)}{4} \phi_\sh^{aaa_3\ldots a_{s'}}(x_1)
\phi_\sh^{bba_3\ldots a_{s'}}(x_2)\Bigl)\,,
\\[10pt]
\label{man02-17072009-05} &&  \hspace{-1cm} w_{s'} =
\frac{\Gamma(s'+d-2)\Gamma(s'+\frac{d-2}{2})}{4^{s-s'}s'!
\Gamma(s+d-2)\Gamma(s+\frac{d-2}{2})} \,.
\eeq

The kernel of the vertex $\Gamma_{12}$ is related to correlation function of
the spin-$s$ conformal current which is described by gauge fields
$\phi_\cur^{a_1\ldots a_{s'}}$, $s'=0,1,\ldots,s$. These gauge fields are
collected in the ket-vector $|\phi_\cur\rangle$ \rf{phikdef01}. Using the
Stueckelberg gauge and light-cone gauge, we now demonstrate relation of the
conformal current correlation function to the gauge fixed two-point vertex
$\Gamma$.

{\bf Stueckelberg gauge frame two-point vertex}. As we have said, standard
approach to $CFT$ is obtained from our approach by using the Stueckelberg
gauge frame. Therefore to illustrate our approach, we begin with the
discussion of Stueckelberg gauge fixed two-point vertex of the spin-$s$
shadow field. The Stueckelberg gauge frame is achieved through the use of the
Stueckelberg gauge and differential constraints. Using the gauge
transformations and differential constraint of the spin-$s$ shadow field
given in Table II, one can demonstrate that, in the Stueckelberg gauge frame,
one has the following relations:
\beq \label{12062000-03xx}
&& \hspace{-1cm} \bar\alpha^2 |\phi_{\sh,\, s'}\rangle =0\,,
\\[15pt]
\label{12062000-04} && \hspace{-1cm} |\phi_{\sh,\, s'}\rangle = X_{s'}
(\albpar)^{s-s'} |\phi_{\sh,\, s}\rangle\,,
\\[15pt]
\label{12062000-04x1} && \hspace{-1cm} X_{s'} \equiv
\frac{(-)^{s-s'}}{(s-s')!}
\nonumber\\[5pt]
&& \hspace{-0.4cm} \times \Bigl(
\frac{2^{s-s'}\Gamma(s+s'+d-3)\Gamma(s+\frac{d-2}{2})}{
\Gamma(2s+d-3)\Gamma(s'+\frac{d-2}{2}) }\Bigr)^{\half}\,,
\eeq
$s'=0,1,\ldots,s$. Relation \rf{12062000-03xx} tells us that all fields
$\phi_\sh^{a_1\ldots a_{s'}}$ are traceless, while from relation
\rf{12062000-04} we learn that the fields $\phi_\sh^{a_1\ldots a_{s'}}$, with
$s'=0,1,\ldots,s-1$, can be expressed in terms of the rank-s tensor field
$\phi_\sh^{a_1\ldots a_s}$. Plugging \rf{12062000-04} in \rf{11062009-10} and
ignoring total derivative, we get
\beq \label{12062009-01}
&& \hspace{-1cm}\Gamma_{12}^{{\rm Stuck.g.fram}} = k_s \Gamma_{12}^{{\rm
stand}}
\\[9pt]
\label{man02-15072009-01} && \hspace{-0.5cm} \Gamma_{12}^{{\rm stand}} = s!
\langle \phi_{\sh,\,s}(x_1)|
\Obf_{12}(\alpha,\bar\alpha)|\phi_{\sh,\,s}(x_2)\rangle \,,\quad
\\[9pt]
&& \hspace{-0.5cm} \Obf_{12}(\alpha,\bar\alpha) \equiv \sum_{k=0}^s
\frac{(-)^k 2^k}{k!} \frac{(\alpha x_{12})^k (\bar\alpha
x_{12})^k}{|x_{12}|^{2(s+d-2+k)}}\,,
\\[7pt]
&& \hspace{-0.5cm} k_s \equiv  \frac{2s+d-3}{2s!(s+d-3)}\,,
\eeq
where $\alpha x_{12} = \alpha^a x_{12}^a$, $\bar\alpha x_{12} = \bar\alpha^a
x_{12}^a$ and $\Gamma_{12}^{{\rm stand}}$ in r.h.s. of \rf{12062009-01}
stands for the two-point vertex of the spin-$s$ shadow field in the standard
approach to $CFT$. Relation \rf{man02-15072009-01} provides oscillator
representation for the $\Gamma_{12}^{{\rm stand}}$. In terms of the tensor
field $\phi_\sh^{a_1 \ldots a_s}$, vertex $\Gamma_{12}^{{\rm stand}}$
\rf{man02-15072009-01} can be represented in the commonly used form,
\be
\label{man02-12072009-02}
\Gamma_{12}^{{\rm stand}} = \phi_\sh^{a_1 \ldots a_s }(x_1)
\frac{O_{12}^{a_1b_1} \ldots O_{12}^{a_s b_s}}{|x_{12}|^{2(s+d-2)}}
\phi_\sh^{b_1\ldots b_s}(x_2) \,,\quad
\ee
where $O_{12}^{ab}$ is defined in \rf{manus2009-02-05}. From
\rf{12062009-01}, we see that our gauge invariant vertex $\Gamma_{12}$ taken
to be in the Stueckelberg gauge frame coincides, up to normalization factor
$k_s$, with the two-point vertex in the standard approach to $CFT$.

The kernel of vertex $\Gamma^{{\rm stand}}$ \rf{man02-12072009-02} defines
two-point correlation function of the spin-$s$ conformal current taken to be
in the Stueckelberg gauge frame. This is to say that, in the Stueckelberg
gauge frame, we obtain $\phi_\cur^{a_1\ldots a_{s'}}=0$ for
$s'=0,1,\ldots,s-1$ and we are left with the traceless divergence-free
conformal current $\phi_\cur^{a_1\ldots a_s}$. Two-point correlation function
of this conformal current is given by the kernel of vertex $\Gamma^{{\rm
stand}}$ \rf{man02-12072009-02}.

As a side of remark we note that $\Gamma_{12}^{{\rm stand}}$
\rf{man02-12072009-02} is invariant under gauge transformation appearing in
the standard approach to $CFT$,
\be \label{man02-12072009-03}  \delta \phi_\sh^{a_1 \ldots a_s} = s!\Pi^\tr
\partial^{(a_1} \xi_\sh^{a_2\ldots a_s)}\,, \ee
where $\xi_\sh^{a_1\ldots a_{s-1}}$ is traceless parameter of gauge
transformation and the projector $\Pi^\tr$ is inserted to respect
tracelessness constraint for the field $\phi_\sh^{a_1\ldots a_s}$.

{\bf Light-cone gauge two-point vertex}. The light-cone gauge frame is
achieved through the use of the light-cone gauge and differential
constraints. We impose the conventional light-cone gauge,
\be \label{man02-17072009-01} \bar\alpha^+ \Pi^\smponetwo |\phi_\sh\rangle
=0\,, \ee
where $\Pi^\smponetwo$ is given in \rf{18052008-08}, and analyze the
differential constraint for the spin-$s$ shadow field $|\phi_\sh\rangle$ (see
Table II). We find that solution to the differential constraint can be
expressed in terms of light-cone ket-vector $|\phi_\sh^{\rm l.c.}\rangle$,
\beq \label{man02-17072009-02}
&& |\phi_\sh\rangle =
\exp\Bigl(-\frac{\alpha^+}{\partial^+}(\bar\alpha^i\partial^i -
\eb_{1\,\sh})\Bigr) |\phi_\sh^{\rm l.c.}\rangle\,,
\\[5pt]
\label{man02-17072009-06} && \bar\alpha^i \bar\alpha^i | \phi_\sh^{\rm
l.c.}\rangle  = 0 \,,
\eeq
where the light-cone ket-vector $|\phi_\sh^{\rm l.c.}\rangle$ is obtained
from $|\phi_\sh\rangle$ \rf{phikdef01sh} by equating $\alpha^+ = \alpha^- =
0$,
\be \label{man02-17072009-07} |\phi_\sh^{\rm l.c.}\rangle =
|\phi_\sh\rangle\Bigr|_{\alpha^+=\alpha^-=0} \,.\ee
We see that we are left with fields $\phi_\sh^{i_1\ldots i_{s'}}$,
$s'=0,1,\ldots,s$, which are traceless $so(d-2)$ algebra tensor fields,
$\phi_\sh^{iii_3\ldots i_{s'}}=0$. These fields constitute the field content
of the light-cone gauge frame. Note that, in contrast to the Stueckelberg
gauge frame, the fields $\phi_\sh^{i_1\ldots i_{s'}}$, with
$s'=0,1,\ldots,s-1$, become independent field D.o.F in the light-cone gauge
frame. Also note that, in contrast to the gauge invariant approach, the
fields $\phi_\sh^{i_1\ldots i_{s'}}$, $s'=0,1,\ldots,s$, are not subject to
any differential constraints.

Using \rf{man02-17072009-02},\rf{man02-17072009-06} in \rf{11062009-10} leads
to light-cone gauge fixed vertex
\beq \label{man02-17072009-08}
\Gamma_{12}^{\rm l.c.} &  = &  \half \langle\phi_\sh^{\rm l.c.}(x_1)|
\frac{f_\nu}{ |x_{12}|^{2\nu + d }} |\phi_\sh^{\rm l.c.}(x_2)\rangle \,,
\qquad
\eeq
where $f_\nu$ is defined in \rf{29102008-01}.

To illustrate the structure of vertex $\Gamma_{12}^{\rm l.c.}$
\rf{man02-17072009-08} we note that, in terms of the fields
$\phi_\sh^{i_1\ldots i_{s'}}$, the vertex can be represented as
\beq \label{man02-17072009-09}
&&  \hspace{-1.2cm} \Gamma_{12}^{\rm l.c.} = \sum_{s'=0}^s
\Gamma_{12}^{(s')\,{\rm l.c.}}\,,
\\[5pt]
\label{man02-17072009-10} &&  \hspace{-1.2cm} \Gamma_{12}^{(s')\,{\rm l.c.}}
= \frac{w_{s'}}{2|x_{12}|^{2(s'+d-2)}} \phi_\sh^{i_1\ldots i_{s'}}
(x_1)\phi_\sh^{i_1\ldots i_{s'}}(x_2)\,,
\eeq
where $w_{s'}$ is given in \rf{man02-17072009-05}. We see that, as in the
case of gauge invariant vertex, light-cone vertex \rf{man02-17072009-08} is
diagonal with respect to the light-cone fields $\phi_\sh^{i_1\ldots i_{s'}}$,
$s'=0,1,\ldots,s$. Note however that, in contrast to the gauge invariant
vertex, the light-cone vertex is constructed out of the light-cone fields
which are not subject to any differential constraints.

As usually, the kernel of light-cone vertex \rf{man02-17072009-08} gives
two-point correlation function of the spin-$s$ conformal current taken to be
in the light-cone gauge. This is to say that using gauge symmetry of the
spin-$s$ conformal current (see Table II) we impose light-cone gauge
condition on the ket-vector $|\phi_\cur\rangle$,
\be \label{man02-17072009-10x1} \bar\alpha^+ \Pi^\smponetwo |\phi_\cur\rangle
=0 \,, \ee
where $\Pi^\smponetwo$ is given in \rf{18052008-08}. Using this gauge
condition in the differential constraint for the conformal spin-$s$ current
(see Table II), we find
\beq \label{man02-17072009-10x2}
&& |\phi_\cur\rangle = \exp\Bigl(
-\frac{\alpha^+}{\partial^+}(\bar\alpha^i\partial^i -
\eb_{1\,\cur}\Box)\Bigr) |\phi_\cur^{\rm l.c.}\rangle\,,\qquad
\\[5pt]
\label{man02-17072009-10x3} && \bar\alpha^i \bar\alpha^i | \phi_\cur^{\rm
l.c.}\rangle  = 0 \,,
\eeq
where a light-cone ket-vector $|\phi_\cur^{\rm l.c.}\rangle$ is obtained from
$|\phi_\cur\rangle$ \rf{phikdef01} by equating $\alpha^+ = \alpha^- = 0$,
\be \label{man02-17072009-10x4} |\phi_\cur^{\rm l.c.}\rangle =
|\phi_\cur\rangle\Bigr|_{\alpha^+=\alpha^-=0} \,.\ee
We see that we are left with light-cone fields $\phi_\cur^{i_1\ldots
i_{s'}}$, $s'=0,1,\ldots,s$, which are traceless tensor fields of $so(d-2)$
algebra, $\phi_\cur^{iii_3\ldots i_{s'}}= 0$. These fields constitute the
field content of the light-cone gauge frame. Note that, in contrast to the
Stueckelberg gauge frame, the fields $\phi_\cur^{i_1\ldots i_{s'}}$, with
$s'=0,1,\ldots,s-1,$ are not equal to zero. Also note that, in contrast to
the gauge invariant approach, the fields $\phi_\cur^{i_1\ldots i_{s'}}$,
$s'=0,1,\ldots,s$, are not subject to any differential constraints.

Defining two-point correlation functions of the fields $\phi_\cur^{i_1\ldots
i_{s'}}$, as the second functional derivative of $\Gamma$ with the respect to
the shadow fields $\phi_\sh^{i_1\ldots i_{s'}}$, we obtain the following
correlation functions:
\beq
&& \langle \phi_\cur^{i_1\ldots i_{s'}}(x_1), \phi_\cur^{j_1\ldots
j_{s'}}(x_2)\rangle
\nonumber\\[5pt]
&&\qquad\qquad  = \frac{w_{s'}}{|x_{12}|^{2(s'+d-2)}} \Pi^{i_1\ldots
i_{s'};j_1\ldots j_{s'}}\,, \qquad
\eeq
$s'=0,1,\ldots, s$, where $w_{s'}$ is defined in \rf{man02-17072009-05}  and
$\Pi^{i_1\ldots i_{s'};j_1\ldots j_{s'}}$ stands for the projector on
traceless spin-$s'$ tensor field of the $so(d-2)$ algebra. Explicit form of
the projector may be found e.g. in Ref.\cite{Erdmenger:1997wy}.

\section{AdS/CFT correspondence}\label{secAdS/CFT}

We now apply our results to the study of $AdS/CFT$ correspondence for free
massless arbitrary spin $AdS$ fields and boundary shadow fields. To this end
we use the gauge invariant $CFT$ adapted description of $AdS$ massless fields
and modified (Lorentz) de Donder gauge found in Ref.\cite{Metsaev:2008ks}%
\footnote{Remarkable feature of our approach is that it can be generalized to
the case of massive fields in a relatively straightforward way. This can be
done by using $CFT$ adapted approach to massive $AdS$ fields developed in
Ref.\cite{Metsaev:2009hp}.}.
Our massless fields are obtained from Fronsdal fields by the invertible
transformation which is described in Sec.V in Ref.\cite{Metsaev:2008ks}. {\it
It is the use of our fields and the modified (Lorentz) de Donder gauge that
leads to decoupled form of gauge fixed equations of motion and surprisingly
simple Lagrangian}. Owing these properties of our fields and the modified
(Lorentz) de Donder gauge, it is possible to simplify significantly the
computation of the effective action. We remind that the bulk action evaluated
on solution of the Dirichlet problem is referred to as effective action in
this paper. Note also that, from the very beginning, our $CFT$ adapted gauge
invariant Lagrangian is formulated in the Poincar\'e parametrization of $AdS$
space,
\be \label{lineelem01} ds^2 = \frac{1}{z^2}(dx^a dx^a + dz\, dz)\,. \ee
Therefore our Lagrangian is {\it explicitly invariant with respect to
boundary Poincar\'e symmetries}, i.e., manifest symmetries of our Lagrangian
are adapted to manifest symmetries of boundary $CFT$.

In this Section, using the modified (Lorentz) de Donder gauge, we are going
to demonstrate that {\it action of massless spin-$s$ $AdS$ field, when it is
evaluated on solution of equations of motion with the Dirichlet problem
corresponding to the boundary shadow field, is equal, up to normalization
factor, to the gauge invariant two-point vertex of spin-$s$ shadow field
which was obtained in Secs. \ref{man02-sec-02x} and \ref{man02-sec04}}. Also
we find the normalization factor.

The modified (Lorentz) de Donder gauge is invariant under the on-shell
leftover gauge symmetries of bulk $AdS$ fields. Note however that, in our
approach, we have gauge symmetries not only at $AdS$ side,
but also at the boundary $CFT$.%
\footnote{ In the standard approach to $CFT$, only the shadow fields are
transformed under gauge transformations, while in our gauge invariant
approach both the currents and shadow fields are transformed under gauge
transformations, i.e., our approach allows us to study the currents and
shadow fields on an equal footing.}
These gauge symmetries are also related via $AdS/CFT$ correspondence. Namely,
in Ref.\cite{Metsaev:2008fs}, we demonstrated that the on-shell leftover
gauge symmetries of non-normalizable solutions of bulk $AdS$ fields match
with the gauge symmetries of shadow fields. It is this matching of on-shell
leftover gauge symmetries of non-normalizable solutions and the gauge
symmetries of shadow fields that explains why the effective action coincides
with the gauge invariant two-point vertex of shadow field.

\subsection{ AdS/CFT correspondence for scalar field}
\label{sec-scal-f-01}

Action of arbitrary spin $AdS$ field taken to be in the modified (Lorentz) de
Donder gauge is similar to the action for a massive scalar $AdS$ field. In
fact, this is main advantage of using  the modified (Lorentz) de Donder gauge
condition. Therefore we begin with brief review of the computation of the
effective action for massive scalar field.

Action and Lagrangian for the
massive scalar field in $AdS_{d+1}$ background take the form%
\footnote{ In Secs. \ref{secAdS/CFT} and \ref{man02-sec-06}, we use the
Euclidian signature.}
\be \label{19072009-01}
S =  \int d^dx dz\,  \LL \,,
\ee
\be \label{19072009-02} \LL = \half \sqrt{|g|}\Bigl(g^{\mu\nu}\partial_\mu
\Phi \partial_\nu \Phi + m^2 \Phi^2 \Bigr)\,. \ee
In terms of the canonical normalized field $\phi$ defined by
\be \label{19072009-03} \Phi= z^{\frac{d-1}{2}}\phi \,, \ee
the Lagrangian takes the form (up to total derivative)
\be \label{19072009-04}
\LL =  \half |d\phi|^2 + \half |\TT_{\nu -\half } \phi|^2\,,
\ee
$|d\phi|^2 \equiv \partial^a\phi \partial^a\phi$, where we use the notation
\be
\label{09072009-04} \TT_\nu \equiv \partial_z + \frac{\nu}{z}\,,
\ee
\be  \label{19072009-05} \nu = \sqrt{m^2 + \frac{d^2}{4}}\,. \ee
We note that $\nu$ is related with the conformal dimension of boundary
conformal spin-0 current, $\phi_\cur$ as
\be \nu=\Delta -\frac{d}{2}\,.\ee
We assume that $\nu>0$. Also note that, for massless scalar field, $m^2=
(1-d^2)/4$.

Equations of motion obtained from Lagrangian \rf{19072009-04} take the form
\beq \label{19072009-06}
&& \Box_\nu  \phi = 0 \,,
\\[5pt]
&& \label{09072009-09}
\Box_\nu \equiv \Box + \partial_z^2 - \frac{1}{z^2}(\nu^2 -\frac{1}{4})\,.
\eeq
It is easy to see that by using equations of motion \rf{19072009-06} in bulk
action \rf{19072009-01} with Lagrangian \rf{19072009-04} we obtain the
effective action given by%
\footnote{ Following commonly used setup, we consider solution of the
Dirichlet problem which tends to zero as $z\rightarrow \infty$. Therefore, in
\rf{19072009-07}, we ignore contribution to $S_\eff$ when $z=\infty$.}
\beq
\label{19072009-07} - S_\eff & = & \int d^dx\,  \LL_\eff\Bigr|_{z\rightarrow
0} \,,
\\[7pt]
\label{19072009-08} \LL_\eff & = & \half \phi \TT_{\nu -\half } \phi
\,.\qquad
\eeq
Following the procedure in \cite{wit}, we note that solution of equations
\rf{19072009-06} with the Dirichlet problem corresponding to boundary shadow
scalar field $\phi_\sh$ takes the form

\beq \label{19072009-09}
\phi(x,z) & = & \sigma \int d^dy\, G_\nu (x-y,z) \phi_\sh(y)\,,
\\[5pt]
\label{10072009-12} && G_\nu(x,z) = \frac{c_\nu z^{\nu+\half}}{ (z^2+
|x|^2)^{\nu + \frac{d}{2}} }\,,
\\[5pt]
\label{10072009-13}&& \ \ c_\nu \equiv \frac{\Gamma(\nu+\frac{d}{2})}{
\pi^{d/2} \Gamma(\nu)} \,.
\eeq
To be flexible, we use normalization factor $\sigma$ in \rf{19072009-09}. For
the case of scalar field, commonly used normalization in \rf{19072009-09} is
achieved by setting $\sigma=1$.

Using asymptotic behavior of the Green function
\be \label{10072009-14}
G_\nu(x,z) \ \ \ \stackrel{z \rightarrow 0}{\longrightarrow} \ \ \ z^{-\nu +
\half} \delta^d(x)\,,\ee
we find the asymptotic behavior of our solution
\be \label{man02-21072009-14}
\phi(x,z) \,\,\, \stackrel{z\rightarrow 0 }{\longrightarrow}\,\,\, z^{-\nu +
\half} \sigma \phi_\sh(x)\,.
\ee
From this expression, we see that our solution has indeed asymptotic behavior
corresponding to the shadow scalar field.

Plugging solution of the Dirichlet problem \rf{19072009-09} into
\rf{19072009-07}, \rf{19072009-08}, we obtain the effective action
\be \label{man02-21072009-15}
-S_\eff  = \nu c_\nu \sigma^2 \int d^dx_1d^dx_2
\frac{\phi_\sh(x_1)\phi_\sh(x_2)}{|x_{12}|^{2\nu + d}}\,.
\ee
Plugging the commonly used value of $\sigma$, $\sigma=1$, in
\rf{man02-21072009-15}, we obtain the properly normalized effective action
found in Refs.\cite{Gubser:1998bc,Freedman:1998tz}. Note however that our
Lagrangian \rf{19072009-04} differs from the one in
Refs.\cite{Gubser:1998bc,Freedman:1998tz} by total derivative with respect to
the radial coordinate $z$, which gives nontrivial contribution to the
effective action. Coincidence of our result and the one in
Refs.\cite{Gubser:1998bc,Freedman:1998tz} is related to the fact that we fix
boundary value of the Dirichlet problem at $z=0$, compute the effective
action at $z=\epsilon$ and then scale $\epsilon \rightarrow 0$ (see also
Ref.\cite{wit}), while authors of Refs.\cite{Gubser:1998bc,Freedman:1998tz}
fix boundary value of the Dirichlet problem at $z = \epsilon$ and, after
computation of the effective action, scale $\epsilon \rightarrow 0$.
Interesting novelty of our computation, as compared to the one in \cite{wit},
is that we use Fourier transform of the Green function. Details of our
computation may be found in Appendix \ref{man02-app-04}.

\subsection{ AdS/CFT correspondence for spin-1 field }\label{subsec-01}

We now discuss $AdS/CFT$ correspondence for bulk massless spin-1 $AdS$ field
and boundary spin-1 shadow field. To this end we use $CFT$ adapted gauge
invariant Lagrangian and the modified Lorentz gauge condition
\cite{Metsaev:1999ui,Metsaev:2008ks,Metsaev:2009hp} (see Appendix
\ref{man02-app-02}).%
\footnote{Discussion of $AdS/CFT$ correspondence for spin-1 Maxwell field by
using the radial gauge may be found in \cite{wit}.}

In $AdS_{d+1}$ space, the massless spin-1 field is described by fields
$\phi^a(x,z)$ and $\phi(x,z)$ which are the respective vector and scalar
fields of the $so(d)$ algebra. $CFT$ adapted gauge invariant action and
Lagrangian for these fields take the form,
\beq \label{09072009-02}
S & = & \int d^dx dz\,  \LL \,,
\\[7pt]
\label{20062009-01}
\LL& = & \half |d\phi^a|^2 +  \half |d\phi|^2
\nonumber\\[5pt]
& + & \half |\TT_{\nu_1 -\half } \phi^a|^2 + \half |\TT_{\nu_0 -\half}
\phi|^2 - \half C^2\,,\qquad
\eeq
where we use the notation
\beq
\label{09072009-03} && C \equiv \partial^a \phi^a + \TT_{ - \nu_1+\half }
\phi\,,
\\[5pt]
\label{09072009-05} && \nu_1= \frac{d-2}{2}\,,\qquad \nu_0 = \frac{d-4}{2}\,,
\eeq
and $\TT_\nu$ is given in \rf{09072009-04}. Lagrangian \rf{20062009-01} is
invariant under gauge transformations
\beq
\label{09072009-06} && \delta \phi^a =\partial^a \xi \,,
\\[5pt]
\label{09072009-07} && \delta \phi =\TT_{\nu_1 - \half } \xi \,,
\eeq
where $\xi$ is a gauge transformation parameter.

Gauge invariant equations of motion obtained from Lagrangian \rf{20062009-01}
take the form
\beq
\label{15052008-32x}&& \Box_{\nu_1} \phi^a -\partial^a C  =0\,,
\\[7pt]
\label{15052008-33x}&& \Box_{\nu_0}\phi -\TT_{\nu_1-\half}  C  =0\,,
\eeq
where the operator $\Box_\nu$ and $\nu$'s are given in \rf{09072009-09} and
\rf{09072009-05} respectively. We see that gauge invariant equations
\rf{15052008-32x},\rf{15052008-33x} are coupled. Using equations of motion
\rf{15052008-32x},\rf{15052008-33x} in bulk action \rf{09072009-02}, we
obtain the following boundary effective action:
\beq
\label{09072009-08} S_\eff & = & - \int d^dx\,  \LL_\eff\Bigr|_{z\rightarrow
0} \,,
\\[7pt]
\label{09072009-10} \LL_\eff & = & \half \phi^a \TT_{\nu_1 -\half } \phi^a +
\half \phi \TT_{\nu_0 -\half} \phi  - \half \phi C  \,.\qquad\quad
\eeq

Now we would like to demonstrate how use of the modified Lorentz gauge
condition provides considerable simplification in solving the equations of
motion and computing effective action \rf{09072009-08}. To this end we note
that it is the quantity $C$ given in \rf{09072009-03} that defines the
modified Lorentz gauge condition,
\be \label{09072009-11} C = 0\,,\qquad \hbox{modified Lorentz gauge} \,.\ee
Using this gauge condition in equations of motion
\rf{15052008-32x},\rf{15052008-33x} gives simple gauge fixed equations of
motion,
\beq
\label{15052008-32xx}&& \Box_{\nu_1} \phi^a=0\,,
\\[7pt]
\label{15052008-33xx}&& \Box_{\nu_0} \phi=0\,.
\eeq
Thus, we see that the gauge fixed equations of motions are decoupled. Using
modified Lorentz gauge \rf{09072009-11} in \rf{09072009-10}, we obtain
\be \label{10072009-09}
\LL_\eff\Bigr|_{C=0} =  \half \phi^a \TT_{\nu_1 -\half } \phi^a + \half \phi
\TT_{\nu_0 -\half} \phi \,,
\ee
i.e. we see that $\LL_\eff$ is also simplified.

In order to find $S_\eff$ we should solve equations of motion
\rf{15052008-32xx},\rf{15052008-33xx} with the Dirichlet problem
corresponding to the boundary shadow field and plug the solution into
\rf{09072009-08},\rf{09072009-10}. We now discuss solution of equations of
motion \rf{15052008-32xx},\rf{15052008-33xx}. Because equations of motion
\rf{15052008-32xx},\rf{15052008-33xx} are similar to the ones for scalar
$AdS$ field \rf{19072009-06} we can simply apply result in Sec.
\ref{sec-scal-f-01}. This is to say that solution of equations
\rf{15052008-32xx},\rf{15052008-33xx} with the Dirichlet problem
corresponding to the spin-1 shadow field takes the form
\beq \label{10072009-10}
\phi^a(x,z) & = & \sigma_{1,\nu_1} \int d^dy\, G_{\nu_1}(x-y,z)
\phi_\sh^a(y)\,,
\\[5pt]
\label{10072009-11} \phi(x,z) \ & = &  \sigma_{1,\nu_0} \int d^dy\,
G_{\nu_0}(x-y,z) \phi_\sh(y)\,,\qquad
\\[5pt]
\label{10072009-10x}&& \sigma_{1,\nu_1} = 1\,,
\\[5pt]
\label{10072009-11x}&& \sigma_{1,\nu_0} = -\frac{1}{d-4}\,,
\eeq
where the Green function is given in \rf{10072009-12}. Note that coefficient
$\sigma_{1,\nu_0}$ \rf{10072009-11x} is singular when $d=4$. In addition to
this singularity, there are other singularities when $d$ is even integer (see
Appendix \ref{man02-app-04}). Therefore to keep the discussion from becoming
unwieldy here and below we restrict our attention to odd $d$. Note however
that our results for the effective action are still valid for the case of
even $d$. This is to say that we can simply use the dimensional
regularization in the intermediate formulas and scale $d$ to even integer in
final expression for the effective action. Because massless higher-spin $AdS$
fields theory is available for arbitrary dimension $d$ (see
Ref.\cite{Vasiliev:2003ev}) the dimensional regularization seems to be
promising method for the study of massless higher-spin fields effective
action.

Using asymptotic behavior of the Green function $G_\nu$ given in
\rf{10072009-14}, we find the asymptotic behavior of our solution
\beq
\label{10072009-15} && \phi^a(x,z) \,\,\, \stackrel{z\rightarrow 0
}{\longrightarrow}\,\,\,   z^{-\nu_1 + \half} \phi_\sh^a(x)\,,
\\[5pt]
\label{10072009-16} && \phi(x,z) \,\,\, \stackrel{z\rightarrow 0
}{\longrightarrow}\,\,\,  - \frac{z^{-\nu_0 + \half}}{d-4} \phi_\sh(x)\,.
\eeq
From these expressions, we see that our solution has indeed asymptotic
behavior corresponding to the spin-1 shadow field. Note that because the
solution has non-integrable asymptotic behavior
\rf{10072009-15},\rf{10072009-16}, such solution is referred to as the
non-normalizable solution in the literature.

We now explain the choice of the normalization factors $\sigma_{1,\nu_1}$,
$\sigma_{1,\nu_0}$ in \rf{10072009-10}-\rf{10072009-11x}. The choice of
$\sigma_{1,\nu_1}$ is a matter of convention. Following commonly used
convention, we set this coefficient to be equal to 1. The remaining
normalization factor $\sigma_{1,\nu_0}$ is then determined uniquely by
requiring that the modified Lorentz gauge condition for the spin-1 $AdS$
field \rf{09072009-11} be amount to the differential constraint for the
spin-1 shadow field (see Table I). With the choice made in
\rf{10072009-10}-\rf{10072009-11x} we find the relations
\beq
&& \partial^a\phi^a  =  \int d^dy\,
G_{\nu_1}(x-y,z)\partial^a\phi_\sh^a(y)\,,
\\[5pt]
&& \TT_{-\nu_1+\half} \phi = \int d^dy\, G_{\nu_1}(x-y,z) \phi_\sh(y)\,.
\eeq
From these relations and \rf{09072009-03}, we see that our choice of
$\sigma_{1,\nu_1}$, $\sigma_{1,\nu_0}$ \rf{10072009-10x},\rf{10072009-11x},
allows us to match modified Lorentz gauge for the spin-1 $AdS$ field
\rf{09072009-11} and differential constraint for the spin-1 shadow field
given in Table I.

All that remains to obtain $S_\eff$ is to plug solution of the Dirichlet
problem for $AdS$ fields, \rf{10072009-10},\rf{10072009-11} into
\rf{09072009-08}, \rf{10072009-09}. Using general formula given in
\rf{man02-21072009-15}, we obtain
\beq  \label{man2009-02-14072009-01}
-S_\eff & = & (d-2)c_{\nu_1} \Gamma \,,
\\[7pt]
&& c_{\nu_1}= \frac{\Gamma(d-1)}{\pi^{d/2}\Gamma(\frac{d-2}{2})}\,,
\eeq
where $\Gamma$ is gauge invariant two-point vertex of the spin-1 shadow field
given in \rf{manus2009-02-01},\rf{manus2009-02-02}.

Thus we see that {\it imposing the modified Lorentz gauge on the massless
spin-1 $AdS$ field and computing the bulk action on the solution of equations
of motion with the Dirichlet problem corresponding to the boundary shadow
field we obtain the gauge invariant two-point vertex of the spin-1 shadow
field}.

Because in the literature $S_\eff$ is expressed in terms of two-point vertex
taken in the Stueckelberg gauge frame, $\Gamma^{{\rm stand}}$
\rf{manus2009-02-04x}, we use \rf{manus2009-02-04} and represent our result
\rf{man2009-02-14072009-01} as
\beq \label{man2009-02-14072009-02}
-S_\eff & = & \half(d-1)c_{\nu_1} \Gamma^{{\rm stand}} \,.
\eeq
This relation, by using the radial gauge for $AdS$ fields, was obtained in
Ref.\cite{wit}. The normalization factor in r.h.s. of
\rf{man2009-02-14072009-02} was found in Ref.\cite{Freedman:1998tz}. Note
that we have obtained more general relation given in
\rf{man2009-02-14072009-01}, while relation \rf{man2009-02-14072009-02} is
obtained from \rf{man2009-02-14072009-02} by using the Stueckelberg gauge
frame. The fact that $S_\eff$ is related to $\Gamma^{{\rm stand}}$ is
expected because of conformal symmetry. What is important for the
systematical study of $AdS/CFT$ correspondence is the computation of the
normalization factor in front of $\Gamma^{{\rm stand}}$
\rf{man2009-02-14072009-02}.

As a side of remark we note that the modified Lorentz gauge and gauge-fixed
equations have left-over on-shell gauge symmetry. Namely, modified Lorentz
gauge \rf{09072009-11} and gauge-fixed equations \rf{15052008-32xx},
\rf{15052008-33xx} are invariant under gauge transformations given in
\rf{09072009-06}, \rf{09072009-07} provided the gauge transformation
parameter satisfies the equation
\be \Box_{\nu_1}\xi= 0\,.\ee
Solution to this equation is given by
\be \xi(x,z)  =  \int d^dy\, G_{\nu_1}(x-y,z)\xi_\sh(y)\,.\ee
Plugging this solution in \rf{09072009-06},\rf{09072009-07} we represent the
gauge transformations of $\phi^a(x,z)$ and $\phi(x,z)$ as
\beq \label{man02-03082009-01}
&& \delta \phi^a  =  \int d^dy\, G_{\nu_1}(x-y,z)\partial^a\xi_\sh(y)\,,
\\[5pt]
\label{man02-03082009-02} && \delta \phi = \frac{1}{d-4} \int d^dy\,
G_{\nu_0}(x-y,z) \Box \xi_\sh(y)\,.\qquad
\eeq
Comparing \rf{man02-03082009-01},\rf{man02-03082009-02} with
\rf{10072009-10},\rf{10072009-11}, we see that the on-shell left-over gauge
symmetries of solution of the Dirichlet problem for $AdS$ spin-1 field amount
to the gauge symmetries of the spin-1 shadow field (see Table I). It is this
{\it matching of the on-shell leftover gauge symmetries of solutions of the
Dirichlet problem and the gauge symmetries of the shadow field that explains
why the effective action coincides with the gauge invariant two-point vertex
for the boundary shadow field}.

\subsection{ AdS/CFT correspondence for spin-2 field }

We now proceed with the discussion of $AdS/CFT$ correspondence for bulk
massless spin-2 $AdS$ field and boundary spin-2 shadow field. To this end we
use $CFT$ adapted gauge invariant Lagrangian and modified de Donder gauge
condition for the massless spin-2 $AdS$ field found in
Ref.\cite{Metsaev:2008ks}.%
\footnote{ Discussion of $AdS/CFT$ correspondence for massless spin-2 field
taken to be in the radial gauge may be found in
\cite{Liu:1998bu,Arutyunov:1998ve,Mueck:1998ug}.}
We begin therefore with the presentation of our result in
Ref.\cite{Metsaev:2008ks}. Some helpful details of the derivation of the
$CFT$ adapted Lagrangian for the massless spin-2 $AdS$ field may be found in
Appendix \ref{man02-app-02}.

In $AdS_{d+1}$ space, the massless spin-2 field is described by fields
$\phi^{ab}(x,z)$, $\phi^a(x,z)$, $\phi(x,z)$. The field $\phi^{ab}$ is rank-2
tensor field of the $so(d)$ algebra, while $\phi^a$ and $\phi$ are the
respective vector and scalar fields of the $so(d)$ algebra. The $CFT$ adapted
gauge invariant Lagrangian for these fields takes the form
\cite{Metsaev:2008ks}
\beq \label{gaufixlag01}
\LL &  = &  \frac{1}{4} |d\phi^{ab}|^2 - \frac{1}{8} |d\phi^{aa}|^2 + \half
|d\phi^a|^2 + \half |d\phi|^2\qquad
\nonumber\\[5pt]
& + & \frac{1}{4} |\TT_{\nu_2 -\half } \phi^{ab}|^2
- \frac{1}{8} |\TT_{\nu_2 -\half} \phi^{aa}|^2
\nonumber\\[5pt]
& + &  \, \half |\TT_{\nu_1 -\half } \phi^a |^2
+ \half |\TT_{\nu_0-\half}\phi|^2
\nonumber \\[5pt]
& - & \half C^a C^a - \half C^2\,,
\eeq
where $\TT_\nu$ is defined in \rf{09072009-04} and we use the notation
\beq
\label{15052008-29} && \hspace{-1cm} C^a = \partial^b \phi^{ab} -\half
\partial^a \phi^{bb} + \TT_{-\nu_2 + \half} \phi^a \,,
\\[5pt]
\label{15052008-30} && \hspace{-1cm} C = \partial^a \phi^a -\half
\TT_{\nu_2-\half} \phi^{aa} + u \TT_{-\nu_1 + \half} \phi\,,
\\[9pt]
&& \hspace{-1cm} \label{15052008-34} \nu_2 =\frac{d}{2}\,,\qquad  \nu_1
=\frac{d-2}{2}\,,\qquad  \nu_0 =\frac{d-4}{2}\,,
\\[5pt]
\label{15052008-34x1} && \hspace{-1cm} u \equiv
\Bigl(2\frac{d-1}{d-2}\Bigr)^{1/2}\,.
\eeq
Lagrangian \rf{gaufixlag01} is invariant under the gauge transformations
\beq
\label{15052008-35} && \hspace{-0.3cm} \delta \phi^{ab} = \partial^a \xi^b
+\partial^b \xi^a + \frac{2}{d-2} \eta^{ab} \TT_{-\nu_2+\half}  \xi\,,
\qquad\quad
\\[5pt]
\label{15052008-36} && \hspace{-0.3cm} \delta \phi^a = \partial^a \xi +
\TT_{\nu_2-\half } \xi^a\,,
\\[5pt]
\label{15052008-37} && \hspace{-0.3cm} \delta \phi = u \TT_{\nu_1-\half}
\xi\,,
\eeq
where $\xi^a$, $\xi$ are gauge transformation parameters.

Gauge invariant equations of motion obtained from Lagrangian \rf{gaufixlag01}
take the form
\beq
\label{15052008-31}&&\hspace{-1cm}  \Box_{\nu_2}\phi^{ab} -\partial^a C^b -
\partial^b C^b - \frac{2\eta^{ab}}{d-2} \TT_{-\nu_2+\half} C =0\,,
\\[7pt]
\label{15052008-32}&& \hspace{-1cm} \Box_{\nu_1} \phi^a -\partial^a C -
\TT_{\nu_2-\half} C^a = 0\,,
\\[7pt]
\label{15052008-33}&& \hspace{-1cm} \Box_{\nu_0}\phi - u\TT_{\nu_1-\half} C =
0\,,
\eeq
where $\Box_\nu$ and $\nu$'s are defined in \rf{09072009-09} and
\rf{15052008-34} respectively. We see that the gauge invariant equations of
motion are coupled.

Using equations of motion \rf{15052008-31}-\rf{15052008-33} in bulk action
\rf{09072009-02} with Lagrangian \rf{gaufixlag01}, we obtain boundary
effective action \rf{09072009-08} with $\LL_\eff$ given by
\beq \label{09072009-12}
\LL_\eff & = & \frac{1}{4} \phi^{ab} \TT_{\nu_2-\half}\phi^{ab} - \frac{1}{8}
\phi^{aa} \TT_{\nu_2 -\half } \phi^{bb}
\nonumber\\[5pt]
& + &  \half \phi^a \TT_{\nu_1 -\half } \phi^a + \half \phi \TT_{\nu_0
-\half} \phi
\nonumber\\[5pt]
& - & \half \phi^a C^a + (\frac{1}{4} \phi^{aa} - \frac{u}{2}\phi) C\,.
\eeq

Now we would like to demonstrate how use of the modified de Donder gauge
condition provides considerable simplification in solving the equations of
motion and computing effective action \rf{09072009-08}. To this end we note
that it is the quantities $C^a$, $C$ given in
\rf{15052008-29},\rf{15052008-30} that define the modified de Donder gauge
condition,
\be \label{09072009-13} C^a=0\,,\quad C = 0\qquad \hbox{modified de Donder
gauge} \,.\ee
Using this gauge condition in equations of motion
\rf{15052008-31}-\rf{15052008-33} gives the following surprisingly simple
gauge fixed equations of motion:
\beq
\label{15052008-31xxx}&& \Box_{\nu_2}\phi^{ab}=0\,,
\\[7pt]
\label{15052008-32xxx}&& \Box_{\nu_1} \phi^a=0\,,
\\[7pt]
\label{15052008-33xxx}&& \Box_{\nu_0} \phi=0\,.
\eeq
Thus, we see that the gauge fixed equations of motions are decoupled. Using
modified de Donder gauge \rf{09072009-13} in \rf{09072009-12}, we obtain
\beq \label{09072009-14}
\LL_\eff\Bigr|_{{\,C^a= 0 \atop C=0}} & = & \frac{1}{4} \phi^{ab}
\TT_{\nu_2-\half}\phi^{ab} - \frac{1}{8} \phi^{aa} \TT_{\nu_2 -\half }
\phi^{bb}
\nonumber\\[5pt]
& + &  \half \phi^a \TT_{\nu_1 -\half } \phi^a + \half \phi \TT_{\nu_0
-\half} \phi\,,
\eeq
i.e. we see that $\LL_\eff$ is also simplified.

In order to find $S_\eff$ we should solve equations of motion
\rf{15052008-31xxx}-\rf{15052008-33xxx} with the Dirichlet problem
corresponding to the boundary shadow field and plug the solution into
$\LL_\eff$. To this end we discuss solution of equations of motion
\rf{15052008-31xxx}-\rf{15052008-33xxx}. Because our equations of motion take
decoupled form and similar to the equations of motion for the massive scalar
$AdS$ field we can apply the procedure described in Sec. \ref{sec-scal-f-01}.
Doing so, we obtain solution of equation
\rf{15052008-31xxx}-\rf{15052008-33xxx} with the Dirichlet problem
corresponding to the spin-2 shadow field,
\beq
\label{10072009-17} \phi^{ab}(x,z) & = & \sigma_{2,\nu_2}\! \int\! d^dy\,
G_{\nu_2}(x-y,z) \phi_\sh^{ab}(y)\,, \qquad
\\[5pt]
\label{10072009-18} \phi^a(x,z) & = & \sigma_{2,\nu_1}\! \int\! d^dy\,
G_{\nu_1}(x-y,z) \phi_\sh^a(y)\,,
\\[5pt]
\label{10072009-19} \phi(x,z) & = & \sigma_{2,\nu_0}\! \int\! d^dy\,
G_{\nu_0}(x-y,z) \phi_\sh(y)\,,
\\[5pt]
\label{10072009-17x}&& \sigma_{2,\nu_2} = 1\,,
\\[5pt]
\label{10072009-18x}&& \sigma_{2,\nu_1} = -\frac{1}{d-2}\,,
\\[5pt]
\label{10072009-19x}&& \sigma_{2,\nu_0} = \frac{1}{(d-2)(d-4)}\,,
\eeq
where the Green function $G_\nu$ is given in \rf{10072009-12}, while $\nu$'s
are defined in \rf{15052008-34}. Choice of normalization factor
$\sigma_{2,\nu_2}$ \rf{10072009-17x} is a matter of convention. The remaining
normalization factors  $\sigma_{2,\nu_1}$, $\sigma_{2,\nu_0}$
\rf{10072009-18x},\rf{10072009-19x} are uniquely determined by requiring that
modified de Donder gauge conditions \rf{09072009-13} be amount to the
differential constraints for the spin-2 shadow field (see Table I). The
derivation of the normalization factor $\sigma_{s,\nu}$ for arbitrary
spin-$s$ $AdS$ field may be found in Appendix \ref{man02-app-03}.

Using asymptotic behavior of the Green function given in $G_\nu$
\rf{10072009-14}, we find the asymptotic behavior of our solution
\beq
&& \label{10072009-20} \hspace{-1.2cm} \phi^{ab}(x,z) \,
\stackrel{z\rightarrow 0 }{\longrightarrow}\,  z^{-\nu_2 + \half}
\phi_\sh^{ab}(x)\,,
\\[5pt]
\label{10072009-21} && \hspace{-1.2cm} \phi^a(x,z) \,\,
\stackrel{z\rightarrow 0 }{\longrightarrow}\,\, -\frac{z^{-\nu_1 +
\half}}{d-2} \phi_\sh^a(x)\,,
\\[5pt]
\label{10072009-22} && \hspace{-1.2cm} \phi(x,z) \,\, \stackrel{z\rightarrow
0 }{\longrightarrow}\,\, \frac{z^{-\nu_0 + \half}}{(d-2)(d-4)} \phi_\sh(x)\,.
\eeq
From these expressions, we see that our solution
\rf{10072009-17}-\rf{10072009-19} has indeed asymptotic behavior
corresponding to the spin-2 shadow field.

Finally, to obtain the effective action we plug solution of the Dirichlet
problem for $AdS$ fields, \rf{10072009-17}-\rf{10072009-19} into
\rf{09072009-08}, \rf{09072009-14}. Using general formula given in
\rf{man02-21072009-15}, we obtain
\beq \label{10072009-22x1}
-S_\eff & = & d c_{\nu_2} \Gamma \,,
\\[5pt]
\label{10072009-22x2}&& c_{\nu_2} =
\frac{\Gamma(d)}{\pi^{d/2}\Gamma(\frac{d}{2})}\,,
\eeq
where $\Gamma$ is gauge invariant two-point vertex of the spin-2 shadow field
given in \rf{manus2009-02-01},\rf{manus2009-02-07}.

To summarize, {\it using the modified de Donder gauge for the massless spin-2
$AdS$ field and computing the bulk action on solution of equations of motion
with the Dirichlet problem corresponding to the boundary shadow field we
obtain the gauge invariant two-point vertex of the spin-2 shadow field}.

Because in the literature $S_\eff$ is expressed in terms of two-point vertex
taken in the Stueckelberg gauge frame, $\Gamma_{12}^{{\rm stand}}$
\rf{manus2009-02-12x}, we use \rf{manus2009-02-12}, \rf{manus2009-02-13} to
represent our result \rf{10072009-22x1} as
\be \label{10072009-22x3} - S_\eff = \frac{d(d+1)}{4(d-1)} c_{\nu_2}
\Gamma^{{\rm stand}} \,.\ee
This relation, by using the radial gauge for $AdS$ fields, was obtained in
Ref.\cite{Liu:1998bu}. Note that we have obtained more general relation given
in \rf{10072009-22x1}, while relation \rf{10072009-22x3} is obtained from
\rf{10072009-22x1} by using the Stueckelberg gauge frame. The fact that
$S_\eff$ is related to $\Gamma^{{\rm stand}}$ is expected because of the
conformal symmetry. What is important for the systematical study of $AdS/CFT$
correspondence is the computation of the normalization factor in front of
$\Gamma^{{\rm stand}}$ \rf{10072009-22x3}. We note that our normalization
factor in \rf{10072009-22x3} coincides with the one found in
Ref.\cite{Liu:1998bu}.

\section{ AdS/CFT correspondence for arbitrary spin field
}\label{man02-sec-06}

We proceed with the discussion of $AdS/CFT$ correspondence for bulk massless
arbitrary spin-$s$ $AdS$ field and boundary spin-$s$ shadow field. To discuss
the correspondence we use the $CFT$ adapted gauge invariant Lagrangian and
{\it modified} de Donder gauge condition for
the massless arbitrary spin $AdS$ field found in Ref.\cite{Metsaev:2008ks}%
\footnote{ In light-cone gauge, $AdS/CFT$ correspondence for arbitrary spin
massless $AdS_{d+1}$ fields was studied in
Ref.\cite{Metsaev:1999ui,Metsaev:2002vr}. Recent interesting applications of
the {\it standard} de Donder-Feynman gauge to the various problems of
higher-spin fields may be found in
Refs.\cite{Guttenberg:2008qe,Manvelyan:2008ks,Fotopoulos:2009iw}. We believe
that our modified de Donder gauge will also be useful for better
understanding of various aspects of AdS/QCD correspondence which are
discussed e.g. in \cite{Andreev:2002aw}-\cite{Abidin:2008ku}.}.
We begin therefore with the presentation of our result in
Ref.\cite{Metsaev:2008ks}%
\footnote{ Representation for the Lagrangian, which we use in this paper, is
different from the one given in Ref.[34]. Namely, in this paper, we use $CFT$
adapted Lagrangian represented in terms of operator ${\cal T}_\nu$. This
operator was introduced in Ref.[35].}.

In $AdS_{d+1}$ space, massless spin-$s$ field is described by the following
scalar, vector, and totally symmetric tensor fields of the $so(d)$ algebra:
\be \label{18052008-02}
\phi^{a_1\ldots a_{s'}}\,, \hspace{1cm} s'=0,1,\ldots,s.
\ee
The fields $\phi^{a_1\ldots a_{s'}}$ with $s' \geq 4$ are double-traceless,%
\footnote{ In this paper, we adopt the formulation in terms of the double
traceless gauge fields \cite{Fronsdal:1978vb}. Discussion of various
formulations in terms of unconstrained gauge fields may be found in
\cite{Francia:2002aa}-\cite{Fotopoulos:2006ci}. For recent review, see
\cite{Fotopoulos:2008ka}. Discussion of other formulations which seem to be
most suitable for the theory of interacting fields may be found e.g. in
\cite{Alkalaev:2003qv}.}
\be \label{18052008-03} \phi^{aabba_5\ldots a_{s'}}=0\,, \hspace{1cm}
s'=4,5,\ldots,s. \ee

In order to obtain the gauge invariant description in an easy--to--use form
we use the oscillators and introduce a ket-vector $|\phi\rangle$ defined by
\beq
\label{18052008-04} && \hspace{-1cm}  |\phi\rangle \equiv \sum_{s'=0}^s
\alpha_z^{s-s'}|\phi_{s'}\rangle \,,
\nonumber\\[5pt]
&& \hspace{-1cm} |\phi_{s'}\rangle \equiv
\frac{\alpha^{a_1} \ldots \alpha^{a_{s'}}}{s'!\sqrt{(s - s')!}}
\, \phi^{a_1\ldots a_{s'}} |0\rangle\,.
\eeq
From \rf{18052008-04}, we see that the ket-vector $|\phi\rangle$ is
degree-$s$ homogeneous polynomial in the oscillators $\alpha^a$, $\alpha^z$,
while the ket-vector $|\phi_{s'}\rangle$ is degree-$s'$ homogeneous
polynomial in the oscillators $\alpha^a$. In terms of the ket-vector
$|\phi\rangle$, double-tracelessness constraint \rf{18052008-03} takes the
form
\beq
 \label{10072009-01} && (\bar{\alpha}^2)^2 |\phi\rangle  = 0 \,.
\eeq

The $CFT$ adapted gauge invariant Lagrangian is given by
\beq \label{07072009-03}
\LL &= &  \half \langle \partial^a \phi|\mubf | \partial^a \phi\rangle +\half
\langle \TT_{\nu-\half} \phi| \mubf | \TT_{\nu-\half} \phi\rangle
\nonumber\\[5pt]
&  - & \half \langle \bar{C}\phi|| \bar{C}\phi\rangle\,,
\eeq
where $\TT_\nu$ is defined in \rf{09072009-04} and we use the notation
\beq
\label{080405-01add} && \hspace{-0.7cm} \Cb \equiv  \albpar - \half \alpar
\bar\alpha^2 - \eb_1\Pi^\smponetwo  + \half e_1 \bar\alpha^2 \,,
\\[5pt]
\label{18052008-09} && e_1 = e_{1,1} \TT_{\nu-\half}\,,
\qquad
\label{18052008-10} \eb_1 = \TT_{-\nu+\half} \eb_{1,1}\,,\qquad
\\[5pt]
\label{18052008-11} && e_{1,1} =  - \alpha^z \ewt_1
\qquad \quad
\eb_{1,1} =  - \ewt_1 \bar\alpha^z\,,
\\[6pt]
\label{18052008-12new} && \ewt_1 =
\Bigl(\frac{2s+d-4-N_z}{2s+d-4-2N_z}\Bigr)^{1/2}\,,
\\[5pt]
 \label{10072009-02} && \mubf \equiv 1- \frac{1}{4}\alpha^2\bar\alpha^2\,,
\\[5pt]
&& \label{18052008-14}  \nu \equiv s + \frac{d-4}{2} - N_z\,,
\eeq
where $\Pi^\smponetwo$ is given in \rf{18052008-08}. Lagrangian
\rf{07072009-03} is invariant under the gauge transformation
\be
\label{18052008-15} \delta \phik =  ( \alpar - e_1  - \alpha^2
\frac{1}{2N_\alpha+d-2}\eb_1 ) \xik \,.
\ee
In terms of the $so(d)$ algebra tensor fields, the ket-vector $\xik$ is
represented as
\beq
\label{08092008-06} && |\xi\rangle \equiv \sum_{s'=0}^{s-1}
\alpha_z^{s-1-s'}|\xi_{s'}\rangle \,,
\\[5pt]
\label{08092008-07} && |\xi_{s'}\rangle \equiv
\frac{\alpha^{a_1} \ldots \alpha^{a_{s'}}}{s'!\sqrt{(s -1 - s')!}}
\, \xi^{a_1\ldots a_{s'}} |0\rangle\,, \ \ \ \ \ \ \ \
\eeq
where gauge transformation parameters are traceless, $\xi^{aaa_3\ldots
a_{s'}}=0$\,.

Gauge invariant equations of motion obtained from Lagrangian \rf{07072009-03}
take the form
\be \label{10072009-04}  \mubf \Box_\nu \phik - C \Cb \phik =0\,, \ee
where $C$ is given by
\be \label{10072009-05}
C \equiv \alpar - \half \alpha^2 \albpar  - e_1 \Pi^\smponetwo + \half \eb_1
\alpha^2\,.\ee
Note that for the derivation of equations of motion \rf{10072009-04} we use
the relations $C^\dagger= - \Cb$ and
\be  \label{10072009-06} \TT_{\nu-\half}^\dagger\TT_{\nu-\half} =
-\partial_z^2 + \frac{1}{z^2}(\nu^2-\frac{1}{4})\,.\ee

Using equations of motion \rf{10072009-04} in bulk action \rf{09072009-02}
with Lagrangian \rf{07072009-03}, we obtain boundary effective action
\rf{09072009-08} with the following $\LL_\eff$:
\beq  \label{10072009-03}
\LL_\eff & = &   \half \langle \phi | \mubf | \TT_{\nu-\half} \phi\rangle
\nonumber\\[5pt]
& + & \langle (\frac{1}{4}\eb_{1,1} - \half e_{1,1}\bar\alpha^2 ) \phi||
\bar{C}\phi\rangle\,.
\eeq

We now, as before, demonstrate how use of the modified de Donder gauge
condition provides considerable simplification in solving equations of motion
\rf{10072009-04} and computing effective action \rf{09072009-08}%
\footnote{ Powerful methods of solving $AdS$ field equations of motion based
on star algebra products in auxiliary spinor variables are discussed in
Refs.\cite{Bolotin:1999fa,Didenko:2009td}. One of interesting features of
these methods is that they do not use any gauge conditions when solving the
equations of motion.}.
To this end we note that it is the operator $\Cb$ given in \rf{080405-01add}
that defines the modified de Donder gauge condition,
\be \label{10072009-07} \Cb\phik=0\,, \qquad \hbox{modified de Donder gauge}
\,.\ee
Using this gauge condition in \rf{10072009-04} gives the following
surprisingly simple gauge fixed equations of motion:
\be \label{10072009-08} \Box_\nu \phik = 0\,, \ee
where $\Box_\nu$ and $\nu$ are given in \rf{09072009-09} and \rf{18052008-14}
respectively. Note that for the derivation of Eq.\rf{10072009-08} we use the
fact that kernel of operator $\mubf$ \rf{10072009-02} is trivial on space of
the double-traceless ket-vectors. Thus, we see that the modified de Donder
gauge leads to the decoupled equations of motion.

Accordingly, using the modified de Donder gauge in \rf{10072009-03} leads to
the simplified expression for $\LL_\eff$,
\beq \label{10072009-08x1}
\LL_\eff\Bigr|_{\Cb\phik=0}  =    \half \langle \phi | \mubf |
\TT_{\nu-\half} \phi\rangle\,.
\eeq

In order to find $S_\eff$ we should solve equations of motion
\rf{10072009-08} with the Dirichlet problem corresponding to the boundary
shadow field and plug the solution into $\LL_\eff$. To this end we discuss
solution of equations of motion \rf{10072009-08}. Solution of equation
\rf{10072009-08} with the Dirichlet problem corresponding to the boundary
shadow field is given by
\beq \label{10072009-08x2}
|\phi(x,z)\rangle & = & \sigma_{s,\nu}\! \int\! d^dy\, G_{\nu}(x-y,z)
|\phi_\sh(y)\rangle\,,\qquad
\\[5pt]
\label{10072009-08x3} && \sigma_{s,\nu} = \frac{(-)^{\nu_s-\nu}
\Gamma(\nu)}{2^{\nu_s-\nu}\Gamma(\nu_s)} \,,
\\[5pt]
\label{10072009-08x5} && \nu_s = s+\frac{d-4}{2}\,,
\eeq
where the Green function $G_\nu$ and $\nu$ are given in \rf{10072009-12} and
\rf{18052008-14} respectively. The derivation of normalization factor
$\sigma_{s,\nu}$ \rf{10072009-08x3} may be found in Appendix
\ref{man02-app-03}.

Using asymptotic behavior of the Green function given in $G_\nu$
\rf{10072009-14}, we find the asymptotic behavior of our solution
\beq \label{10072009-08x4}
&& |\phi(x,z)\rangle \,\, \stackrel{z\rightarrow 0 }{\longrightarrow}\,\,
\sigma_{s,\nu}  z^{-\nu + \half} |\phi_\sh(x)\rangle\,.\qquad
\eeq
From this expression, we see that our solution \rf{10072009-08x2} has indeed
asymptotic behavior corresponding to the spin-$s$ shadow field. Finally,
plugging our solution \rf{10072009-08x2} into \rf{10072009-08x1} and using
general formula given in \rf{man02-21072009-15}, we obtain  the effective
action
\beq \label{man02-20072009-01}
-S_\eff & = & (2s+d-4) c_{\nu_s} \Gamma \,,
\\[9pt]
\label{man02-20072009-02} && c_{\nu_s} = \frac{\Gamma(s+d-2)}{\pi^{d/2}
\Gamma(s+\frac{d-4}{2})}\,,
\eeq
where $\Gamma$ stands for two-point gauge invariant vertex \rf{11062009-10}
of the spin-$s$ shadow field.

To summarize, {\it imposing the modified de Donder gauge on the massless
spin-$s$ $AdS$ field and computing the bulk action on solution of equations
of motion with the Dirichlet problem corresponding to the boundary shadow
field we obtain the gauge invariant two-point vertex of the spin-$s$ shadow
field}.

Because in the literature $S_\eff$ is expressed in terms of two-point vertex
taken in the Stueckelberg gauge frame, $\Gamma^{{\rm stand}}$
\rf{man02-12072009-02}, we use \rf{12062009-01} and represent our result
\rf{man02-20072009-01} as
\beq \label{man2009-02-14072009-03}
-S_\eff & = & \frac{(2s+d-3)(2s+d-4)}{2s!(s+d-3)} c_{\nu_s} \Gamma^{{\rm
stand}} \,.\qquad
\eeq
For the particular values $s=1$ and $s=2$, our normalization factor in front
of $\Gamma^{{\rm stand}}$ \rf{man2009-02-14072009-03} coincides with the
respective normalization factors given in
\rf{man2009-02-14072009-02},\rf{10072009-22x3}. Thus, our result agrees with
the previously reported results for the particular values $s=1,2$ (see
Refs.\cite{Freedman:1998tz,Liu:1998bu,Mueck:1998ug}) and gives the
normalization factor for arbitrary values of $s$ and $d$.%
\footnote{ For massless arbitrary spin $AdS_5$ fields, the effective action,
by using radial gauge, was studied in \cite{Germani:2004jf}. Normalization
factor in Ref.\cite{Germani:2004jf} disagrees with our result and results
reported in Refs.\cite{Freedman:1998tz,Liu:1998bu,Mueck:1998ug}.}
Knowledge of the normalization factor for arbitrary values of $s$ is
important for the systematical study of $AdS/CFT$ correspondence because
higher-spin gauge field theories \cite{Vasiliev:1990en,Vasiliev:2003ev}
involve infinite tower of $AdS$ fields with all values of $s$,
$s=0,1,\ldots,\infty$.

As a side of remark we note that the modified de Donder gauge and the
gauge-fixed equations of motion have the on-shell left-over gauge symmetry.
Namely, modified de Donder gauge \rf{10072009-07} and gauge-fixed equations
of motion \rf{10072009-08} are invariant under gauge transformation
\rf{18052008-15} provided the gauge transformation parameter satisfies the
equation
\be \label{man02-24072009-01} \Box_\nu |\xi\rangle= 0\,,\ee
where $\Box_\nu$ and $\nu$ are given in \rf{09072009-09} and \rf{18052008-14}
respectively.  Solution to Eq.\rf{man02-24072009-01} is given by
\be |\xi(x,z)\rangle  =  \sigma_{s,\nu}\int d^dy\, G_\nu (x-y,z)
|\xi_\sh(y)\rangle\,.\ee
Plugging this solution in \rf{18052008-15} we find the following expression
for the on-shell left-over gauge transformation of solution of the Dirichlet
problem:
\be \label{man02-24072009-02}
\delta |\phi\rangle  =  \sigma_{s,\nu} \int d^dy\, G_\nu (x-y,z) \delta
|\phi_\sh(y)\rangle\,,
\ee
where $\delta |\phi_\sh\rangle$ is the gauge transformation of the spin-$s$
shadow field (see Table II). From \rf{man02-24072009-02}, we see that the
on-shell left-over gauge symmetry of solution of the Dirichlet problem for
spin-$s$ $AdS$ field amounts to the gauge symmetry of the spin-$s$ shadow
field (see Table II). It is {\it matching of the on-shell leftover gauge
symmetry of solution of the Dirichlet problem for $AdS$ field and the gauge
symmetry of the shadow field that explains why the effective action coincides
with the gauge invariant two-point vertex for the boundary shadow field}.

\section{Conformal fields}\label{man02-sec-07}

The kernel of two-point vertex \rf{11062009-10} is not well-defined when $d$
is even integer and $\nu$ takes integer values (see e.g.
\cite{Aref'eva:1998nn}). However this kernel can be regularized and after
that it turns out that the leading logarithmic divergence of the two-point
vertex $\Gamma$ leads to Lagrangian of conformal fields. To explain what has
just been said we note that the kernel of $\Gamma$ can be regularized by
using dimensional regularization. This is to say that using the dimensional
regularization and denoting the integer part of $d$ by $[d]$, we introduce
the regularization parameter $\epsilon$ as
\be  \label{man02-20072009-29} d- [d]= - 2\epsilon\,,\qquad [d]-\hbox{even
integer} \,.\ee
With this notation we have the following behavior of the regularized
expression for the kernel in \rf{11062009-10}:
\beq \label{man02-20072009-30}
&& \frac{1}{|x|^{2\nu+d}}\,\,\, \stackrel{\epsilon \sim
0}{\mbox{\Large$\sim$}}\,\,\, \frac{1}{\epsilon} \varrho_\nu \Box^\nu
\delta(x)\,,
\\[5pt]
&&  \label{man02-20072009-31} \varrho_\nu = \frac{\pi^{d/2}}{4^\nu \Gamma(\nu
+ 1)\Gamma(\nu + \frac{d}{2})}\,,
\eeq
when $\nu$ is integer. Note that, in view of \rf{18052008-14}, $\nu$ takes
integer values when $d$ is even integer. Using \rf{man02-20072009-30} in
\rf{manus2009-02-01},\rf{11062009-10}, we obtain
\be  \label{man02-20072009-32} \Gamma \,\,\, \stackrel{ \epsilon \sim
0}{\mbox{\Large$\sim$}}\,\,\, \frac{1}{\epsilon} \varrho_{\nu_s} \int
d^dx\,\, \LL\,, \ee
where $\nu_s$ is defined in \rf{10072009-08x5} and $\LL$ is a
higher-derivative Lagrangian for conformal spin-$s$ field. We now discuss the
Lagrangian for spin $s=1,2$ and arbitrary spin-$s$ conformal fields in turn.

\subsection{Spin-1 conformal filed}

Using two-point vertex for spin-1 shadow field \rf{manus2009-02-02} and
adopting relation \rf{man02-20072009-30} to the case of $s=1$, we obtain the
following gauge invariant Lagrangian for the spin-1 conformal field:
\be \label{26062009-01} \LL = \half \phi^a \Box^{k+1}\phi^a  + \half \phi
\Box^k \phi \,, \qquad k = \frac{d-4}{2}\,, \ee
where we have made the identification
\be \label{man02-21072009-01} \phi^a = \phi_\sh^a\,,\qquad
\phi=\phi_\sh\,.\ee
Using this identification, we note that the differential constraint for
spin-1 shadow field (see Table I) implies the same differential constraint
for $\phi^a$ and $\phi$,
\be \label{15052008-02new} \partial^a \phi^a + \phi =  0\,. \ee
Lagrangian \rf{26062009-01} and constraint \rf{15052008-02new} are invariant
under gauge transformations
\beq
\label{14052008-08sh} && \delta \phi^a = \partial^a \xi\,,
\\[3pt]
\label{14052008-09sh} && \delta \phi=  - \Box \xi \,.
\eeq
To check gauge invariance of the Lagrangian we use the notation  $C$ for the
left hand side of \rf{15052008-02new} and note that gauge variation of
Lagrangian \rf{26062009-01} takes the form
\be \delta \LL =  - \xi \Box^{k+1} C\,, \ee
i.e. we see that Lagrangian \rf{26062009-01} is indeed invariant when $C=0$.

{\bf Interrelation between our approach and standard approach.} Standard
formulation of the spin-1 conformal field is obtained from our approach by
solving the differential constraint. This is to say that using differential
constraint \rf{15052008-02new} we express the scalar field in terms of the
vector field,
\be \label{man02-21072009-02} \phi= - \partial^a \phi^a\,, \ee
and plug $\phi$ \rf{man02-21072009-02} in Lagrangian \rf{26062009-01}. By
doing so, we obtain the standard Lagrangian for the spin-1 conformal field,
\be \LL = -\frac{1}{4} F^{ab}\Box^k F^{ab}\,,\qquad F^{ab} =
\partial^a\phi^b - \partial^b \phi^a \,. \ee

{\bf Light-cone gauge Lagrangian}. We now consider light-cone gauge
Lagrangian for the spin-1 conformal field. As usually, the light-cone gauge
frame is achieved through the use of the light-cone gauge and differential
constraints. Gauge transformations \rf{14052008-08sh},\rf{14052008-09sh} and
differential constraint \rf{15052008-02new} of the spin-1 conformal field
take the same form as the ones for spin-1 shadow field (see Table I).
Therefore to discuss the light-cone gauge for the spin-1 conformal field we
can use results obtained for the spin-1 shadow field in Sec.
\ref{man02-sec-02}. This is to say that light-cone gauge and solution for
differential constraint for the spin-1 conformal field are obtained from the
respective expressions \rf{man02-16072009-01} and \rf{man02-16072009-02} by
using identification \rf{man02-21072009-01}. Doing so, we are left with
$so(d-2)$ algebra vector field $\phi^i$ and scalar field $\phi$. These fields
constitute the field content of the light-cone gauge frame. Note that, in
contrast to the standard approach, the scalar field $\phi$ becomes an
independent field D.o.F in the light-cone gauge frame. Making use of the
light-cone gauge in gauge invariant Lagrangian \rf{26062009-01}, we obtain
the light-cone gauge Lagrangian
\be \label{man02-21072009-03} \LL^{{\rm l.c.}} = \half \phi^i
\Box^{k+1}\phi^i + \half \phi \Box^k \phi \,, \qquad k = \frac{d-4}{2}\,. \ee

\subsection{Spin-2 conformal filed}

We proceed with the discussion of spin-2 conformal field. Using two-point
vertex for the spin-2 shadow field \rf{manus2009-02-07} and adopting relation
\rf{man02-20072009-30} to the case of $s=2$, we obtain the gauge invariant
Lagrangian for the spin-2 conformal field,
\beq
\label{26062009-02}
&& \hspace{-1cm} \LL  =  \frac{1}{4} \phi^{ab} \Box^{k+1}\phi^{ab} -
\frac{1}{8} \phi^{aa} \Box^{k+1}\phi^{bb}
\nonumber\\[5pt]
&&\hspace{-0.5cm}  + \half \phi^a \Box^k \phi^a + \half \phi \Box^{k-1} \phi
\,, \qquad k = \frac{d-2}{2}\,,\qquad
\eeq
where we have made the identification
\be \label{man02-21072009-04} \phi^{ab} = \phi_\sh^{ab}\,, \qquad \phi^a =
\phi_\sh^a\,, \qquad \phi=\phi_\sh\,.\ee
Using this identification, we note that the differential constraints for the
spin-2 shadow field (see Table I) imply the same differential constraints for
the fields $\phi^{ab}$, $\phi^a$, and $\phi$,
\beq
\label{05152008-10new} && \partial^b \phi^{ab} - \half \partial^a \phi^{bb} +
\phi^a = 0 \,,
\\[7pt]
\label{05152008-11new}&& \partial^a \phi^a + \half \Box \phi^{aa} + u \phi =
0 \,,
\eeq
\be u =\Bigl(2\frac{d-1}{d-2}\Bigr)^{1/2}\,.\ee
The Lagrangian and the constraints are invariant under the gauge
transformations
\beq
\label{14052008-11} && \delta \phi^{ab} =\partial^a \xi^b +
\partial^b \xi^a + \frac{2}{d-2} \eta^{ab} \xi\,,
\\[7pt]
\label{14052008-12} && \delta \phi^a = \partial^a \xi - \Box \xi^a \,,
\\[7pt]
\label{14052008-13} && \delta \phi = - u \Box \xi\,.
\eeq
To demonstrate gauge invariance of the Lagrangian we use the notation $C^a$
and $C$ for the respective left hand sides of \rf{05152008-10new} and
\rf{05152008-11new} and find that gauge variation of Lagrangian
\rf{26062009-02} takes the form
\be \label{man02-21072009-05} \delta \LL = - \xi^a \Box^{k+1} C^a  - \xi
\Box^k C\,, \ee
i.e. we see that Lagrangian \rf{26062009-02} is indeed invariant when
$C^a=0$, $C=0$.

{\bf Interrelation between our approach and standard approach.} Standard
formulation of the spin-2 conformal field is obtained from our approach as
follows. First, we use differential constraints \rf{05152008-10new},
\rf{05152008-11new} and express the vector field and scalar field in terms of
the field $\phi^{ab}$,
\beq
\label{man02-21072009-06} &&  \phi^a = - \partial^b \phi^{ab} + \half
\partial^a \phi^{bb}\,,
\\[5pt]
\label{man02-21072009-07} && \phi= \frac{1}{u} ( \partial^a  \partial^b
\phi^{ab} - \Box \phi^{aa})\,. \eeq
Second, we plug $\phi^a$, $\phi$ \rf{man02-21072009-06},
\rf{man02-21072009-07} in Lagrangian \rf{26062009-02} and obtain the standard
Lagrangian for the spin-2 conformal field,
\beq \label{07072009-01}
\LL & = & \frac{1}{4} \phi^{ab}\Box^{k+1} \phi^{ab} -\frac{1}{4(d-1)}
\phi^{aa}\Box^{k+1} \phi^{bb}
\nonumber\\[5pt]
& + & \half (\partial\phi)^a\Box^k (\partial\phi)^a + \frac{1}{2(d-1)}
\phi^{aa} \Box^k (\partial\partial\phi)
\nonumber\\[5pt]
& + & \frac{d-2}{4(d-1)} (\partial\partial\phi) \Box^{k-1}
(\partial\partial\phi)\,,
\\[5pt]
&&  k =\frac{d-2}{2}\,,
\eeq
\be (\partial\phi)^a \equiv
\partial^b \phi^{ab}\,,\qquad (\partial\partial\phi)=
\partial^a\partial^b\phi^{ab}\,. \ee
Lagrangian \rf{07072009-01} is invariant under gauge transformation of
$\phi^{ab}$ given in \rf{14052008-11}.

As is well known, Lagrangian \rf{07072009-01} can be represented in terms of
the linearized Ricci tensor and Ricci scalar
\be \label{07072009-02} \LL = R^{ab} \Box^{k-1} R^{ab} - \frac{d}{4(d-1)}
R\Box^{k-1}R \,, \ee
or equivalently in terms of the Weyl tensor
\be \label{Lconfie2lag01}  \LL= \frac{1}{ q^2}C^{abce} \Box^{k-1}
C^{abce}\,,\qquad  q^2 \equiv 4\frac{d-3}{d-2}\,. \ee
For the derivation of relation \rf{07072009-02}, we use the following
expressions for the linearized Ricci tensor and Ricci scalar:
\beq
&&\hspace{-1.1cm}  R^{ab} = \half ( -\Box\phi^{ab} + \partial^a
(\partial\phi)^b +
\partial^b (\partial\phi)^a - \partial^a \partial^b \phi^{cc}),
\\[5pt]
&& \hspace{-1.1cm} R = (\partial\partial\phi) -\Box \phi^{aa}\,,
\eeq
while for the derivation of relation \rf{Lconfie2lag01} we use the fact that
the Gauss-Bonnet combination taken at second order in the field $\phi^{ab}$
is a total derivative,
\be R^{abce}R^{abce} - 4R^{ab}R^{ab} +R^2= 0 \ \ \ \ (\hbox{up to total
deriv}.)\,.\quad \ee

{\bf Light-cone gauge Lagrangian}. As before, the light-cone gauge frame is
achieved through the use of light-cone gauge and differential constraints.
Gauge transformations \rf{14052008-11}-\rf{14052008-13} and differential
constraints \rf{05152008-10new},\rf{05152008-11new} for the spin-2 conformal
field take the same form as the ones for the spin-2 shadow field (see Table
I). Therefore to discuss the light-cone gauge for the spin-2 conformal field
we can use results obtained for the spin-2 shadow field in Sec.
\ref{man02-sec-01}. This is to say that the light-cone gauge fixing and
solution for differential constraints for the spin-2 conformal field are
obtained from the respective expressions \rf{man02-16072009-04} and
\rf{man02-16072009-05}-\rf{man02-16072009-08} by using identification
\rf{man02-21072009-04}. Doing so, we are left with traceless tensor field
$\phi^{ij}$, vector field $\phi^i$ and scalar field $\phi_\sh$. These fields
constitute the field content of the light-cone gauge frame. Note that, in
contrast to the standard approach, the vector field $\phi^i$ and scalar field
$\phi$ become independent field D.o.F in the light-cone gauge frame. Making
use of the light-cone gauge in gauge invariant Lagrangian \rf{26062009-02},
we obtain the light-cone gauge Lagrangian for the spin-2 conformal field,
\be
\LL^{{\rm l.c.}}  =  \frac{1}{4} \phi^{ij} \Box^{k+1}\phi^{ij}
+ \half \phi^i \Box^k \phi^i + \half \phi \Box^{k-1} \phi \,,
\ee
where $k = \frac{d-2}{2}$.

%%%%%%%%%%%%%%%%%%%%%%%%%%%%%%%%%%%%%%%%%%%%%%%%%%%%%%%%%%%%%%%%%%%%%%
%%%%%%%%%%%%%%%%%%%%%%%%%%%%%%%%%%%%%%%%%%%%%%%%%%%%%%%%%%%%%%%%%%%%%%
\subsection{Arbitrary spin-$s$ conformal filed}

%%%%%%%%%%%%%%%%%%%%%%%%%%%%%%%%%%%%%%%%%%%%%%%%%%%%%%%%%%%%%%%%%%%%%%
%%%%%%%%%%%%%%%%%%%%%%%%%%%%%%%%%%%%%%%%%%%%%%%%%%%%%%%%%%%%%%%%%%%%%%

We now discuss arbitrary spin-$s$ conformal field. Using two-point vertex for
spin-$s$ shadow field \rf{11062009-10} and relation \rf{man02-20072009-30},
we obtain the Lagrangian for the spin-$s$ conformal field,
\beq
&& \label{27062009-01} \hspace{-1cm} \LL = \half \phibr \mubf \Box^\nu \phik
\,,
\\[5pt]
\label{man02-21072009-08} && \hspace{-0.4cm} \mubf \equiv 1
-\frac{1}{4}\alpha^2\bar\alpha^2\,,
\\[5pt]
&& \hspace{-0.4cm} \nu \equiv s+ \frac{d-4}{2}-N_z \,,
\eeq
where we have made the identification
\be \label{man02-21072009-09} \phik   =  |\phi_\sh\rangle\,.\ee
Using this identification, we note that the differential constraint for the
spin-$s$ shadow field (see Table II) implies the same differential constraint
for the ket-vector $\phik$,
\beq
\label{17052008-05sh} && \hspace{-1cm} \Cb \phik  =  0 \,,
\\[7pt]
\label{17052008-07sh} && \Cb  =  \Cb_\perp  - \eb_1 \Pi^\smponetwo + \half
e_1 \bar\alpha^2 \Box\,,
\\[7pt]
\label{17052008-06sh} && \bar{C}_\perp = \albpar - \half \alpar
\bar\alpha^2\,,
\\[7pt]
\label{17052008-08sh} &&  \Pi^\smponetwo = 1 -\alpha^2 \frac{1}{2(2N_\alpha
+d)}\bar\alpha^2\,,
\\[7pt]
\label{17052008-09sh} && e_1 =  \alpha^z \ewt_1\,,
\qquad
\eb_1 =  - \ewt_1 \bar\alpha^z\,,
\\[7pt]
\label{masewtdef01sh} && \ewt_1 =
\Bigl(\frac{2s+d-4-N_z}{2s+d-4-2N_z}\Bigr)^{1/2}\,.
\eeq
Lagrangian \rf{27062009-01} and constraint \rf{17052008-05sh} are invariant
under the gauge transformation
\be \label{gautraarbspi01sh}
\delta |\phi\rangle  =   ( \alpar - e_1 \Box  -  \alpha^2\frac{1}{2N_\alpha +
d- 2} \eb_1) |\xi\rangle \,,
\ee
where ket-vector $\xik$ takes the same form as the ket-vector
$|\xi_\sh\rangle$ given in \rf{man02-22072009-02}.

To demonstrate gauge invariance of Lagrangian \rf{27062009-01} we find that
variation of the Lagrangian under gauge transformation \rf{gautraarbspi01sh}
takes the form
\be \label{27062009-02} \delta \LL = - \xibr\Box^\nu \Cb \phik\,,\ee
i.e. we see that $\LL$ is indeed gauge invariant provided the ket-vector
$\phik$ satisfies differential constraint \rf{17052008-05sh}.

To illustrate the structure of the Lagrangian we note that, in terms of
tensor fields $\phi^{a_1\ldots a_{s'}}$ defined as
\beq
&&  |\phi\rangle \equiv \sum_{s'=0}^s \alpha_z^{s-s'}|\phi_{s'}\rangle \,,
\nonumber\\[5pt]
\label{22072009-03} && |\phi_{s'}\rangle \equiv
\frac{\alpha^{a_1} \ldots \alpha^{a_{s'}}}{s'!\sqrt{(s-s')!}}
\, \phi^{a_1\ldots a_{s'}} |0\rangle\,,
\eeq
Lagrangian \rf{27062009-01} takes the form
\beq \label{man02-21072009-10}
&& \hspace{-1cm}  \LL = \sum_{s'=0}^s \LL_{s'}\,,
\\[5pt]
\label{man02-21072009-11} && \hspace{-1cm} \LL_{s'} = \frac{1}{2 s'!}\Bigl(
\phi^{a_1\ldots a_{s'}} \Box^{\nu_{s'}} \phi^{a_1\ldots a_{s'}}
\nonumber\\[7pt]
&& \hspace{-0.4cm} - \frac{s'(s'-1)}{4} \phi^{aaa_3\ldots a_{s'}}
\Box^{\nu_{s'}} \phi^{bba_3\ldots a_{s'}}\Bigr)\,,
\\[7pt]
\label{man02-21072009-12} && \hspace{-0.4cm} \nu_{s'} = s'+\frac{d-4}{2}\,.
\eeq

{\bf Stueckelberg gauge frame}. Standard formulation of the conformal field
is obtained from our approach by using the Stueckelberg gauge frame.
Therefore to illustrate our approach, we now present Stueckelberg gauge fixed
Lagrangian of the spin-$s$ conformal field. The Stueckelberg gauge frame is
achieved through the use of the Stueckelberg gauge and differential
constraints. Gauge transformation \rf{gautraarbspi01sh} and differential
constraint \rf{17052008-05sh} for the spin-$s$ conformal field take the same
form as the ones for the spin-$s$ shadow field (see Table II). Therefore to
discuss the Stueckelberg gauge frame for the spin-$s$ conformal field we can
use results obtained for the spin-$s$ shadow field in Sec. \ref{man02-sec04}.
This is to say that, by imposing the Stueckelberg gauge, solution to
differential constraint for the spin-$s$ conformal field is obtained from
\rf{12062000-03xx}-\rf{12062000-04x1} by using identification
\rf{man02-21072009-09} and $|\phi_{\sh,s'}\rangle = |\phi_{s'}\rangle$,
$s'=0,1,\ldots,s$. After that, plugging $|\phi\rangle$ in
\rf{27062009-01} we obtain the Stueckelberg gauge frame Lagrangian,%
\footnote{ In the Stueckelberg gauge frame, Lagrangian of the arbitrary spin
conformal field for $d=4$ and $d\geq 4$ was discussed in
Ref.\cite{Fradkin:1985am} and Ref.\cite{Segal:2002gd} respectively. Our
expression \rf{22072009-01} provides the explicit representation for
Lagrangian of the conformal arbitrary spin field. Note that Lagrangian in
\cite{Fradkin:1985am,Segal:2002gd} and the one given in \rf{22072009-01}
involve the higher derivatives. Discussion of ordinary-derivative Lagrangian
of the conformal arbitrary spin field may be found in \cite{Metsaev:2007fq}.}
\beq
\label{22072009-01} \LL & = & \half \sum_{s'=0}^s
\frac{2^{s-s'}(s'+\frac{d-2}{2})_{s-s'}}{(s-s')! (s+s'+d-3)_{s-s'}}\qquad
\nonumber\\[5pt]
& \times & \langle (\albpar)^{s-s'}\phi_s|\Box^{\nu_{s'}} |(\albpar)^{s-s'}
\phi_s\rangle\,,
\eeq
where $\nu_{s'}$ is defined in \rf{man02-21072009-12} and  we use the
notation $(p)_q$ to indicate the Pochhammer symbol, $(p)_q \equiv
\frac{\Gamma(p+q)}{\Gamma(p)}$. To illustrate the structure of Lagrangian
\rf{22072009-01} we note that, in terms of tensor fields $\phi^{a_1\ldots
a_{s'}}$ defined in \rf{22072009-03}, Lagrangian \rf{22072009-01} takes the
form
\beq
\label{22072009-02} \LL & = & \half \sum_{s'=0}^s
\frac{2^{s-s'}(s'+\frac{d-2}{2})_{s-s'}}{s'!(s-s')! (s+s'+d-3)_{s-s'}}\qquad
\nonumber\\[5pt]
&\times& (\partial^{s-s'}\phi)^{a_1\ldots a_{s'}}\Box^{\nu_{s'}}
(\partial^{s-s'}\phi)^{a_1\ldots a_{s'}}\,, \eeq
\be (\partial^{s-s'}\phi)^{a_1\ldots a_{s'}} \equiv
\partial^{b_1} \ldots \partial^{b_{s-s'}} \phi^{b_1\ldots b_{s-s'}a_1\ldots
a_{s'}}\,.\ee
For the readers convenience, we write down leading terms in Lagrangian
\rf{22072009-02}.
\beq
\LL & = & \frac{1}{2s!} \phi^{a_1\ldots a_s} \Box^{\nu_s} \phi^{a_1\ldots
a_s}
\\[5pt]
& +  & \frac{1}{2(s-1)!} (\partial\phi)^{a_1\ldots a_{s-1}}\Box^{\nu_{s-1}}
(\partial\phi)^{a_1\ldots a_{s-1}}
\nonumber\\[5pt]
& +  & \frac{1}{4(s-2)!} \frac{2s+d-6}{2s+d-5}
\nonumber\\[5pt]
&\times & (\partial^2\phi)^{a_1\ldots
a_{s-2}}\Box^{\nu_{s-2}}(\partial^2\phi)^{a_1\ldots a_{s-2}} + \ldots \,.
\nonumber \eeq

{\bf Light-cone gauge Lagrangian}.  The light-cone gauge frame is achieved
through the use of the light-cone gauge and differential constraints. Because
the gauge transformation and differential constraint of the spin-$s$
conformal field take the same form as the ones for the spin-$s$ shadow field
we can use the results obtained for the spin-$s$ shadow field. This is to say
that the light-cone gauge condition and solution for the differential
constraint for the spin-$s$ conformal field are obtained from the respective
expressions \rf{man02-17072009-01} and \rf{man02-17072009-02},
\rf{man02-17072009-06} by using identification \rf{man02-21072009-09}. Doing
so, we are left with traceless $so(d-2)$ algebra fields $\phi^{i_1\ldots
i_{s'}}$, $s'=0,1,\ldots, s$. These fields constitute the field content of
light-cone gauge frame. Note that, in contrast to the standard approach, the
fields $\phi^{i_1\ldots i_{s'}}$, with $s'=0,1,\ldots, s-1$, become
independent field D.o.F in the light-cone gauge frame. Making use of the
light-cone gauge in gauge invariant Lagrangian \rf{27062009-01}, we obtain
the light-cone gauge Lagrangian for the spin-$s$ conformal field
\be \label{man02-21072009-13}  \LL^{{\rm l.c.}} =\half \langle \phi^{{\rm
l.c.}}|\Box^\nu |\phi^{{\rm l.c.}}\rangle\,. \ee
To illustrate structure of the light-cone gauge Lagrangian we note that, in
terms of the tensor fields $\phi^{i_1\ldots i_{s'}}$, Lagrangian
\rf{man02-21072009-13} takes the form
\beq
\LL^{{\rm l.c.}} = \sum_{s'=0}^s
\frac{1}{2 s'!} \phi^{i_1\ldots i_{s'}} \Box^{\nu_{s'}} \phi^{i_1\ldots
i_{s'}}\,,
\eeq
where $\nu_{s'}$ is given in \rf{man02-21072009-12}.

%%%%%%%%%%%%%%%%%%%%%%%%%%%%%%%%%%%%%%%%%%%%%%%%%%%%%%%%%%%%%%%%%%%%%%%%%%%%%%
\section{Conclusions}\label{conl-sec-01}
%%%%%%%%%%%%%%%%%%%%%%%%%%%%%%%%%%%%%%%%%%%%%%%%%%%%%%%%%%%%%%%%%%%%%%%%%%%%%%

In this paper, we have further developed the gauge invariant Stueckelberg
approach to $CFT$ initiated in Ref.\cite{Metsaev:2008fs}. The Stueckelberg
approach turned out to be efficient for the study of massive fields and
therefore we believe that this approach might also be useful for the study of
$CFT$. In this paper, we studied the two-point gauge invariant vertices of
the shadow fields and applied our approach to the discussion of $AdS/CFT$
correspondence. In our opinion, use of the Stueckelberg approach to conformal
currents and shadow fields turns out to be efficient for the study of
$AdS/CFT$ correspondence and therefore this approach seems to be very
promising. The results obtained should have a number of the following
interesting applications and generalizations.

(i) In this paper we considered the gauge invariant approach for the
conformal currents and shadow fields which, in the framework of $AdS/CFT$
correspondence, are related to the respective normalizable and
non-normalizable solutions of massless $AdS$ fields. It would be interesting
to generalize our approach to the case of anomalous conformal currents and
shadow fields. In the framework of $AdS/CFT$ correspondence, the anomalous
conformal currents and shadow fields are related to solution of equations of
motion for massive $AdS$ fields. Therefore such generalization might be
interesting for the study of $AdS/CFT$ duality between string massive states
and the boundary conformal currents and shadow fields.

(ii) We studied the {\it bosonic} conformal currents and shadow fields.
Generalization of our approach to the case of fermionic conformal currents
and shadow fields will make it possible to involve the supersymmetry and
apply our approach to the type IIB supergravity in $AdS_5\times S^5$
background and then to the string in this background;

(iii) This paper was devoted to the study of the two-point gauge invariant
vertices. Generalization of our approach to the case of 3-point and 4-point
gauge invariant vertices will give us the possibility to the study of various
applications of our approach along the lines of
Refs.\cite{Liu:1998ty,Leonhardt:2002ta}

(iv) In recent years, mixed symmetry fields have attracted a considerable
interest (see e.g. Refs.\cite{Bekaert:2002dt}-\cite{Alkalaev:2008gi}). We
think that generalization of our approach to the case of mixed symmetry
conformal currents and shadow fields might be useful for the study of
$AdS/CFT$ correspondence because the mixed symmetry fields appear in string
theory and higher-spin gauge fields theory.

(v) In this paper, we have discussed $AdS/CFT$ correspondence between the
massless arbitrary spin $AdS$ fields and the boundary shadow fields. By now
it is known that to construct self-consistent interaction of massless higher
spin fields it is necessary to introduce, among other things, a infinite
chain of massless $AdS$ fields which consists of every spin just once
\cite{Vasiliev:1990en,Vasiliev:2003ev}. This implies that to maintain
$AdS/CFT$ correspondence for such interaction equations of motion we should
also introduce an infinite chain of the boundary shadow fields. We have
demonstrated that use of the modified de Donder gauge provides considerably
simplifications in the analysis of free equations of motion of $AdS$ fields.
In this respect it would be interesting to apply the modified de Donder gauge
to the study  of the consistent equations for interacting gauge fields of all
spins \cite{Vasiliev:1990en} and extend the analysis of this paper to the
case of infinite chain of interacting massless fields and the corresponding
boundary shadow fields.

\begin{acknowledgments}
This work was supported by the RFBR Grant No.08-02-01118, RFBR Grant for
Leading Scientific Schools, Grant No. 1615.2008.2, by the Dynasty Foundation
and by the Alexander von Humboldt Foundation Grant PHYS0167.
\end{acknowledgments}

\appendix

%%%%%%%%%%%%%%%%%%%%%%%%%%%%%%%%%%%%%%%%%%%%%%%%%%%%%%%%%%%%%%%%%%%%%
%%%%%%%%%%%%%%%%%%%%%%%%%%%%%%%%%%%%%%%%%%%%%%%%%%%%%%%%%%%%%%%%%%%%%%%
\section{ Derivation of two-point vertex }
\label{man02-app-01}
%%%%%%%%%%%%%%%%%%%%%%%%%%%%%%%%%%%%%%%%%%%%%%%%%%%%%%%%%%%%%%%%%%%%%%%
%%%%%%%%%%%%%%%%%%%%%%%%%%%%%%%%%%%%%%%%%%%%%%%%%%%%%%%%%%%%%%%%%%%%%%%

In this Appendix, we demonstrate that two-point vertex \rf{11062009-10} is
uniquely determined by requiring the vertex to be invariant under the shadow
field gauge symmetries, Poincar\'e algebra symmetries, and dilatation
symmetry. We proceed in the following way.

{\bf i}). Taking into account double-tracelessness constraint, the shadow
field differential constraint, and Poincar\'e algebra symmetries, we note
that the general form of two-point vertex is given by
\beq
\label{05072009-01} \Gamma & = &\int d^dx_1 d^dx_2 \Gamma_{12} \,,
\\[5pt]
\label{man02-03082009-03} \Gamma_{12} & =  & \half \langle \phi_1
|H_{12}|\phi_2 \rangle\,,
\\[5pt]
\label{man02-03082009-04} H_{12}& = & h_1 + \alpha^2 h_2 + h_3\bar\alpha^2
+ \alpha^2 h_4 \bar\alpha^2\,,
\\[10pt]
&& h_1 =  h_1(|x_{12}|,N_z)\,,
\\
&& h_2 =  h_2(|x_{12}|, \alpha x_{12},\alpha^z,\bar\alpha^z)\,,
\\[5pt]
&& h_3 =  h_3(|x_{12}|, \bar\alpha x_{12},\alpha^z,\bar\alpha^z)\,,
\\[5pt]
&& h_4  =   h_4 ( |x_{12}|, \alpha x_{12}, \bar\alpha x_{12}, \alpha^z,
\bar\alpha^z)\,,\qquad
\eeq
where $N_z=\alpha^z\bar\alpha^z$, $\alpha x_{12}=\alpha^a x_{12}^a$
$\bar\alpha x_{12}=\bar\alpha^a x_{12}^a$  and we use the shortcuts $\langle
\phi_1|$ and $|\phi_2\rangle$ for the respective ket-vectors $\langle
\phi_\sh(x_1)|$ and $|\phi_\sh(x_2)\rangle$. All that is required is to find
the functions $h_a$, $a=1,2,3,4$. To this end we note that variation of
$\Gamma_{12}$ under gauge transformation
\be \label{man02-20072009-17} \delta |\phi_\sh\rangle  = \alpar
|\xi_\sh\rangle + \ldots \,, \ee
takes the form (up to total derivative)
\beq \label{05072009-20}
\delta \Gamma_{12} & = & \half \langle \phi_1| h_1 \alpha^2\albpar | \xi_2
\rangle
+ \langle \phi_1 | h_2 \alpha^2 \alpar | \xi_2 \rangle
\nonumber\\[5pt]
& +  & \langle \phi_1 | h_3 \albpar | \xi_2 \rangle + 2 \langle \phi_1 | h_4
\alpha^2 \albpar | \xi_2 \rangle + \ldots \qquad
\eeq
where dots in \rf{man02-20072009-17} and \rf{05072009-20} stand for the
contributions that are independent of the derivative $\partial^a$ and we use
the shortcut $|\xi_2\rangle$ for ket-vector $|\xi_\sh(x_2)\rangle$. Note that
to derive \rf{05072009-20} we use the differential constraint for the shadow
field $|\phi_\sh\rangle$ (see Table II),
\be \label{man02-20072009-18} \Cb_\perp |\phi_\sh\rangle  + \ldots = 0\,, \ee
where dots stand for contributions that are independent of the derivative
$\partial^a$. From \rf{05072009-20}, we see that requiring the variation
$\delta \Gamma_{12}$ to vanish gives constraints
\be \label{man02-20072009-19} h_4 = -\frac{1}{4}\alpha^2\bar\alpha^2 h_1
\,,\qquad h_2=h_3=0\,. \ee
Plugging \rf{man02-20072009-19} in \rf{man02-03082009-04} we obtain
\be \label{man02-20072009-20} H_{12} = \mubf h_1 \,, \ee
where $\mubf$ is given in \rf{11062009-10x1}. We now note that requiring
invariance of the vertex under shadow field dilatation symmetry (see Table
II) we are led to the following solution to $h_1$:
\be \label{man02-04082009-01} h_1 = f_\nu \rho\,,\qquad \rho \equiv
|x_{12}|^{-(2\nu +d)}\,, \ee
where $f_\nu$ depends only on the operator $N_z$. Taking into account
\rf{18052008-14} we consider $f_\nu$ as function of $\nu$. Thus, restrictions
imposed by Poincar\'e symmetries, dilatation symmetry and {\it some
restrictions} imposed by gauge symmetries lead to the following expression
for $H_{12}$:
\be \label{man02-20072009-21} H_{12} = \mubf f_\nu \rho\,.\ee

{\bf ii}) We now consider {\it all restrictions} on $f_\nu$ imposed by gauge
symmetries. In other words, taking into account the gauge transformation of
the shadow field,
\beq
\label{man02-20072009-22} && \delta |\phi_\sh \rangle = G_\sh
|\xi_\sh\rangle\,,
\\[5pt]
\label{man02-20072009-23} && G_\sh = \alpar - e_{1\,\sh}\Box
-\alpha^2\frac{1}{2N_\alpha+d-2}\eb_{1\,\sh} \,,\qquad
\eeq
we consider restrictions imposed on $f_\nu$ by equation (up to total
derivative)
\be \label{man02-20072009-24} \langle G_\sh \xi_1 | \mubf f_\nu \rho
|\phi_2\rangle = 0\,. \ee
To this end we note the following helpful relation (up to total derivative):
\beq \label{man02-20072009-25}
&& -\langle G_\sh\xi_1| \mubf f_\nu \rho |\phi_2\rangle
\nonumber\\[5pt]
&& \quad = \langle\xi_1|\Cb_\perp - \eb_{1\,\sh} \Box + \half
e_{1\,\sh}\bar\alpha^2\Bigl) f_\nu \rho |\phi_2\rangle\,.\qquad\quad
\eeq
Using differential constraint for the shadow field (see Table II), we
transform $\Cb_\perp$-term in \rf{man02-20072009-25} as
\beq \label{man02-20072009-26}
&& \langle\xi_1 |\Cb_\perp f_\nu \rho |\phi_2\rangle
= \langle\xi_1 |f_\nu \rho \bar{C}_\perp |\phi_2\rangle
\nonumber\\[5pt]
&& = \langle\xi_1 |f_\nu \rho \Bigr(\eb_{1\,\sh} - \half e_{1\,\sh}
\bar\alpha^2\Box\Bigl) |\phi_2\rangle
\nonumber\\[5pt]
&& = \langle\xi_1 |f_\nu \rho \eb_{1\,\sh} |\phi_2\rangle
\nonumber\\[5pt]
&& -  \langle\xi_1 | \frac{(\nu+1)(2\nu+d)f_\nu}{|x_{12}|^{2\nu+d+2}}
e_{1\,\sh} \bar\alpha^2 |\phi_2\rangle\,.\qquad
\eeq
The $\eb_{1\,\sh}$- and $e_{1\,\sh}$-terms in \rf{man02-20072009-25} can be
transformed as
\beq \label{man02-20072009-27}
&& \langle\xi_1 | \Bigr( - \eb_{1\,\sh} \Box + \half e_{1\,\sh}
\bar\alpha^2\Bigl) f_\nu \rho |\phi_2\rangle
\nonumber\\[5pt]
&& = - \langle\xi_1 |f_{\nu-1} \rho 2\nu (2\nu +d-2) \eb_{1\,\sh}
|\phi_2\rangle
\nonumber\\[6pt]
&& + \langle \xi_1 | \frac{f_{\nu+1}}{2|x_{12}|^{2\nu+d+2}}
e_{1\,\sh}\bar\alpha^2 |\phi_2 \rangle\,.\qquad
\eeq
Taking into account \rf{man02-20072009-26},\rf{man02-20072009-27} we see that
requiring the contribution of $\eb_{1\,\sh}$-terms in \rf{man02-20072009-25}
to vanish we find the following equation for $f_\nu$:
\be \label{20072009-01}  \frac{f_\nu}{f_{\nu-1}} = 2\nu (2\nu +d-2)\,.\ee
Using \rf{man02-20072009-26},\rf{man02-20072009-27}, we see that  that
requiring the contribution of $e_{1\,\sh}$-terms in \rf{man02-20072009-25} to
vanish also leads to \rf{20072009-01}. Thus, equations \rf{20072009-01}
amount to requiring that gauge variation of two-point vertex vanishes
\rf{man02-20072009-24}. Solution to Eq. \rf{20072009-01} with initial
condition $f_{\nu_s}=1$ is given in \rf{29102008-01}.

%%%%%%%%%%%%%%%%%%%%%%%%%%%%%%%%%%%%%%%%%%%%%%%%%%%%%%%%%%%%%%%%%%%%%
%%%%%%%%%%%%%%%%%%%%%%%%%%%%%%%%%%%%%%%%%%%%%%%%%%%%%%%%%%%%%%%%%%%%%%%
\section{ Invariance of two-point vertex
under conformal boost transformations }\label{app-04082009-01}
%%%%%%%%%%%%%%%%%%%%%%%%%%%%%%%%%%%%%%%%%%%%%%%%%%%%%%%%%%%%%%%%%%%%%%%
%%%%%%%%%%%%%%%%%%%%%%%%%%%%%%%%%%%%%%%%%%%%%%%%%%%%%%%%%%%%%%%%%%%%%%%

We now demonstrate invariance of the shadow field two-point vertex
\rf{11062009-10} under conformal boost transformations ($K^a$
transformations). We start with expression for two-point vertex
\rf{05072009-01} with $\Gamma_{12}$ given by
\beq
\label{man02-04082009-02} && \Gamma_{12}  = \half \langle \phi_1 | H_{12}
|\phi_2 \rangle\,,
\qquad
H_{12} = \mubf f_\nu \rho \,, \qquad \eeq
where $\mubf$, $f_\nu$ and $\rho$ are defined in \rf{11062009-10x1},
\rf{29102008-01} and \rf{man02-04082009-01} respectively. In
\rf{man02-04082009-02}, we use the shortcuts $\langle \phi_1|$ and
$|\phi_2\rangle$ for the respective shadow field ket-vectors $\langle
\phi_\sh(x_1)|$ and $|\phi_\sh(x_2)\rangle$. We proceed in the following way.

{\bf i}) Before analyzing restrictions imposed on the two-point vertex by the
conformal boost symmetries we find explicit form of the restrictions imposed
on the two-point vertex by the Poincar\'e algebra and dilatation symmetries.
Invariance with respect to the Poincar\'e translations implies that $H_{12}$
depends on $x_{12}^a\equiv x_1^a-x_2^a$. Requiring the vertex $\Gamma$ to be
invariant under dilatation and Lorentz algebra symmetries
\be \delta_D \Gamma = 0 \,, \qquad \delta_{J^{ab}} \Gamma = 0 \,,\ee
amounts to the following respective equations for $H_{12}$:
\beq
\label{05072009-05} && (x^a_{12} \partial_{x_{12}}^a +2d - \Delta_\sh )H_{12}
- H_{12} \Delta_\sh = 0 \,,\qquad
\\[5pt]
\label{05072009-06} && ( l_{12}^{ab} + M^{ab}) H_{12} - H_{12} M^{ab} = 0\,,
\\[10pt]
\label{05072009-07} && \hspace{1cm} l_{12}^{ab} \equiv x_{12}^a
\partial_{x_{12}}^b - x_{12}^b \partial_{x_{12}}^a\,,
\\[5pt]
\label{05072009-12} &&  \hspace{1cm} \Delta_\sh \equiv 2 - s + N_z \,,
\eeq
where $M^{ab}$ is given in \rf{mabdef0001}. It easy to see that $H_{12}$
given in \rf{man02-20072009-21} satisfies Eqs.\rf{05072009-05},
\rf{05072009-06}.

{\bf ii}) We now prove invariance of $\Gamma$ under the $K^a$
transformations. To this end we analyze variation of the $\Gamma$ under the
$K^a$ transformations. Taking into account \rf{conalggenlis04}, we see that
variation of $\Gamma$ can be represented as
\beq
&&  \label{05072009-02} \delta_{K^a} \Gamma = \delta_{K_{\Delta,M}^a} \Gamma
+ \delta_{R^a}\Gamma\,.
\eeq
Taking into account Eqs.\rf{05072009-05},\rf{05072009-06}, one can make sure
that the variation of $\Gamma_{12}$ under $K_{\Delta,M}^a$ transformations
takes the form (up to total derivative)
\beq
&& \label{05072009-03} \delta_{K_{\Delta,M}^a} \Gamma_{12} = \half \langle
\phi_1|\Bigl( \half |x_{12}| \PP_{x_{12}}^a  - M^{ab} x_{12}^b )
H_{12}\qquad\quad
\nonumber\\[5pt]
&&\hspace{1.6cm}  + \ \half x_{12}^a [\Delta_\sh ,H_{12}]\Bigr)
|\phi_2\rangle\,,
\\[10pt]
&& \hspace{1.6cm} \PP_x^a \equiv |x|\partial^a - \frac{x^a}{|x|}
x^b\partial^b\,.
\eeq
Using the notation $R_\sh^a$ for the shadow field operator $R^a$ given in
Table II, it is easy to see that variation of the vertex $\Gamma$ under
action of the operator $R^a$ is given by
\be
\label{05072009-04} \delta_{R^a} \Gamma  = \int d^dx_1 d^dx_2 \, \langle
R_\sh^a \phi_1 |H_{12} |\phi_2\rangle \,. \ee
We note that relations \rf{05072009-03} and \rf{05072009-04} are valid for
arbitrary $H_{12}$ which satisfies Eqs.\rf{05072009-05},\rf{05072009-06}.
Making use of expression for $H_{12}$ in \rf{man02-04082009-02}, it is easy
to see that requiring the vertex $\Gamma$ to be invariant under $K^a$
transformations,
\be \delta_{K^a} \Gamma = 0\,, \ee
amounts to the following equation (up to total derivative):
\be \label{05072009-13} - \half \langle \phi_1| M^{ab} x_{12}^b H_{12}
|\phi_2\rangle + \langle R_\sh^a \phi_1| H_{12}|\phi_2\rangle  = 0\,.
\ee
Note that for the derivation of this equation we use the relations
$[\PP_x^a,|x|]=0$ and $[H_{12},\Delta_\sh]=0$.

Thus, all that remains to be done is to prove Eq.\rf{05072009-13} with
$H_{12}$ given in \rf{man02-04082009-02}. To this end we note the relations
\beq
\label{05072009-16} && x_{12}^a H_{12} = -\frac{1}{2\nu
+d-2}\partial_{x_{12}}^a (|x_{12}|^2 H_{12})\,,\qquad
\\[7pt]
\label{05072009-16x1} &&  M^{ab} \partial^b + G_\sh \Cb_\perp^a
\nonumber\\[7pt]
&& \hspace{1cm} = \alpha^a \Cb_\sh - e_{1\,\sh}\Box \bar\alpha^a + \Iwt^a
\eb_{1\,\sh}\,,
\\[10pt]
&&\hspace{1cm}  \Iwt^a \equiv \alpha^a -\alpha^2
\frac{1}{2N_\alpha+d-2}\bar\alpha^a \,,
\\
&& \hspace{1cm} \Cb_\perp^a \equiv \bar\alpha^a - \half
\alpha^a\bar\alpha^2\,,
\eeq
where $\nu$ and $G_\sh$ are given in \rf{29102008-01xxx1} and
\rf{man02-20072009-23} respectively, while $\Cb_\sh$, $e_{1\,\sh}$,
$\eb_{1\,\sh}$ are given in Table II. Using \rf{05072009-16},
\rf{05072009-16x1}, the constraint $\Cb_\sh|\phi_\sh\rangle=0$, and the
relation $M^{ab\dagger} = -M^{ab}$, we obtain (up to total derivative)
\beq
\label{05072009-17} && - \langle \phi_1| M^{ab} x_{12}^b H_{12}
|\phi_2\rangle
\nonumber\\[5pt]
&& \quad =  \langle \frac{1}{2\nu +d-2} M^{ab}
\partial_{x_1}^b \phi_1 | |x_{12}|^2 H_{12} |\phi_2\rangle
\nonumber\\[5pt]
&&\quad = \langle \Bigl( - G_\sh \Cb_\perp^a - e_{1\,\sh} \Box \bar\alpha^a +
\Iwt^a \eb_{1\,\sh} \Bigr) \phi_1 |\qquad
\nonumber\\[5pt]
&&\quad \times  \frac{|x_{12}|^2 H_{12}}{2\nu +d-2} |\phi_2\rangle\,.
\eeq
We now consider expressions appearing in \rf{05072009-17} in turn.
$G_\sh$-term can be transformed as (up to total derivative)
\beq \label{man02-04082009-03}
&& - \langle G_\sh \Cb_\perp^a \phi_1 | \frac{|x_{12}|^2 H_{12}}{2\nu+d-2}
|\phi_2\rangle
\nonumber\\[5pt]
&& = \langle \Cb_\perp^a \phi_1 | \Cb_\cur \frac{|x_{12}|^2 f_\nu \rho}{
2\nu+d-2 }|\phi_2\rangle
\nonumber\\[5pt]
&& =  \langle \Cb_\perp^a \phi_1 | \frac{|x_{12}|^2 f_\nu \rho}{2\nu+d-2}
\Cb_\sh |\phi_2\rangle
\nonumber\\[5pt]
&& +  \langle \Cb_\perp^a \phi_1 |\Bigl( 2\eb_{1\,\sh} f_\nu \rho + f_\nu
\rho e_{1\,\sh} \bar\alpha^2 \Bigr) | \phi_2\rangle
\nonumber\\[5pt]
&& =  \langle (-2 \bar\alpha^a + 2\alpha^a \bar\alpha^2 ) e_{1\,\sh} \phi_1 |
H_{12} | \phi_2\rangle\,,
\eeq
where the operator $\Cb_\cur$ is given in Table II. Also, note that we use
the differential constraint $\Cb_\sh |\phi_\sh\rangle = 0$. The $e_{1\,\sh}$-
and $\eb_{1\,\sh}$- terms in \rf{05072009-17} can be transformed as
\beq \label{man02-04082009-04}
&& - \langle e_{1\,\sh} \Box \bar\alpha^a \phi_1 | \frac{|x_{12}|^2
H_{12}}{2\nu+d-2} |\phi_2\rangle
\nonumber\\[7pt]
&& \qquad \ \ =  - \langle 2\nu e_{1\,\sh} \bar\alpha^a \phi_1 | H_{12}
|\phi_2\rangle\,,
\\[15pt]
\label{man02-04082009-05}
&& \langle \Iwt^a \eb_{1\,\sh} \phi_1 | \frac{|x_{12}|^2 H_{12}}{2\nu+d-2}
|\phi_2\rangle
\nonumber\\[7pt]
&& \qquad  \ \ = - \langle 2(\nu+1)e_{1\,\sh} \bar\alpha^a \phi_2 | H_{12}
|\phi_1\rangle\,.\qquad
\eeq
Plugging \rf{man02-04082009-03}, \rf{man02-04082009-04} and
\rf{man02-04082009-05} in \rf{05072009-17} and taking into account $R_\sh^a$
given in Table II we make sure that relation \rf{05072009-13} holds true.

%%%%%%%%%%%%%%%%%%%%%%%%%%%%%%%%%%%%%%%%%%%%%%%%%%%%%%%%%%%%%%%%%%%%%%%
%%%%%%%%%%%%%%%%%%%%%%%%%%%%%%%%%%%%%%%%%%%%%%%%%%%%%%%%%%%%%%%%%%%%%%%
\section{ Derivation of effective action }\label{man02-app-04}
%%%%%%%%%%%%%%%%%%%%%%%%%%%%%%%%%%%%%%%%%%%%%%%%%%%%%%%%%%%%%%%%%%%%%%%
%%%%%%%%%%%%%%%%%%%%%%%%%%%%%%%%%%%%%%%%%%%%%%%%%%%%%%%%%%%%%%%%%%%%%%%

We now discuss details of the derivation of effective action
\rf{man02-21072009-15}. We are going to prove the following relations:
\be \label{man02-21072009-16}
\int\!\! d^dx \phi(x,z) \partial_z \phi(x,z) \, \stackrel{z\rightarrow
0}{\longrightarrow}\,
\frac{\cwt_\nu^2}{c_\nu}\!\! \int\!\! d^dx_1 d^dx_2 \frac{\phi_1 \phi_2}{
|x_{12}|^{2\nu+d}}\,,
\ee

\be \label{man02-21072009-17}
\int\!\! d^dx \frac{1}{z} \phi(x,z) \phi(x,z)
\, \stackrel{z\rightarrow\, 0}{\longrightarrow} \frac{2\cwt_\nu^2}{c_\nu}\!\!
\int\!\! d^dx_1 d^dx_2 \frac{\phi_1 \phi_2}{ |x_{12}|^{2\nu+d}}\,,
\ee
\be  \label{man02-21072009-18} \cwt_\nu \equiv \sigma c_\nu\,, \ee
where $\phi(x,z)$ is solution of the Dirichlet problem given in
\rf{19072009-09} and we use the shortcuts $\phi_1$ and $\phi_2$ for the
respective boundary shadow fields $\phi_\sh(x_1)$ and $\phi_\sh(x_2)$. The
$|x_{12}|$ is given in \rf{manus2009-02-03}. Taking into account expressions
for the effective action given in \rf{19072009-07}, \rf{19072009-08} it is
easy to see that relations \rf{man02-21072009-16}, \rf{man02-21072009-17} do
indeed lead to effective action \rf{man02-21072009-15}. Note that in r.h.s.
of \rf{man02-21072009-16},\rf{man02-21072009-17} we keep, as usually, only
non-local contributions.

We now prove relations \rf{man02-21072009-16}, \rf{man02-21072009-17}. To
this end we use expression for solution $\phi(x,z)$ in \rf{19072009-09} to
find the relations
\beq  \label{man02-21072009-19}
&& \int\!\! d^dx\, \phi(x,z) \partial_z \phi(x,z)
\nonumber\\[5pt]
&& \qquad = \cwt_\nu^2 ( (\nu+\half)X_1 - (2\nu+d) X_2 )\,,
\\[5pt]
\label{man02-21072009-20} && \int\!\! d^dx\, \frac{1}{z} \phi(x,z) \phi(x,z)
= \cwt_\nu^2 X_1\,,
\eeq
where we use the notation
\beq  \label{man02-21072009-21}
&& X_1 \equiv \int\!\! d^dx_1 d^dx_2 d^dx_3 \phi_1 \phi_2 \frac{ z^{2\nu}}{
f_{13}^{\nu+\frac{d}{2}} f_{23}^{\nu+\frac{d}{2}} }\,,
\\[5pt]
\label{man02-21072009-22} && X_2 \equiv \int\!\! d^dx_1 d^dx_2 d^dx_3 \phi_1
\phi_2
\frac{ z^{2\nu+2} }{ f_{13}^{\nu+\frac{d}{2}} f_{23}^{\nu+\frac{d}{2}+1} }\,,
\\[10pt]
\label{man02-21072009-23} && f_{mn} \equiv z^2 + |x_{mn}|^2 \,,\qquad
x_{mn}^a \equiv x_m^a-x_n^a \,.\qquad
\eeq
Using the Fourier transform of the kernels in \rf{man02-21072009-21},
\rf{man02-21072009-22}
\beq \label{man02-21072009-24}
&& \frac{z^\nu}{(z^2+|x|^2)^{\nu+\frac{d}{2}}} = \omega_\nu\!\! \int\!\!
d^dk\, e^{{\rm i} k \cdot x }k^\nu K_\nu(kz)\,,
\\[5pt]
\label{man02-21072009-24x1} && \hspace{2cm}  \omega_\nu^{-1} \equiv
\pi^{d/2}2^{\nu+d-1} \Gamma(\nu+\frac{d}{2})\,,\qquad \eeq
where $K_\nu$ is the modified Bessel, and integrating over $x_3$, we cast
$X_1$ and $X_2$ into the form
\beq \label{man02-21072009-25}
&&  X_1 = (2\pi)^d\!\! \int \!\! d^dx_1 d^dx_2 d^dk\, \phi_1 \phi_2  e^{{\rm
i} k\cdot x_{12}}
\nonumber\\[5pt]
&& \hspace{1cm} \times\, \omega_\nu^2 k^{2\nu} (K_\nu(kz))^2\,,
\\[10pt]
\label{man02-21072009-26} && X_2 = (2\pi)^d\!\! \int\!\! d^dx_1 d^dx_2 d^dk\,
\phi_1 \phi_2 e^{{\rm i} k \cdot x_{12}}
\nonumber \\[5pt]
&& \hspace{1cm} \times\,  \omega_\nu\omega_{\nu+1} z k^{2\nu+1} K_\nu(kz)
K_{\nu+1}(kz)\,.\qquad
\eeq
We now consider the asymptotic behavior, as $z\rightarrow 0$, of $X_1$ and
$X_2$. As usually, we are interested in non-local contributions to
\rf{man02-21072009-25},\rf{man02-21072009-26}. To this end we use the
definition of $K_\nu$,
\beq
&& K_\nu(z) = \frac{\pi}{2\sin\pi \nu} (I_{-\nu}(z) - I_\nu(z))\,,
\\[3pt]
&& I_\nu(z) = \sum_{k=0}^\infty \frac{1}{k! \Gamma(k+\nu+1)}
\left(\frac{z}{2}\right)^{\nu+2k}\,, \qquad \eeq
to obtain the following well-known formula:
\beq \label{man02-21072009-27}
&& \hspace{-1cm} K_\nu(kz)\stackrel{z\rightarrow 0}{\longrightarrow}\,\,
\frac{2^{\nu-1} \Gamma(\nu)}{(kz)^\nu}  + \frac{(kz)^\nu \Gamma(-\nu)
}{2^{\nu+1}}+\ldots\,,
\eeq
where dots stand for the terms which are not relevant for the analysis of
non-local contributions to \rf{man02-21072009-25},\rf{man02-21072009-26}.
Making use of \rf{man02-21072009-27}, we obtain
\beq \label{man02-21072009-28}
&&  X_1 \, \stackrel{z\rightarrow 0}{\longrightarrow}\, (2\pi)^d\!\! \int\!\!
d^dx_1 d^dx_2 d^dk\, \phi_1\phi_2 k^{2\nu} e^{{\rm i} k \cdot x_{12} }\qquad
\nonumber\\[15pt]
&& \hspace{0.7cm} \times\, \half \omega_\nu^2 \Gamma(\nu) \Gamma(-\nu)
\nonumber\\[5pt]
&& \hspace{0.7cm} = \frac{2}{c_\nu}\!\! \int\!\! d^dx_1 d^dx_2 \frac{\phi_1
\phi_2}{ |x_{12}|^{2\nu+d}}\,,
\\[10pt]
\label{man02-21072009-29}
&& X_2 \, \stackrel{z\rightarrow 0}{\longrightarrow}\, (2\pi)^d\!\! \int\!\!
d^dx_1 d^dx_2 d^dk\, \phi_1\phi_2 k^{2\nu} e^{{\rm i} k \cdot x_{12}}\qquad
\nonumber\\[5pt]
&& \hspace{0.7cm} \times\, \half \omega_\nu\omega_{\nu+1}\Gamma(-\nu)
\Gamma(\nu+1)
\nonumber\\[15pt]
&&\hspace{0.7cm} = \frac{2\nu}{c_\nu(2\nu+d)}\!\! \int\!\! d^dx_1 d^dx_2
\frac{\phi_1 \phi_2}{ |x_{12}|^{2\nu+d}}\,.
\eeq
For the derivation of relations
\rf{man02-21072009-28},\rf{man02-21072009-29}, we use the formulas
\be \label{18062009-01}
\int\!\! d^d k k^{2\nu} e^{{\rm i} k \cdot x}  =  \frac{2^{2\nu+d} \pi^{d/2}
\Gamma(\nu + \frac{d}{2})}{\Gamma(-\nu)\, |x|^{2\nu+d}}\,, \ee

\be\label{man02-21072009-30} \frac{\omega_{\nu+1}}{\omega_\nu} =
\frac{1}{2\nu+d}\,, \ee
where \rf{man02-21072009-30} is simply obtained by using
\rf{man02-21072009-24x1}. Making use of
\rf{man02-21072009-28},\rf{man02-21072009-29} in
\rf{man02-21072009-19},\rf{man02-21072009-20}, we arrive at desired relations
\rf{man02-21072009-16},\rf{man02-21072009-17}.

\section{ CFT adapted Lagrangian for massless spin-1 and spin-2 fields
in $AdS_{d+1}$ } \label{man02-app-02}

In this Appendix, we explain some details of the derivation of the $CFT$
adapted gauge invariant Lagrangian for massless spin-1 and spin-2 fields
given in \rf{20062009-01} and \rf{gaufixlag01}. Presentation in this Appendix
is given by using Lorentzian signature. Euclidean signature Lagrangian in
Sec. \ref{secAdS/CFT}, is obtained from the Lorentzian signature Lagrangian
by simple substitution $\LL\rightarrow -\LL$.

{\bf Spin-1 massless field}. We use field $\Phi^A$ carrying flat Lorentz
algebra $so(d,1)$ vector indices $A,B=0,1,\ldots, d-1,d$. The field $\Phi^A$
is related with field carrying the base manifold indices $\Phi^\mu$,
$\mu=0,1,\ldots,d$, in a standard way $\Phi^A= e_\mu^A\Phi^\mu$, where
$e_\mu^A$ is vielbein of $AdS_{d+1}$ space. For the Poincar\'e
parametrization of $AdS_{d+1}$ space \rf{lineelem01}, vielbein $e^A=e^A_\mu
dx^\mu$ and Lorentz connection, $de^A+\omega^{AB}\wedge e^B=0$, are given by
\be\label{eomcho01} e_\mu^A=\frac{1}{z}\delta^A_\mu\,,\qquad
\omega^{AB}_\mu=\frac{1}{z}(\delta^A_z\delta^B_\mu
-\delta^B_z\delta^A_\mu)\,, \ee
where $\delta_\mu^A$ is Kronecker delta symbol. We use a covariant derivative
with the flat indices $\DD^A$,
\be \DD_A \equiv e_A^\mu \DD_\mu\,,\qquad \DD^A = \eta^{AB}\DD_B\,,\ee
where $e_A^\mu$ is inverse of $AdS$ vielbein, $e_\mu^A e_B^\mu = \delta_B^A$
and $\eta^{AB}$ is flat metric tensor. With choice made in \rf{eomcho01}, the
covariant derivative takes the form
\be
\DD^A \Phi^B = \hat{\partial}^A \Phi^B + \delta_z^B \Phi^A -
\eta^{AB}\Phi^z\,,\quad \ \hat{\partial}^A \equiv z\partial^A\,,
\ee
where we adopt the following conventions for the derivatives and coordinates:
$\partial^A=\eta^{AB}\partial_B$, $\partial_A=\partial/\partial x^A $, $x^A
\equiv \delta_\mu^A x^\mu$, $x^A=x^a,x^d$, $x^d\equiv z$.

In arbitrary parametrization of $AdS$, Lagrangian of the massless spin-1
field takes the standard form
\be \label{man02-13072009-01} e^{-1} \LL = - \frac{1}{4} F^{AB}F^{AB}\,,
\quad F^{AB}= D^A \Phi^B - \DD^B\Phi^A\,,\quad
\ee
where $e\equiv \det e_\mu^A$. Using the notation
\be \label{man02-13072009-02} C_\st  =  \DD^A \Phi^A\,, \ee
we note that it is the relation $C_\st=0$ that defines the standard Lorentz
gauge. Lagrangian \rf{man02-13072009-01} can be represented as
\beq
\label{man02-13072009-03} e^{-1}\LL & = &  \half \Phi^A (\DD^2 + d) \Phi^A +
\half C_\st^2 \,.
\eeq
We now use the Poincar\'e parametrization of $AdS$ and introduce the
following quantity:
\be \label{man02-13072009-04} \Cbf = \DD^A \Phi^A + 2\Phi^z\,. \ee
We note that it is the relation $\Cbf=0$ that defines the modified Lorentz
gauge. Using the relations (up to total derivative)
\beq
\label{man02-13072009-05} e\Phi^A \DD^2 \Phi^A & = & e\Bigl(\Phi^A
(\Box_{_{0\,AdS}} -1) \Phi^A
\nonumber\\[5pt]
&  + & 4 \Phi^z \Cbf + (d-7) \Phi^z \Phi^z\Bigr)\,,
\\[7pt]
\label{man02-13072009-06} C_\st^2 & = &  \Cbf^2 - 4 \Phi^z \Cbf + 4 \Phi^z
\Phi^z\,,
\\[7pt]
\label{man02-13072009-07} \Box_{_{0\,AdS}} & \equiv & z^2(\Box+\partial_z^2)
+(1-d)z\partial_z \,,
\eeq
we represent Lagrangian \rf{man02-13072009-03} and $\Cbf$
\rf{man02-13072009-04} as
\beq
\label{man02-13072009-08} e^{-1} \LL & = &  \half \Phi^A (\Box_{0\,AdS}  +
d-1 ) \Phi^A
\nonumber\\[5pt]
& + & \frac{d-3}{2}\Phi^z\Phi^z  + \half \Cbf^2 \,,
\\[5pt]
\label{man02-13072009-09} \Cbf  & = &   \hat\partial^A \Phi^A + (2-d)\Phi^z
\,.\eeq
In terms of $so(d-1,1)$ tensorial components of the field $\Phi^A$ given by
$\Phi^a$, $\Phi^z$, Lagrangian \rf{man02-13072009-08} takes the form
\beq \label{man02-13072009-10} e^{-1} \LL & = & \half \Phi^a (\Box_{0\, AdS}
+ d-1) \Phi^a
\nonumber\\[5pt]
& + & \half \Phi^z (\Box_{0\, AdS} + 2d-4) \Phi^z + \half \Cbf^2\,,\qquad
\\[5pt]
\label{man02-13072009-11} \Cbf  & = &   z\partial^a \Phi^a + z
\TT_{2-d}\Phi^z\,.
\eeq
Introducing the canonically normalized field $\phi^A$,
\be \label{man02-13072009-12} \Phi^A  =  z^{\frac{d-1}{2}}\phi^A \,, \ee
and using the identification $\phi^z=\phi$ we make sure that Lagrangian
\rf{man02-13072009-10} takes the form
\be  \label{man02-13072009-13} \LL =  \half \phi^a \Box_{\nu_1}\phi^a +\half
\phi \Box_{\nu_0}\phi + \half C^2 \,, \ee
where $\Box_\nu$, $C$, and $\nu$'s are defined in \rf{09072009-09},
\rf{09072009-03}, and \rf{09072009-05} respectively.  Note that $\Cbf =
z^{(d+1)/2} C$. Finally, taking into account expression for operator
$\Box_\nu$ \rf{09072009-09} and relation \rf{10072009-06}, we see that
Lagrangian \rf{man02-13072009-13} is equal, up to total derivative and
overall sign, to the one given in \rf{20062009-01}.

Lagrangian \rf{man02-13072009-01} is invariant under the gauge
transformations $\delta \Phi^A = \hat{\partial}^A \Xi$. Making the rescaling
$\Xi = z^{(d-3)/2}\xi$, we check that these gauge transformations lead to the
ones given in \rf{09072009-06}, \rf{09072009-07}.

{\bf Spin-2 massless field}. We use a tensor field $\Phi^{AB}$ with flat
indices. This field is related with the tensor field carrying the base
manifold indices in a standard way $\Phi^{AB} = e_\mu^A e_\nu^B
\Phi^{\mu\nu}$. In arbitrary parametrization of $AdS$, Lagrangian of massless
spin-2 field takes the standard form
\beq \label{man02-13072009-15}
e^{-1} \LL & = & \frac{1}{4} \Phi^{AB} (E_{_{EH}} \Phi)^{AB} + \half
\Phi^{AB} \Phi^{AB}
\nonumber\\[5pt]
& + & \frac{d-2}{4} \Phi^2 \,,
\\[7pt]
(E_{_{EH}}\Phi)^{AB} & = & \DD^2 \Phi^{AB} - \DD^A (\DD\Phi)^B - \DD^B
(\DD\Phi)^A
\nonumber\\[7pt]
& + & \DD^A \DD^B \Phi +  \eta^{AB}( \DD^C \DD^E \Phi^{CE} - \DD^2 \Phi) \,,
\nonumber\\[7pt]
&&\!\!\!\!\!\!\! \Phi\equiv \Phi^{AA}\,, \quad (\DD\Phi)^A\equiv
\DD^B\Phi^{AB} \,.
\eeq
Using the notation
\be
\label{man02-13072009-16} C_\st^A \equiv \DD^B\Phi^{AB} -\half \DD^A \Phi \,,
\ee
we note that it is the relation $C_\st^A=0$ that defines the standard de
Donder gauge condition. Using $C_\st^A$, Lagrangian \rf{man02-13072009-15}
can be represented as
\beq \label{man02-13072009-17}
e^{-1} \LL & = &   \frac{1}{4} \Phi^{AB}(\DD^2 + 2)\Phi^{AB}
\nonumber\\[5pt]
&  - & \frac{1}{8}\Phi(\DD^2 - 2d + 4)\Phi + \half C_\st^A C_\st^A \,.
\eeq
We now use the Poincar\'e parametrization of $AdS$, introduce the notation
\be \label{man02-13072009-18} \Cbf^A \equiv C_\st^A  + 2 \Phi^{zA} -
\delta_z^A \Phi \,,\ee
and note that it is the relation $\Cbf^A=0$ that defines the modified de
Donder gauge condition. Using the relations (up to total derivative)
\beq \label{man02-13072009-19}
&& \frac{1}{4}e \Phi^{AB}\DD^2 \Phi^{AB}  = e\Bigl( \frac{1}{4} \Phi^{AB}
(\Box_{0\,AdS} -2)\Phi^{AB}
\nonumber\\[5pt]
&& + \frac{d-5}{2} \Phi^{zA}\Phi^{zA} + 2\Phi^{zz}\Phi
-\frac{d}{4}\Phi^2\qquad
\nonumber\\[5pt]
&& +  2\Phi^{zA} \Cbf^A - \Phi \Cbf^z\Bigr)\,,
\\[10pt]
\label{man02-13072009-20} && \half C_\st^A C_\st^A  =   \half \Cbf^A \Cbf^A -
2\Phi^{zA} \Cbf^A + \Phi \Cbf^z
\nonumber\\[5pt]
&& + 2\Phi^{zA}\Phi^{zA} - 2\Phi^{zz}\Phi + \half \Phi^2\,,
\eeq
we represent Lagrangian \rf{man02-13072009-17} and $\Cbf^A$ as
\beq \label{man02-13072009-21}
e^{-1} \LL & = & \frac{1}{4}\Phi^{AB} \Box_{_{0\,AdS}} \Phi^{AB} -
\frac{1}{8}\Phi \Box_{_{0\,AdS}} \Phi \qquad
\nonumber\\[5pt]
& + &  \frac{d-1}{2}\Phi^{zA}\Phi^{zA} + \half \Cbf^A \Cbf^A\,,
\\[5pt]
\label{man02-13072009-21x1} \Cbf^A & = & \hat\partial^B \Phi^{AB} - \half
\hat\partial^A \Phi + (1-d) \Phi^{zA}\,,
\eeq
where $\Box_{_{0\,AdS}}$ is given in \rf {man02-13072009-07}. In terms of
canonically normalized fields $\widetilde\Phi^{AB}$, defined by
\be \label{man02-13072009-23} \Phi^{AB} =
z^{\frac{d-1}{2}}\widetilde\Phi^{AB}\,, \ee
Lagrangian \rf{man02-13072009-21} takes the form
\beq \label{man02-13072009-22}
\LL & =  & \frac{1}{4}\widetilde\Phi^{AB} \Box_{\nu_2}\widetilde\Phi^{AB} -
\frac{1}{8}\widetilde\Phi^{AA} \Box_{\nu_2} \widetilde\Phi^{BB}
\nonumber\\[5pt]
& + &  \frac{d-1}{2z^2}\widetilde\Phi^{zA}\widetilde\Phi^{zA} + \half
\widetilde\Cbf^a \widetilde\Cbf^a + \half \widetilde\Cbf^z
\widetilde\Cbf^z\,,\qquad
\eeq
\beq \label{man02-13072009-25}
\widetilde\Cbf^a & = & \partial^b \widetilde\Phi^{ab} - \half \partial^a
\widetilde\Phi + \TT_{-\frac{d-1}{2}} \widetilde\Phi^{za}\,,
\\[5pt]
\label{man02-13072009-26} \widetilde\Cbf^z & = & \partial^a
\widetilde\Phi^{za} - \half \TT_{\frac{d-1}{2}} \widetilde\Phi + \TT_{-
\frac{d-1}{2}}\widetilde\Phi^{zz}\,,
\eeq
where $\Box_\nu$ and $\nu$'s are defined in \rf{09072009-09} and
\rf{15052008-34} respectively. We note the relation $\Cbf^A= z^{(d+1)/2}
\widetilde\Cbf^A$.

Introducing new fields $\phi^{ab}$, $\phi^a$, $\phi$ by the relations
\beq
&& \widetilde\Phi^{ab} = \phi^{ab} - \frac{u}{d-1}\eta^{ab} \phi\,,
\\[7pt]
&& \widetilde\Phi^{za} =  \phi^a\,,
\\[7pt]
&& \widetilde\Phi^{zz} = \frac{2}{u} \phi\,,
\eeq
where $u$ is given in \rf{15052008-34x1}, we represent Lagrangian
\rf{man02-13072009-22} as
\beq \label{man02-app001}
\LL & =  & \frac{1}{4} \phi^{ab} \Box_{\nu_2} \phi^{ab} - \frac{1}{8}
\phi^{aa} \Box_{\nu_2} \phi^{bb} + \half \phi^a \Box_{\nu_1} \phi^a
\nonumber\\[5pt]
&  + & \half \phi \Box_{\nu_0} \phi
+ \half C^a C^a + \half C^2\,,
\eeq
where $C^a$ and $C$ are defined in \rf{15052008-29},\rf{15052008-30}. We note
the relations $C^a=\widetilde\Cbf^a$, $C=\widetilde\Cbf^z$. Taking into
account \rf{10072009-06}, we see that Lagrangian \rf{man02-app001} is equal,
up to total derivative and overall sign, to the one given in
\rf{gaufixlag01}.

Lagrangian \rf{man02-13072009-15} is invariant under gauge transformations
\be \label{21052008-01}\delta \Phi^{AB} = \DD^A \Xi^B + \DD^B \Xi^A\,.\ee
Introducing gauge transformation parameter $\xi^A$ by the relation $\Xi^A =
z^{(d-3)/2}\xi^A$, and making the identification for the $so(d-1,1)$ algebra
scalar mode $\xi \equiv \xi^z$ we check that gauge transformations
\rf{21052008-01} lead to the ones given in \rf{15052008-35}-\rf{15052008-37}.

\section{ Derivation of normalization factor
$\sigma_{s,\nu}$}\label{man02-app-03}

In this Appendix, we outline the derivation of the normalization factor
$\sigma_{s,\nu}$ given in \rf{10072009-08x2},\rf{10072009-08x3}. To this end
we note that \rf{10072009-08x2} can be represented as
\beq
\label{man02-20072009-03} |\phi(x,z)\rangle & = &  \int\!\! d^dy\, p_\nu
F(x-y) |\phi_\sh(y)\rangle\,,
\\[5pt]
&&  \label{man02-20072009-04} F(x)\equiv \frac{z^{\nu + \half}}{(z^2 +
|x|^2)^{\nu + \frac{d}{2}}}\,,
\eeq
where $p_\nu$ depends on operator $\nu$ \rf{18052008-14}. Comparing
\rf{man02-20072009-03} and \rf{10072009-08x2}, we see that $p_\nu$ and
$\sigma_{s,\nu}$ are related as
\be  \label{man02-20072009-16} p_\nu = \sigma_{s,\nu} c_\nu \,, \ee
where $c_\nu$ is defined in \rf{10072009-13}, i.e. we see that if we find
$p_\nu$ then we fix the coefficient $\sigma_{s,\nu}$.

The coefficient $p_\nu$ is uniquely determined by the following two
requirements:

{\bf i}) Modified de Donder gauge condition for $AdS$ field $\phik$
\rf{10072009-07} should lead to the differential constraint for the shadow
field $|\phi_\sh\rangle$ (see Table II).

{\bf ii}) For $\nu = \nu_s$ (see \rf{29102008-01xxx2}), the $p_{\nu_s}$ is
normalized to be
\be   \label{man02-20072009-13} p_{\nu_s} = c_{\nu_s}\,, \ee
where $c_{\nu_s}$ is given in \rf{man02-20072009-02}.

We note that the choice of normalization condition \rf{man02-20072009-13} is
a matter of convenience. This condition implies that solution
\rf{man02-20072009-03} leads to the following asymptotic behavior for the
leading rank-$s$ tensor field $\phi^{a_1\ldots a_s}$ (see \rf{18052008-04}):
\be \phi^{a_1\ldots a_s}(x,z)\,\,\stackrel{z \rightarrow 0}{\longrightarrow}
z^{-\nu_s + \half} \phi_\sh^{a_1\ldots a_s}(x)\,, \ee
where $\phi_\sh^{a_1\ldots a_s}$ is the leading rank-$s$ tensor field in
$|\phi_\sh\rangle$ (see \rf{phikdef01sh}).

We now analyze restrictions imposed by the first requirement. To this end we
note the relation
\beq  \label{man02-20072009-09}
\Cb |\phi(x,z)\rangle & = &  \int\!\! d^dy\, p_\nu F(x-y) W
|\phi_\sh(y)\rangle\,,
\\[5pt]
\label{man02-20072009-09x}
W  & \equiv  & \Cb_\perp - \frac{ p_{\nu+1}}{2p_\nu(2\nu+d)} e_{1,\sh}
\bar\alpha^2 \Box
\nonumber\\[5pt]
& + & \frac{p_{\nu-1}}{p_\nu} (2\nu+d-2) \eb_{1,\sh} \Pi^\smponetwo\,,
\eeq
where $\Cb$ is modified de Donder operator \rf{080405-01add}. Matching of
modified de Donder gauge condition for $AdS$ field $\phik$ and the
differential constraint for the shadow field $|\phi_\sh\rangle$ implies the
relation
\be  \label{man02-20072009-10}
\Cb|\phi(x,z)\rangle =  \int\!\! d^dy\, p_\nu  F(x-y) \Cb_\sh
|\phi_\sh(y)\rangle\,.\qquad
\ee
Comparison of \rf{man02-20072009-09} and \rf{man02-20072009-10} gives the
equation
\be \label{man02-app-d-01}\Cb_\sh = W\,. \ee
Comparing $\Cb_\sh$ given in Table II and $W$ given in
\rf{man02-20072009-09x} we see that Eq.\rf{man02-app-d-01} amounts to the
following equation for $p_\nu$:
\be  \label{man02-20072009-11} p_{\nu-1}(2\nu+d-2) = - p_\nu\,. \ee
Solution to this equation is given by
\be  \label{man02-20072009-12} p_\nu = (-2)^\nu \Gamma(\nu+\frac{d}{2})
p_0\,, \ee
where $p_0$ does not depend on $\nu$. Requiring normalization condition
\rf{man02-20072009-13}, we find

\be  \label{man02-20072009-14} p_0
=\frac{c_{\nu_s}}{(-2)^{\nu_s}\Gamma(\nu_s+\frac{d}{2})}\,. \ee
Plugging this $p_0$ in \rf{man02-20072009-12} we get
\be  \label{man02-20072009-15}
p_\nu = \frac{ (-)^{\nu_s-\nu} \Gamma(\nu+\frac{d}{2}) }{ 2^{\nu_s-\nu}
\Gamma(\nu_s+\frac{d}{2})} c_{\nu_s} \,.\ee
Taking into account \rf{man02-20072009-16}, \rf{man02-20072009-15} and
$c_\nu$ \rf{10072009-13}, we obtain solution for $\sigma_{s,\nu}$ given in
\rf{10072009-08x3}.

For the readers convenience, we note the formulas which are helpful for the
derivation of relation \rf{man02-20072009-09},
\beq \label{man02-20072009-07}
&& \hspace{-1cm}  e_1 |\phi(x,z) \rangle  =  -\!\! \int\!\! d^dy\, F(x-y)
\frac{p_{\nu+1}}{2\nu+d} e_{1\sh}\Box |\phi_\sh (y)\rangle\,,
\nonumber\\[-2pt]
\\[-5pt]
&& \label{man02-20072009-08} \hspace{-1cm} \eb_1 |\phi(x,z) \rangle  = -\!\!
\int\!\! d^dy\, F(x-y)
\nonumber\\[5pt]
&& \hspace{0.7cm} \times \,  p_{\nu-1}  (2\nu+d-2) \eb_{1\sh}|\phi_\sh
(y)\rangle\,,
\eeq

\beq
&& e_1 F(x)  = -\frac{1}{2\nu+d} \Box F(x) e_{1,\sh} \,,
\\[5pt]
&& \bar{e}_1 F(x)  = - F(x)(2\nu+d -2) \eb_{1,\sh} \,,
\eeq
where the operators $e_1$ and $\eb_1$ are defined in \rf{18052008-10}.

\end{document}